\documentclass[aip,jcp,reprint]{revtex4-1}
\usepackage{graphicx}
\usepackage{enumerate}
\usepackage{color}
\usepackage{multirow}
\usepackage{setspace}
\usepackage{amsmath,amssymb,gensymb}
\usepackage{upgreek,booktabs} 
\newcommand{\uml}{\upmu\mathrm{l}}

\begin{document}

\title {Controlling the crystal polymorph by exploiting the
time dependence of nucleation rates}

\author{Laurie J. Little}

\affiliation{Department of Physics, University of Surrey, Guildford, Surrey GU2 7XH, United Kingdom}

\author{Alice A. K. King}

\affiliation{Department of Physics, University of Surrey, Guildford, Surrey GU2 7XH, United Kingdom}
\affiliation{Department of Physics and Astronomy, University of Sussex, Brighton, BN1 9RH, United Kingdom}

\author{Richard P. Sear}

\email{r.sear@surrey.ac.uk}

\author{Joseph L. Keddie}
\affiliation{Department of Physics, University of Surrey, Guildford, Surrey GU2 7XH, United Kingdom}

\keywords{Crystallisation $|$ Polymorphism $|$ Crystal nucleation $|$ Crystal growth} 

\begin{abstract}
Most substances can crystallise into two or more different crystal lattices, called polymorphs. Despite this, there are no systems in which we can quantitatively predict the probability of one competing polymorph forming, instead of the other. We address this problem using large scale (hundreds of events) studies of the competing nucleation of the alpha and gamma polymorphs of glycine. In situ Raman spectroscopy is used to identify the polymorph of each crystal. We find that the nucleation kinetics of the two polymorphs is very different. Nucleation of the alpha polymorph starts off slowly but accelerates, while nucleation of the gamma polymorph starts off fast but then slows. We exploit this difference to increase the purity with which we obtain the gamma polymorph by a factor of ten. The statistics of the nucleation of crystals is analogous to that of human mortality, and using a result from medical statistics we show that conventional nucleation data can say nothing about what, if any, are the correlations between competing nucleation processes. Thus we can show that, with data of our form, it is impossible to disentangle the competing nucleation processes. We also find that the growth rate  and the shape of a crystal depends on when it nucleated. This is new evidence that nucleation and growth are linked.
\end{abstract}

\maketitle

\section{Introduction}

Most molecules can crystallise into more than one crystal lattice,
these crystal structures are called polymorphs of that molecule. 
A crystal's polymorph
defines its properties, and so polymorph control is crucial for intended applications. For pharmaceuticals,
polymorph control is also required by regulators; without it a crystalline
drug cannot be sold \cite{bauer01,morissette03}. 
We have almost no understanding of why one polymorph forms
and not another.
We cannot directly observe the process of crystal
nucleation, and we have very little quantitative data on the nucleation
of competing polymorphs.
Here, we provide quantitative nucleation data, use that data
to predict how to improve polymorph purity,
and successfully test this prediction.
We also study crystal growth, and find that its
rate depends on nucleation time; crystals of a given polymorph that nucleate at
late times are different from those that nucleate at early times.
This observation of a link between crystal properties and nucleation
time is novel, and potentially allows the control of crystal properties
via nucleation.

Our work is a contribution to a growing literature that studies nucleation
quantitatively \cite{little15,sear_rev14,diao12,toldy12,kim13,brandel15,akella14,duft04,laval09,javid16}.
In a previous publication, \cite{little15}, we measured glycine's nucleation rate but
did not quantitatively study the competition between glycine's polymorphs.
All earlier quantitative work on crystal nucleation, with the exception of that of Diao {\it et al.}\cite{diao12}, did
not consider polymorphism.
Diao {\it et al.}\cite{diao12} studied the polymorphs
of a molecule called ROY and  found that one polymorph dominated nucleation at early times, but competing polymorphs were more likely to nucleate at later times.

We report the results of large-scale (hundreds of nucleation
events), quantitative studies of the nucleation of crystals of glycine.
As nucleation is a random process, studying just one sample is not enough to characterise
a nucleation rate \cite{sear_rev14}. So we and others study hundreds of samples
\cite{little15,sear_rev14,akella14}, at each set
of conditions.
The novelty in our approach lies in our combination of
Raman spectroscopy to identify the polymorph of each crystal in situ, with
models taken from medical statistics.
We show that the statistics of nucleation, and the statistics of mortality
are analogous. Both nucleation and death are irreversible processes,
and both can have multiple causes, polymorphs in the case of
crystal nucleation, competing mortality causes such as lung cancer, heart
disease, etc, in the case of death.


\subsection{Previous work on glycine}

We choose glycine because it is a well studied model system for
studies of nucleation of crystals from solution
\cite{yang08a,little15, toldy12,han12, han15,
srinivasan07,srinivasan08,kim11myerson,forsyth16,poornachary08,toldy12,chew07,chen11,
jawor13,han13,he06,kim09glycine,nicholson05,nicholson11,javid16}.
Glycine has three polymorphs: $\alpha$, $\gamma$ and $\beta$.
$\alpha$ glycine is the most common polymorph to crystallise from
aqueous solution at neutral pH, but the $\gamma$ polymorph
is the most stable of the glycine polymorphs \cite{rivera08}. The $\beta$ polymorph is the least stable.
Adding salt favours
the $\gamma$ polymorph \cite{han15,srinivasan07,duff14,srinivasan08,yang08a}.

\begin{figure}[tb!]
(a)\hspace*{-0.0mm}\includegraphics[scale =0.68]{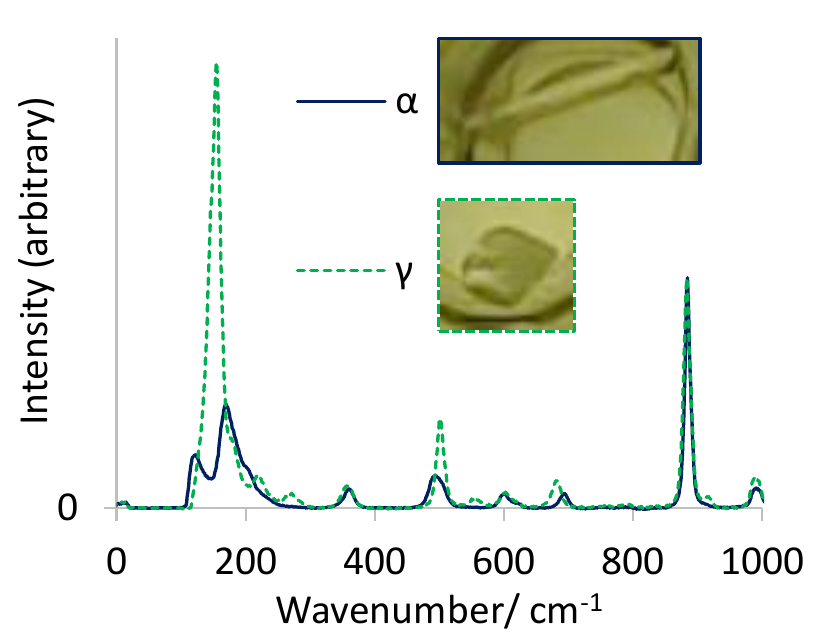}
\hspace*{-4.0mm}(b)\includegraphics[scale =0.08]{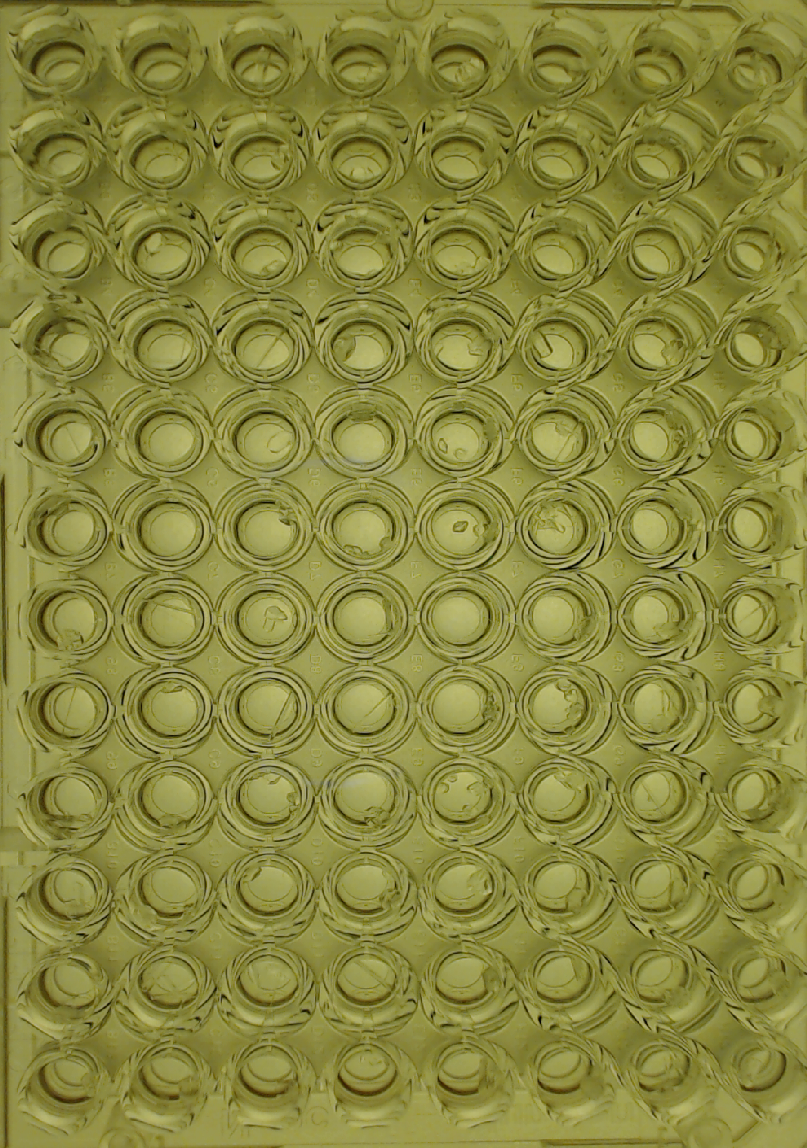}
\caption{(a) Raman spectra for $\alpha$ and $\gamma$ crystals, showing
that the spectra are distinct. We show a needle-like $\alpha$ crystal and
non-needle-like $\gamma$ crystal in the insets. 
(b) Image of one our 96-well plates, at end of the crystallisation. This is
for the first of three runs at a salt concentration of 250 mg/ml.
Crystallisation has occurred in 94 wells, with 43 $\gamma$ and 51 $\alpha$.
Each well is circular and has an internal diameter
of 6.8~mm.
}
\label{Raman_xtal}
\end{figure}

Glycine also has the advantage that the $\alpha$ and $\gamma$ polymorphs 
have clearly different Raman spectra\cite{shi05,sultana12,baran05}, in particular 
in the spectral region between 100 and 200~cm$^{-1}$, as
we can see in Fig.~\ref{Raman_xtal}(a).
The Supplementary Material has further details
and sets of Raman spectra for both polymorphs. The characteristic peaks of the $\beta$ polymorph
are not found.
This allows us to use Raman spectroscopy to identify polymorphs.
We can do this in situ on
crystals a millimetre or less in size, which allows us to identify the polymorphs
of the many hundreds of crystals we needed to study to obtain robust statistics.
See Fig.~\ref{Raman_xtal}(b) for an image of one of the 96-well plates
we used to study nucleation in a large number of samples in parallel.
We are not the first to use Raman spectroscopy to identify crystals. For example
Sultana and Jensen \cite{sultana12} used it in microfluidic studies of seeded crystallistion,
and Cui {\it et al.}\cite{cui16} used it to study the results of contact-induced nucleation.
Diao and coworkers \cite{diao12} have previously identified the polymorph of large numbers
of single crystals in situ, but their work
relied on the different polymorphs of the molecule ROY having very different visual
appearances.

There have been many studies of the growth rate of glycine crystals.
Measured growth
rates\cite{han15,han12,han10,li92,dowling10,toldy12,sultana12} for
glycine vary from $\mu$m/h to mm/s, a dynamic range of $10^6$;
see Fig.S1 of the Supplementary Material.
Growth rates increase rapidly with increasing supersaturation,
and measured growth rates also depend on the experimental set up.
The $\alpha$ polymorph consistently grows faster than the
$\gamma$ polymorph.  

In the next section, we will introduce the statistical tools needed to analyse the
nucleation of competing polymorphs. We then have two results sections, one on
nucleation and one on crystal growth.

\section{Materials and Methods}

Solutions of glycine and sodium chloride were made by adding deionized water
(Milli-Q, 18.2 M\ohm~cm) to solid glycine ($\geq$ 99\% HPLC from Sigma, cat. no. G7126)
and sodium chloride ($\geq$ 99.999\% from Sigma, cat. no. 38979).
This solid glycine consisted of mainly the $\gamma$
polymorph but had a small amount of the $\alpha$ polymorph.
We fully dissolved the glycine during preparation of the solution.
The solution was heated to 70 $\degree$C in a sealed vial and stirred at 1200 rpm for 1 h
using a hot plate and magnetic stirrer bar.
At all NaCl concentrations, the concentration of the glycine solution was fixed
at 320 mg of glycine per ml of water.
Solutions were held in the pipette for a time $t_H = 15$ s before distributing into the wells. 


\subsection{Crystallisation experiments}\label{sec:crystallisation_experiments}

Each experiment used a new microplate (Nunclon Delta Surface) with
96 wells (arranged as 8 rows of 12 wells, see Fig.~\ref{Raman_xtal}(b)).
Each well initially contained 100 $\uml$ of tridecane
($\ge 99\%$ from Sigma-Aldrich).
Then, 100 $\uml$ of the heated glycine solution was transferred simultaneously into each well in successive rows of wells using a 12-channel multipipette (Scipette).
The glycine solution was deposited on top of the tridecane after which it
would immediately sink below the oil because of its higher density.
The plate and samples cooled to room temperature within the first hour of the experiment.
The experiments were carried out in a temperature controlled room,
with a setpoint at 21$^{\circ}$C.
There are slight variations in temperature during the experiment,
as was discussed in our earlier work \cite{little15},
where there are further details
of our experimental setup.

\subsection{Determination of nucleation times}

Images of samples were recorded with a Logitech HD Pro Webcam C920 placed underneath the microplate, as in our earlier work \cite{little15}. An example image is shown
in Fig.~\ref{Raman_xtal}(b).
Images of the microplate were recorded at a resolution of
$1080\times 1920$ pixels.   
An image was recorded every 10 min for
the first 3 h, then every 30 min for the following 6 h, and then at every hour for the remainder of the experiment. The images were then
analysed by eye to determine the time at which each sample
crystallized. 

\subsection{Crystal sizes and growth rates}

We define the size of a crystal as being
the largest distance across a crystal,
as illustrated in Fig.~S2 of our earlier
work \cite{little15}.
The wells are 6.8 $\pm$ 0.2 mm in diameter, which is 90 pixels in our images,
so one pixel is approximately 0.076 mm across.
The smallest crystal we observed was two pixels diagonally across.
This corresponds to a center-to-center distance of $\sqrt{2} \times$ 0.076 = 0.11 mm.
This as an approximation of the smallest size at which a crystal can be detected.

We define needle-like crystals as follows.
The aspect ratio of crystals can change over time,
and initially needle-like crystals may become less needle-like
as time progresses.
Here, we define needle-like crystals,
as crystals that have an aspect ratio above 5:1 for at least
the first five hours after nucleation.

\subsection{Experiments heated after 18 h}

For one set of experiments, we heated
the filled microplate 18 hours into each experiment.
The microplates were heated in a Sanyo Mov-112F oven. The temperature
within the oven was measured to be 30.4 $\degree$C, which
remained constant to within 0.1$\degree$C. The samples were kept at this temperature in the oven for 48 h. The mass of the samples was measured before and after experiments to check for
evaporation of the water (tridecane is much less volatile).
The microplate was found to be 0.1 g lighter after the samples had been
in the oven for 48 h. This corresponds to evaporation
of 1\% of the total mass of glycine solution.

\subsection{Raman spectroscopy of glycine}

To identify the polymorph of a crystal, we used
an NTEGRA Raman microscope (NTEGRA, NT-MDT) equipped with a 20x objective lens,
and a 473 nm laser. The laser exposure time was 20~s for all samples.
The crystals remained in solution within the wells while the Raman spectroscopy was carried out. If a well contained more than one crystal, we obtained spectra from
two crystals in the well. The two spectra were always identified as the
same polymorph.
Raman spectra, together with the XRD patterns we used to validate
our use of Raman spectra to identify polymorphs,
are in the Supplementary Material.

\section{Statistics of the nucleation of competing polymorphs} \label{sec:med_stats}

Our droplets of supersaturated solution are sufficiently small that
in each droplet we only ever observe the nucleation
of one polymorph, either $\alpha$ or $\gamma$. In other words
nucleation of the two competing polymorphs are mutually exclusive events.
Thus each droplet contributes a time and a polymorph, e.g., nucleation
after 2 h, of the $\alpha$ polymorph.
Our data is therefore of exactly the form studied over many years
by medical statisticians who study competing causes of
mortality \cite{tsiatis75,peterson76,slud88,beyersmann09}.
Their data is also a time and a label, e.g., death
32 weeks after surgery, due to heart disease,
death after 22 weeks, due to cancer, etc.

Medical statisticians have developed powerful techniques and best practices
for analysing data of this form. We propose that
the study of competing polymorphs will benefit greatly from using the
methods developed in medical statistics.
The study of competing irreversible processes
is called survival data analysis
\cite{cox_book,lee_book,geskus_book}. Survival here means
the survival either of
a patient, or here of the solution state, i.e., survival until death or nucleation.
Here we
introduce the standard way of modelling data of our form.

Survival data in the presence of competing processes are typically plotted as what are called cumulative incidence functions (CIFs) \cite{geskus_book,beyersmann09}.
We record two CIFs,
one for the $\alpha$ polymorph, $I_{\alpha}$,
and one for the $\gamma$ polymorph, $I_{\gamma}$.
The CIF  $I_i(t)$ is defined as the probability
that nucleation has occurred at or before time $t$, and that
polymorph $i$ has nucleated. 
The survival probability $P(t)$ is the probability that nucleation has
{\em not} occurred at or before time $t$, so $P(t)=1-I_{\alpha}(t)-I_{\gamma}(t)$.

The effective nucleation rates for the two polymorphs,
$h_{\alpha}(t)$ and $h_{\gamma}(t)$, are examples of a type of function
called
cause-specific hazard functions (CSHs) \cite{geskus_book,beyersmann09}.
The CSH for the nucleation of polymorph $i$, $h_{i}$, is \cite{dignam12,beyersmann09}
\begin{equation}\label{eq_hazard}
h_i(t)=\frac{1}{P(t)}\frac{{\rm d}I_i(t)}{{\rm d} t} ~~~~~~i=\alpha, \gamma
\end{equation}
In words: the effective nucleation rate of a polymorph is the time
derivative of the polymorph's CIF,
divided by the fraction of wells remaining
uncrystallised, $P(t)$.

Both CIFs and CSHs are observables:
CIFs are what we measure, and CSHs are
time derivatives of the CIFs.
It is important to note that although we refer to $h_{\alpha}$ as an effective
nucleation rate for the $\alpha$ polymorph,
$h_{\alpha}$ also depends on $\gamma$ nucleation.
We only observe one polymorph in a well, so once a $\gamma$ crystal has formed
in a well, $\alpha$ nucleation will not occur.
The two competing processes are entangled, we cannot separate them out, and
so care is needed in interpreting $h_{\alpha}$ and $I_{\alpha}$.
This problem is discussed by Geskus \cite{geskus_book}
and by Beyersmann {\it et al.}~\cite{beyersmann09}.


\subsection*{Models that include only observables}

As CSHs are observables, if we directly model these then we have a model that
uses only experimentally observable functions.
We model the CSHs (the effective nucleation rates)
as power laws, with exponents $\beta_i$
and characteristic timescales $\tau_i$. These CSHs have the form
of the hazard function of the widely used \cite{sear_rev14} Weibull function:
\begin{equation}
h_i(t)=
\beta_i\left(t^{\beta_i-1}/\tau_{i}^{\beta_i}\right) ~~~~
~~~~~~i=\alpha,\gamma
\label{obs_model2}
\end{equation}
This model includes the time-independent nucleation rate as a
special case ($\beta_i=1$).

\subsection{Models with latent nucleation times}

Another approach to modelling our competing nucleation processes is
to associate two nucleation times with every
droplet \cite{tsiatis75,beyersmann09,peterson76,slud88,geskus_book}:
a nucleation time for the $\alpha$ polymorph, $t_{nuc,\alpha}$,
and a nucleation time for
the $\gamma$ polymorph, $t_{nuc,\gamma}$.
We only observe the shorter of these two times. The first nucleation event
preempts nucleation of the other polymorph, and so we do not observe the longer
time. This longer time is then latent, i.e., hidden, and
so is not an experimental observable.

Models with latent nucleation times are natural choices for studying possible
correlations between the competing processes, and hence for making
predictions of what we would observe if we could somehow prevent
one polymorph nucleating.
Correlations would arise if both polymorphs are nucleating
on the same impurity, or if the less stable polymorph forms as a transient
intermediate to the formation of a more stable polymorph.

\begin{figure}[tb!]
\centering
\includegraphics[scale = 0.85]{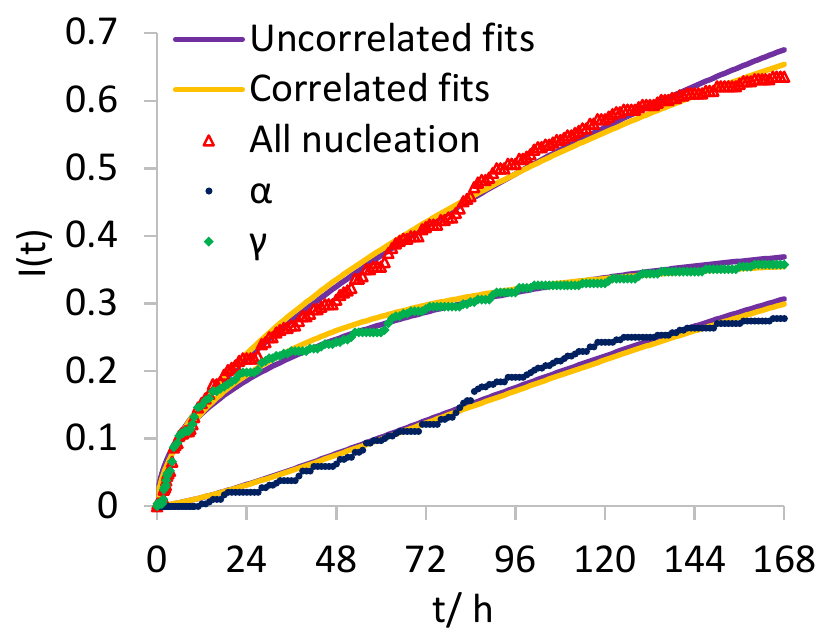}
\caption{Plot of CIFs at the salt concentrations of 250 mg/ml.
The points are our data:
$I_{\alpha}$ (blue), $I_{\gamma}$ (green) and
$I_{\alpha}+I_{\gamma}$ (red), respectively.
Purple curves 
are fits to a model in which the
values of $t_{nuc,\alpha}$ and $t_{nuc,\gamma}$ for a single well are uncorrelated,
and are taken from a Weibull distribution. This is equivalent to a fit
to Weibull CSHs.
The yellow curves are from a fit of a model in which the values of $t_{nuc,\alpha}$ and
$t_{nuc,\gamma}$ for a single well are correlated with a Spearman's rank coefficient
of 0.95.
} \label{Induction_curves}
\end{figure}

\section{Results for the competing nucleation of glycine's polymorphs}

\subsection{Time dependence of the nucleation rates}

In Fig.~\ref{Induction_curves},
we have plotted the CIFs for experiments at a salt concentration of 250 mg/ml.
The $\alpha$ and $\gamma$ nucleation rates
have opposite time dependencies: $\gamma$ nucleation starts fast,
$I_{\gamma}$ has an initial steep rise, but then it slows, while
$\alpha$
nucleation starts slow but accelerates.
Toward the end of the experiment, $I_{\alpha}$
almost catches up with $I_{\gamma}$.

\begin{figure}[ht!]
\centering
(a)\includegraphics[width=3.3cm]{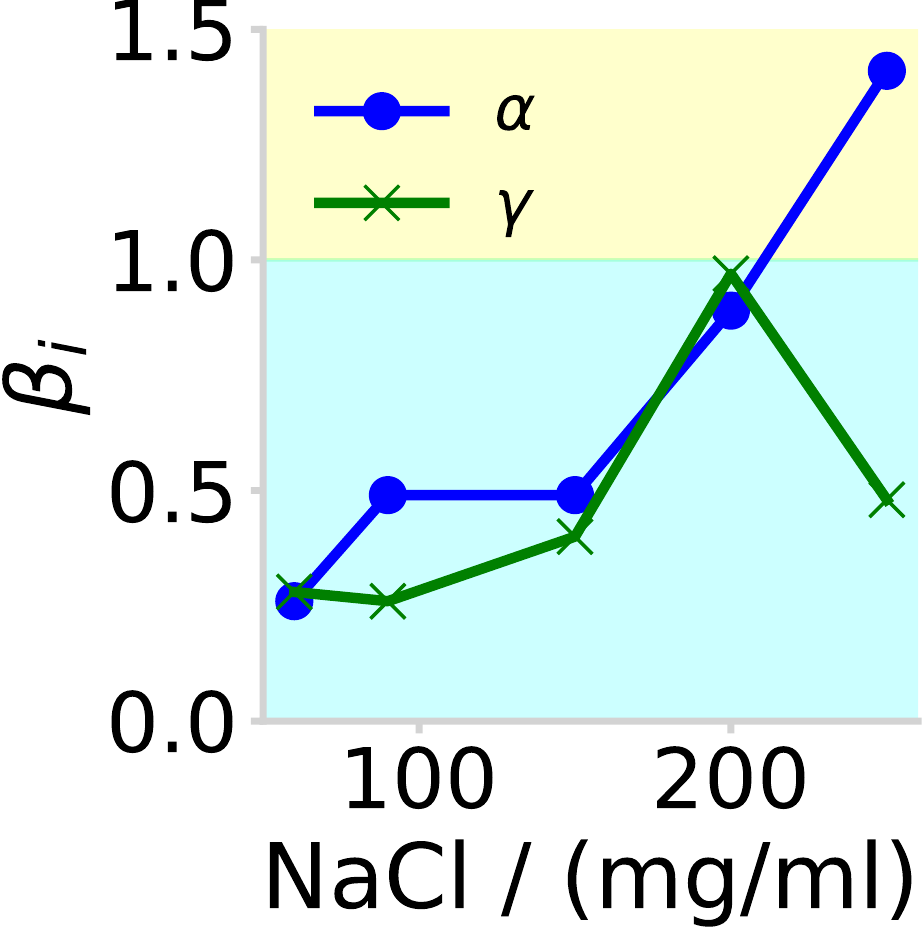}
(b)\includegraphics[width=3.3cm]{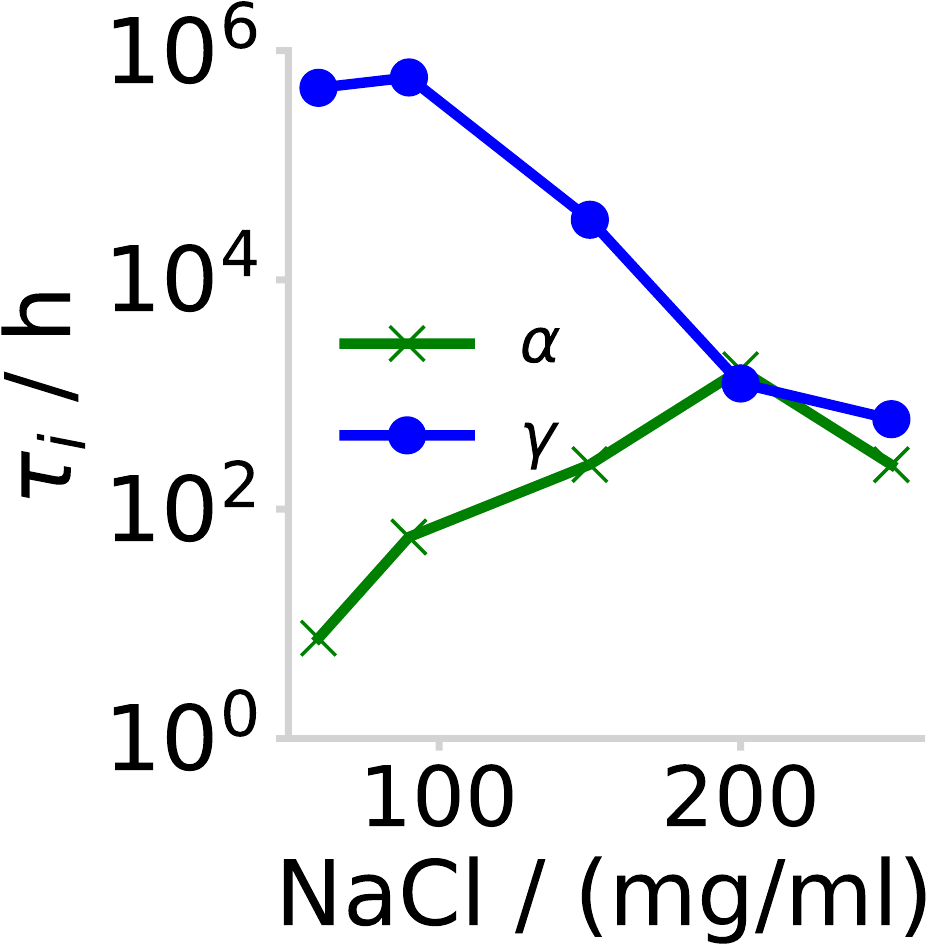}
\caption{The Weibull exponents $\beta_i$ (a), and
$\tau_i$ (b) for fits to the nucleation rates
of both polymorphs, as functions of the concentration of NaCl.
In (a) we have shaded in yellow the region where the nucleation
rate increases with time, and in blue the region where it decreases.
}\label{beta_salt}
\end{figure}

To quantify the variation in the nucleation rates with time, we have
fitted models with Weibull-type CSHs (equation (\ref{obs_model2}))
to the experimental CIF curves. The
best-fit parameter values are plotted in Fig.~\ref{beta_salt},
the numbers are in Table S2.
A value of $\beta<1$ gives an effective nucleation rate that slows down
as time passes, while $\beta>1$ means that the rate
accelerates with time.
At 250 mg/ml salt,
$\beta_\gamma=0.48$ and
$\beta_\alpha=1.41$, which confirms that $\gamma$ nucleation
slows with time while $\alpha$ nucleation accelerates.

The nucleation rate for the $\alpha$ polymorph
only accelerates with time at high salt.
At low salt the effective nucleation rate
for $\alpha$, the CSH $h_{\alpha}$, actually decreases with time;
the best-fit values of $\beta_{\alpha}$ in Fig.~\ref{beta_salt}
are less than one except
at the highest salt concentration.
Increasing the salt concentration not only
slows down the timescale for $\alpha$ nucleation, $\tau_{\alpha}$, it
also qualitatively changes the kinetics of nucleation, from initially fast nucleation
that slows, to initially slow nucleation that accelerates.
Plots of CIFs for five salt concentations are shown in Fig.~S7 of the Supplementary Material.
Increasing the salt concentration dramatically decreases the
apparent timescale
for $\gamma$ nucleation, as shown in Fig.~\ref{beta_salt}(b).
At high salt concentrations
a larger fraction of the crystals are of $\gamma$ polymorph
(Fig.~S8).

We note that our lowest salt concentrations of 60 mg/ml
significantly accelerates nucleation with respect
to no added salt, which is why we work
at a glycine concentration of 320 mg/ml,
lower than the
333 mg/ml where we obtained
most of our data in our earlier work \cite{little15}.
But then at higher salt concentrations the rate slows down again.
The nucleation
rate does not vary monotonically with salt concentration.


The CIFs in
Fig.~\ref{Induction_curves} are obtained
by adding together three individual runs.
The CIFs for individual
runs are shown in
Fig.~S9(e) in the Supplementary Information.
The shapes of the CIFs for individual runs are the same as
in Fig.~\ref{Induction_curves}, so our conclusions
on the time dependencies of $\alpha$ and $\gamma$
nucleation are highly reproducible.
There is however significant run-to-run variability
in the total amount of nucleation we observe,
and to a lesser extent in the fractions of the
polymorphs that form. This was also found in earlier
work \cite{little15} and is common in crystal nucleation.
For the three runs that make up the data in
Fig.~\ref{Induction_curves} the final fractions
of the crystals that formed
that were of the $\gamma$ polymorph that form are 47\%, 40\% and 72\%,
while the final fraction of wells where crystallisation occurred
was 98\%, 45\% and 48\%. Reproducibility is comparable
at other salt concentrations, see Supplementary Material.

\subsection{We cannot
determine if the nucleation of the $\alpha$ and $\gamma$ polymorphs is correlated}

One obvious question about the nucleation of the
competing polymorphs is: Are $\alpha$ and $\gamma$ nucleation correlated?
By correlated, we mean that if
in an individual well the $\alpha$ polymorph is likely to nucleate early, is the $\gamma$ polymorph also likely to nucleate early?
We expect that the nucleation in our samples is heterogeneous, i.e., occurring on impurities, and so if the same impurities tend to induce nucleation of both polymorphs, we would expect nucleation of the polymorphs to be correlated.


\begin{table}[ht]
\centering
\small
\caption{The best-fit values for fits of Fig.~\ref{Induction_curves}.
 Note that the uncorrelated values refer to both
the case of uncorrelated latent nucleation times, and to the
case of Weibull CSHs, the two cases are mathematically equivalent.
$R_i^2$, $i=\alpha$, $\gamma$, is 
the $R^2$ value for comparison of the fit $I_i$ to the data.}
    \begin{tabular}{ c | c c c c | c c }

\midrule
	& $\beta_{\alpha}$ & $\beta_{\gamma}$ & $\tau_{\alpha}$~(h) &
$\tau_{\gamma}$~(h) & $R_{\alpha}^2$ & $R^2_{\gamma}$ \\
 \midrule
	Uncorrelated 	& 1.41 & 0.48 & 244 & 610 & 0.99  & 0.99 \\ \hline
		Correlated 	 & 0.92 & 0.60 & 164 & 260 & 0.98  & 0.99  \\

 \bottomrule
	\end{tabular}

\label{r_table}
\end{table}

Unfortunately, Tsiatis \cite{tsiatis75} has
rigorously proved that from data of our form
(nucleation time plus polymorph for each well),
we cannot determine if the competing processes are independent of each other, or are correlated.
We have illustrated this in Fig.~\ref{Induction_curves},
where there are fits of our data using models with and without correlations.
The two fits are essentially as good as each other;
the $R^2$ values in Table \ref{r_table}, are almost identical.
The model with correlations was based on latent nucleation times;
details are in the Supplementary Information.

We can see in Table
\ref{r_table} that when the hazard functions are correlated, the best fit $\beta$ and $\tau$ parameters are significantly changed.
Both $\beta_{\alpha}$ and $\beta_{\gamma}$ are now less than one, so
if there are strong correlations, the data is consistent with both
nucleation rates decreasing with time.
We do not know the degree of correlation between $\alpha$ nucleation and $\gamma$ nucleation, and multiple $\beta$ and $\tau$ parameters will fit the data equally well, but with different correlations.
This means that with data of this type,
we can say almost nothing about what we would observe in the hypothetical cases where we could prevent nucleation of one polymorph,
and just observe nucleation of the other \cite{tsiatis75,beyersmann09,peterson76,slud88,geskus_book}.

\begin{figure}[ht]
\centering
\includegraphics[scale = 0.85]{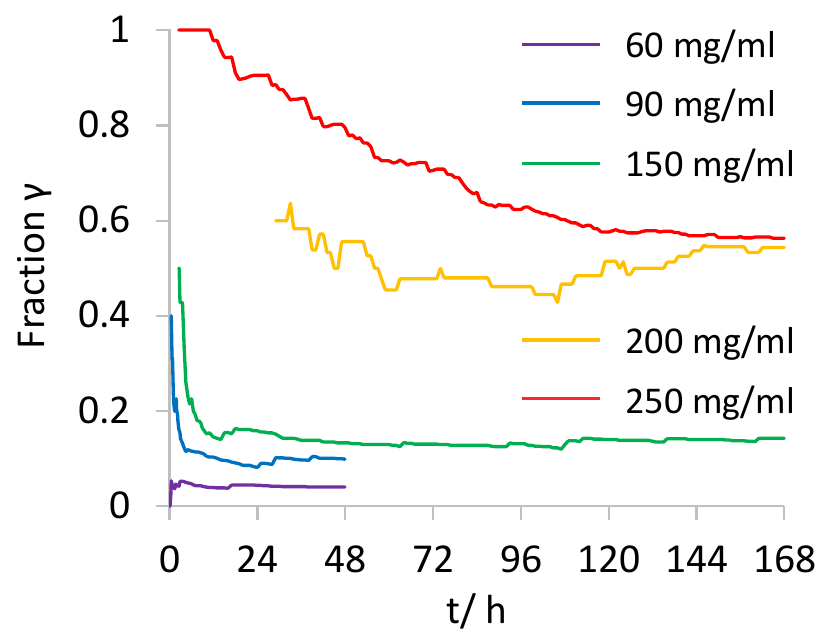}
\caption{Polymorph composition as a function of time. The composition is the
fraction of the wells where crystallisation has occurred, that contain the $\gamma$ polymorph.
The lines start at the point when a total of ten nucleation events have occurred.
Each curve is obtained from all runs at that salt concentration.
} \label{purity_vs_time}
\end{figure}

%

\subsection{Polymorph composition vs time} \label{sec:polymorph_composition}

At high salt, the two hazard functions
$h_\alpha$ and $h_\gamma$ have opposite time dependencies,
which implies that the polymorph composition varies with time.
We have plotted the polymorph composition as a function of time in
Fig.~\ref{purity_vs_time}.
We plot $n_\gamma (t)/[n_\alpha (t) + n_\gamma (t)]$,
where $n_\alpha (t)$ and $n_\gamma (t)$ are the number of wells containing $\alpha$ crystals and $\gamma$ crystals at time $t$, respectively.

If we consider the curve for 250 mg/ml NaCl, we see that
the fraction of $\gamma$ starts at one (pure $\gamma$) and
decreases towards its final value of 0.56. This is consistent
with our Weibull fits that have exponents $\beta_{\alpha}>\beta_{\gamma}$
(see Fig.~\ref{beta_salt}), and so give a $h_{\alpha}(t)/h_{\gamma}(t)$
that is
an increasing function of time. We have many fewer $\gamma$ nucleation events
at other salt concentrations, so our conclusions are less reliable there, but
it seems likely that at all salt concentrations the fraction of the $\gamma$
polymorph decreases with time.
We are not the first to observe a polymorph purity that changes with
the time, Diao {\it et al.}~\cite{diao12} found this
behaviour for the molecule ROY (see their Figure~8).


\subsection{Time-dependent supersaturation increases polymorph purity}
\label{sec:isothermal_vs_non-isothermal}

When producing crystals, the aim is often to obtain a particular polymorph with high purity.
We observed that at a salt concentration of 250 mg/ml, $\gamma$ glycine nucleation dominated at early times, but then the $\alpha$ nucleation rate increased.
This suggests that stopping
nucleation early in the experiment will
increase $\gamma$-glycine purity. To test this hypothesis we
performed additional experiments at 250 mg/ml NaCl.
In these experiments we reduced the supersaturation after 18 hours, by increasing the temperature from room temperature (21 $\degree$C) to 30.4 $\degree$C.
This had the effect of almost completely halting nucleation after 18 hours;
in two runs only one crystal nucleated after 18 hours.

The final fraction of crystals
that are in the $\gamma$ polymorph increases from
$0.56\pm0.04$ to $0.94\pm0.02$. Here, the first number is the final fraction
of $\gamma$ crystals at the end (168 h) of three isothermal runs,
and the second number is the fraction of $\gamma$ crystals after 18 h, of all
five runs.
The CIFs for these experiments are
in Fig.~S10.



\begin{figure}[tb]
\centering
\includegraphics[scale=0.85]{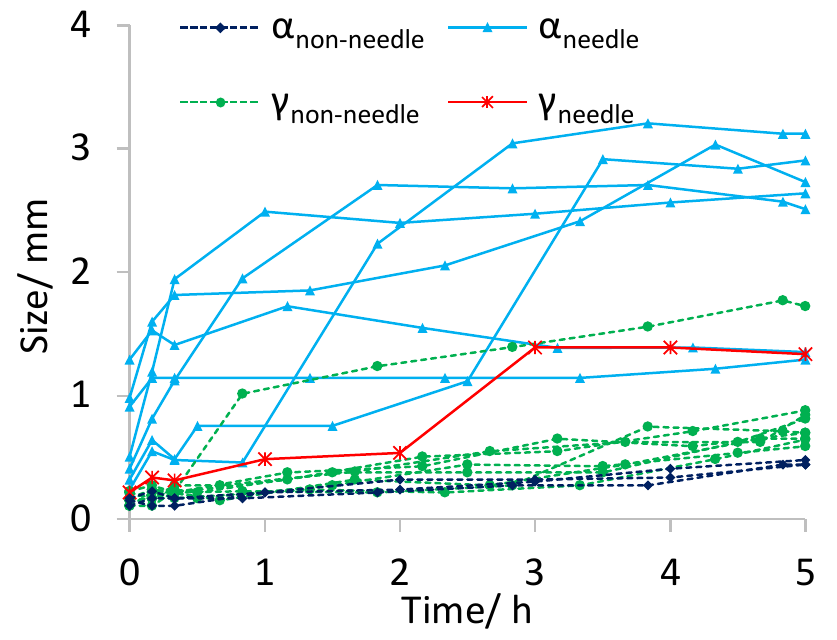}
\caption{Plot of the maxium distance
across a crystal, as a function of time since it nucleated.
Data is shown for ten $\alpha$ and ten $\gamma$ crystals.
These are seven $\alpha$ needle crystals plus three $\alpha$ non-needle crystals,
and one $\gamma$ needle plus nine $\gamma$ non-needles.
For each crystal $t = 0$ is defined as the time of the first image in which there is a visible crystal.
The salt concentration is 250 mg/ml.
} \label{growths}
\end{figure}

\section{Results for the growth rates and crystal habits of glycine crystals}

Having considered nucleation, we will now turn to consider
crystal growth. We will present results that show
that nucleation and growth are linked.
We have plotted the sizes of 10 $\alpha$ and 10 $\gamma$ crystals,
as a function of time, in Fig.~\ref{growths}.

\begin{figure}[tbh]
\centering
(a)\includegraphics[scale=1.1]{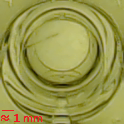}
(b)\includegraphics[scale=1.1]{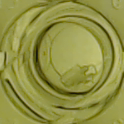}
\caption{Here we show some typical crystal habits.
(a) is a needle-like $\alpha$ crystal,
observed less than an hour after nucleation.
(b) is a non needle-like $\alpha$ crystal 
observed 5 h after nucleation.
The scale bar in (a) applies to both images.
}
\label{habit}
\end{figure}

Note that we estimate the size of a crystal, by using the maximum length across the crystal;
see the Materials and Methods section,
and our earlier work \cite{little15} for details of how we measure size.
Some crystals grow as needle-like shapes, while others
have more compact habits.
We show example crystals with
needle and non-needle habits in Fig.~\ref{habit}.

Three things are clear from Fig.~\ref{growths}:
i) The growth rates of
 the linear dimension of both polymorphs vary between one crystal and another.
ii) Much of this variability is related to crystal habit.
Needle-like crystals have much faster growth rates 
along their long axis
than do non-needle-like crystals.
iii) Growth of the $\alpha$ polymorph
is generally faster than the $\gamma$ polymorph
(in agreement with other studies \cite{han12}).
 Note that growth of the needles perpendicular
to their long axis is quite slow, approximately comparable to the growth rate of
non-needle-like crystals. Therefore, although one axis of the needles
grows very rapidly, the growth of the volume of these crystals
may not be faster than the growth of the volume of the non-needle-like
crystals.

We do not know why the growth rate of crystals growing under identical conditions
is so variable.
Crystal growth often relies on defects \cite{davey_book},
and so it may be due to different crystals acquiring different defects as they
grow. But whatever the cause, this variability has consequences. For example,
any crystallisation model that assumes a single well-defined growth rate at
a given supersaturation is clearly very far from the truth for glycine.


\begin{figure}[ht]
\centering
\includegraphics[scale=0.75]{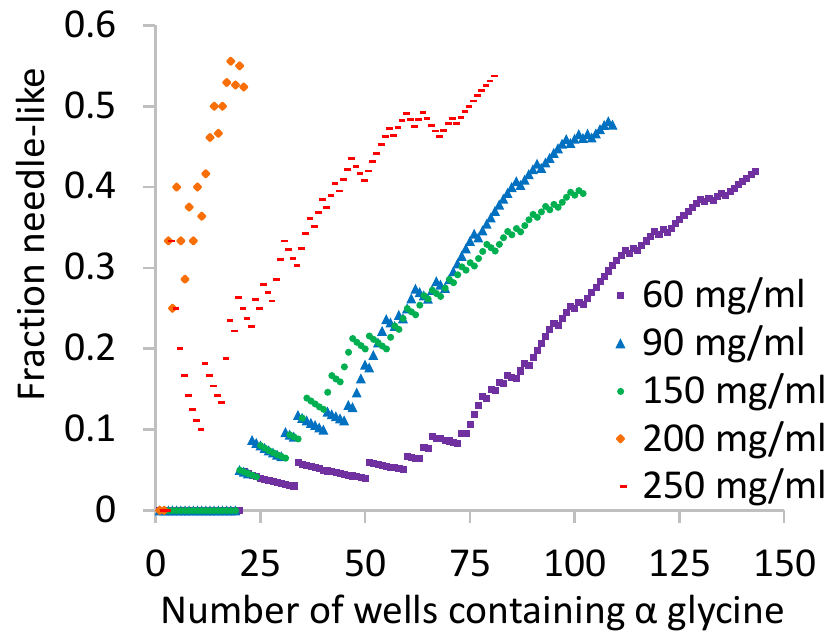}
\caption{For the $\alpha$ polymorph, we plot the fraction of crystals
 with a needle-like habit, as a function of the total
number of wells that contain $\alpha$ crystals.
}
\label{needle-like_induction_times}
\end{figure}

\subsection{Nucleation time and crystal habit}

Next we look at whether the habit of a crystal depends on the time at which it nucleates.
In Fig.~\ref{needle-like_induction_times}
we plot the fraction of crystals of the $\alpha$ polymorph
that are needle-like, as a function of the total number of wells with $\alpha$ crystals.
The $x$-axis is effectively a time axis, as the number crystallised increases with time,
but using the number crystallised as the $x$-axis allows easier comparison between
different salt concentrations.

At early times almost all crystals
have the non-needle-like habit. However, at later times most crystals that nucleate are needles,
so the fraction of needle-like crystals increases rapidly.
There is a very strong dependence of crystal habit
on nucleation time. This is due to these late-nucleating crystals
having much faster growth along one axis (only).
 For example at 60 mg/ml, for the first 70
$\alpha$ nucleation events,
six are needle-like crystals, but for the following 70 nucleation events,
51 are needle-like.
For $\gamma$ crystals at high salt concentrations,
we see the same trend, see Fig.~S14,
but as we have few needle-like $\gamma$ crystals, our statistics are poorer.

 We were surprised when we found that typical late-nucleating crystals have a very
different growth rate along one axis than typical early-nucleating crystals.
We analysed the Raman spectra for needles and non-needles, see Supplementary Material.
Although there is crystal-to-crystal variability in the Raman spectra,
there is no sign of any systematic difference between the set
of the spectra of needles, and the set of spectra of the non-needles.
So via Raman spectroscopy we separated our crystals into two sets: one set with
spectra chracteristic of the $\alpha$ polymorph, and one
with spectra characteristic of the $\gamma$ polymorph \cite{shi05}. Both sets
of crystals contain needles and non-needle-like crystals.
Our XRD patterns (Fig.~S3(a) and Fig.~S4) confirm this finding.

 Finally, note
that in these XRD patterns we only ever observe peaks characteristic
of the $\alpha$ and $\gamma$ polymorphs; characteristic peaks for
$\beta$ are absent (Fig.~S4). In addition, the $\beta$
polymorph is not typical for crystallisation from aqueous solutions, and indeed
under these conditions is expected
to be unstable with respect to conversion to the $\alpha$ form \cite{ferrari03}.
We do not observe visible changes in habit that might suggest polymorph-to-polymorph conversion.

We do not know what causes the growth and habit difference between early- and late-nucleating crystals.
However, if it is due to differences in defects that control crystal growth, then this
suggests that nucleation may involve more than one step, and that at least one of these
rate limiting steps involves nucleation of a growth defect. It
is possible that our initially slow nucleation of the $\alpha$ polymorph at high salt concentration is
associated with slow or multistep defect nucleation, which is needed before that
polymorph can grow.



\section{Conclusions}

We have studied the competing nucleation of two crystal polymorphs of glycine, using
Raman spectroscopy to identify the polymorph of individual crystals.
We found that at high salt,
the nucleation behaviour of the competing polymorphs is analogous to
the competition of the tortoise and the hare in Aesop's story. In the story,
at first the hare runs faster than the tortoise, but
then the hare goes to sleep, allowing the tortoise to overtake it.
Like the hare, $\gamma$ nucleation is initially fast but then slows,
while $\alpha$ nucleation accelerates during the experiment,  and so
the rate of $\alpha$ nucleation overtakes that of $\gamma$ nucleation.

This observation can be exploited.
Drawing on this observation, we predicted that reducing the supersaturation
after 18 h would increase the purity of the $\gamma$ polymorph.
It did, the fraction of the $\alpha$ impurity was reduced from 44\% to 4\%.
Thus, the complex time dependence of competing polymorphs can be exploited
to greatly increase the purity of a polymorph.

By using a result from medical statistics \cite{tsiatis75}, we were able
to prove that with our data it is impossible to disentangle
the nucleation of $\alpha$ and $\gamma$
polymorphs \cite{tsiatis75,peterson76,slud88,beyersmann09,geskus_book}.
These two processes may or may not be strongly correlated --- we simply do not know.
This means that care must be taken when interpreting results on competing
polymorphs. For example, if we alter an experimental parameter,
such as salt concentration, pH, etc, and observe more of one polymorph,
that does not allow us to infer that the nucleation rate of that polymorph
has increased, the rate of the competing polymorph may have
decreased \cite{slud88}. Work by medical statisticians over the last
40 years has found that interpretation of data of our form is subtle, and
so we will need to learn from their work, in order to correctly
understand the competition between polymorphs.

A possible way around this problem is to use the experimental approach of
Laval {\it et al.}~\cite{laval09}, who cycled their system,
to repeatedly induce crystallisation and dissolution
of crystals, in a set of droplets produced by microfluidics.
They did not consider
polymorphism, but if their approach can be applied
to a system with competing
polymorphs, then the two competing processes could
then be disentangled.

All our results are obtained from system volumes of 0.1~ml, and
in common with most earlier work, we do not know how
varying the volume will change what is seen. However, we
do note that our droplets are large enough
that we often observe two or more
crystals, and that in all cases studied (via Raman) all the crystals
in the same droplet
are of the same polymorph. This strongly suggests that
the nucleation events that produce multiple crystals in a
single well are not independent. The second, third, etc,
crystals may be forming via secondary nucleation \cite{botsaris76,agrawal15},
i.e., arising due to the already-existing crystal.

In addition to nucleation, we also studied crystal growth.
For both polymorphs we observed fast growing needle-like crystals,
and much slower growing non-needle-like crystals.
Nucleation time, crystal habit and growth rate are all correlated.
Crystals that nucleate at late times tend to
have needle-like habits, and to grow rapidly.
This appearance of the fastest growing crystals at late times is the opposite
to what we would expect. Further work will be needed to understand the
mechanism that underlies this interesting observation.

\section*{Supplementary Material}

The Supplementary Material contains: (I) Literature data on the growth rates of glycine crystals.
(II) The solubility of glycine in water, as a function of temperature.
(III) Raman spectra and XRD patterns of our crystals, as well as reference XRD patterns of all
three polymorphs, for comparison.
(IV) Details of the statistical techniques we borrow from the field of medical statistics,
to model our data.
(V) Additional experimental results for nucleation.
(VI) Additional experimental results for growth.

\acknowledgements{We thank Violeta Doukova (Surrey) for technical assistance, Dan Driscoll (University of Surrey)
for help with the X-ray diffraction, and Sharon Cooper (University of Durham), Baron Peters
(University of California Santa Barbara) and Jan Sefcik (University of Strathclyde)
for helpful discussions.
We acknowledge EPSRC for funding a PhD studentship for LJL.}

\section*{Data Availability}
The authors confirm that data underlying the findings are available
without restriction at https://doi.org/10.6084/m9.figshare.c.3898480.


%

\onecolumngrid
\setcounter{section}{0}
\renewcommand{\figurename}{Figure}
\renewcommand{\tablename}{Table}

\setcounter{figure}{0} \renewcommand{\thefigure}{S\arabic{figure}}
\setcounter{table}{0} \renewcommand{\thetable}{S\arabic{table}}

\begin{center}
\large{\bf Supplementary Material for:
Controlling the crystal polymorph by exploiting the time dependence of
nucleation rates
}
\end{center}

This supplementary material has six sections:
\begin{enumerate}[(I)]
\item We discuss literature values of glycine crystal growth rates
in the main text. In this section we plot the
rates as a function of supersaturation and we give
details of the experimental conditions
in these studies.
\item This contains plots of the solubility of glycine in water, as a function of temperature.
\item This section contains Raman spectra of the
$\alpha$ and $\gamma$ polymorphs of glycine. We also show the
X-ray diffraction (XRD) results that
validate Raman spectroscopy's ability to distinguish
between the two polymorphs of glycine.
\item Background to the statistics needed to understand
the nucleation of competing polymorphs, and
details of the models we used.
\item Additional experimental results for nucleation.
\item Additional experimental results for growth.
\end{enumerate}


\begin{figure}[h]
\singlespace
\centering
\includegraphics[scale=1]{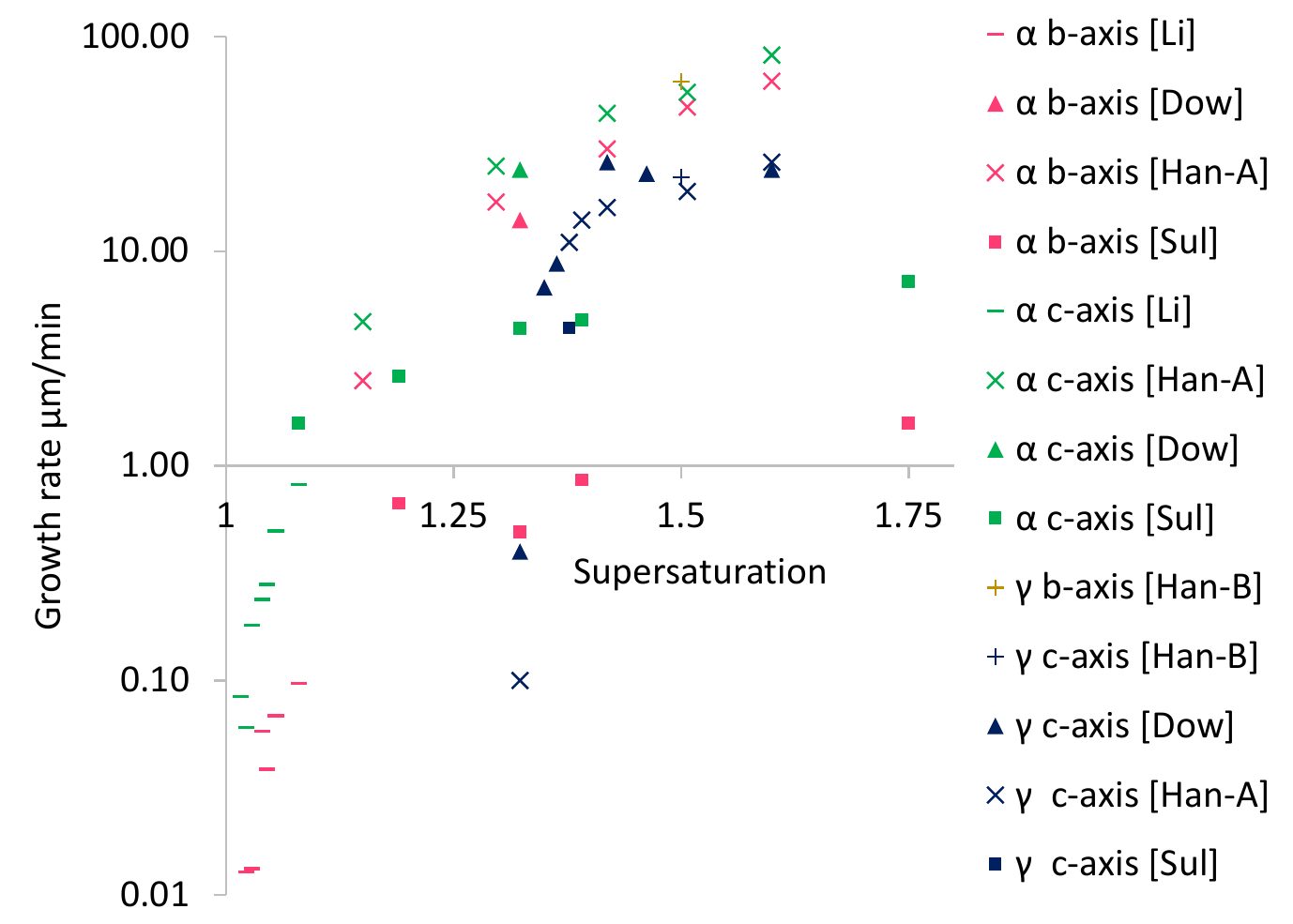}
\caption{Glycine crystal
growth rates, are plotted as a function of supersaturation.
This is data from five earlier studies:
[Li] (Li et al.) \cite{li92}, [Han-A] (Han  et al.)  \cite{han12}, [Han-B] (Han et al.)\cite{han15}, [Dow] (Dowling et al.) \cite{dowling10} and [Sul] (Sultana et al.) \cite{sultana12}. These growth rates have been obtained from graphs, so are approximate. Colour coding is as follows, $\alpha$ b-axis, $\alpha$ c-axis, $\gamma$ b-axis and $\gamma$ c-axis are pink, green, gold and dark blue respectively. [Sul] used an anti-solvent, other studies are purely aqueous.} \label{literature_growths}
\end{figure}

\linespread{1.15}

\section{Literature studies of the growth of glycine crystals}

In
the main text we discussed literature values for the growth rates of glycine crystals. These rates are plotted in Figure \ref{literature_growths}. There was one growth rate \cite{toldy12},
of $\approx 80,000$~$\mu$m/min, which was too large to fit on our graph.
The experimental details for these experiments are summarised in
Table \ref{table:growths}.

\begin{table}[thb!]
\caption{The experimental details of a number of studies on glycine growth. The growth rates from these studies are shown in
Figure \ref{literature_growths}.}\label{table:growths}
\small
\begin{center}
	\begin{tabular}{| p{2.8cm} | p{2.6cm} | p{10cm} |} \hline
Authors & Supersaturation & Experimental set-up \\
 & range & \\ \hline

\multirow{2}{*}{Toldy et al.\cite{toldy12}} & \multirow{2}{*}{$3.5 - 6.5$} & Crystals nucleate in supersaturated droplets within an emulsion at 84$\degree$C. Crystals sizes are of the order of tens of $\mu$m. \\ \hline

\multirow{3}{*}{Dowling et al. \cite{dowling10}} &  \multirow{3}{*}{$1.3 - 1.45$} & Individual crystal seeds are placed in supersaturated solution, and observed under a microscope at 20$\degree$C. Crystal sizes are of the order of mm. 
Solubility $c_s$ is given as 212 g/l. \\ \hline

\multirow{3}{*}{Han et al. \cite{dowling10}} & \multirow{3}{*}{$1.15 - 1.6$} & Individual crystal seeds are placed in supersaturated solution and observed under a microscope at 23$\degree$C. Crystal sizes are of the order of mm.
Solubility $c_s$ is given as 226 g/l. \\ \hline

\multirow{4}{*}{Sultana et al. \cite{sultana12}} & \multirow{4}{*}{$1.08 - 1.75$} & Supersaturated solutions containing crystal seeds flow through a PDMS (poly-dimethylsiloxane) microfluidic device. Solutions are supersaturated using the anti-solvent methanol. Crystal sizes are of the order tens of $\mu$m \\ \hline

\multirow{3}{*}{Li et al. \cite{li92}} & \multirow{3}{*}{$1.01 - 1.08$} & Crystal seeds are placed in a glass cell at 23$\degree$C while supersaturated solution flows through the cell, seed sizes are of the order of hundreds of $\mu$m\\ \hline

Han et al.\cite{han15} & {1.5} & Set-up is the same as in
Dowling et al.  \cite{dowling10}  \\ \hline
	\end{tabular}
\end{center}
\end{table}

Crystal growth rates appear to be very sensitive
to crystal and solution properties, such as crystal size and solvent.
The growth rates of $\alpha$ glycine measured by Sultana et al. are lower than those measured by  Han et al. and Dowling et al. In the work of
Sultana et al., the glycine solution is supersaturated by the addition of the anti-solvent methanol, while in the experiments of Han et al. and Dowling et al. only water and glycine are present. Also, both Dowling and Han work with crystals with sizes of the order of mm, while Sultana et al.~work with crystals of sizes of the order tens of $\mu$m.
It may be that the presence of methanol is affecting the growth of crystals
in the experiments of Sultana et al, and the crystal growth rate may also
be changing with crystal size.

\section{Glycine solubility}

\begin{figure}[ht!]
\centering
(a)\includegraphics[trim={45mm 114mm 145mm 30mm}, clip, scale=0.78]{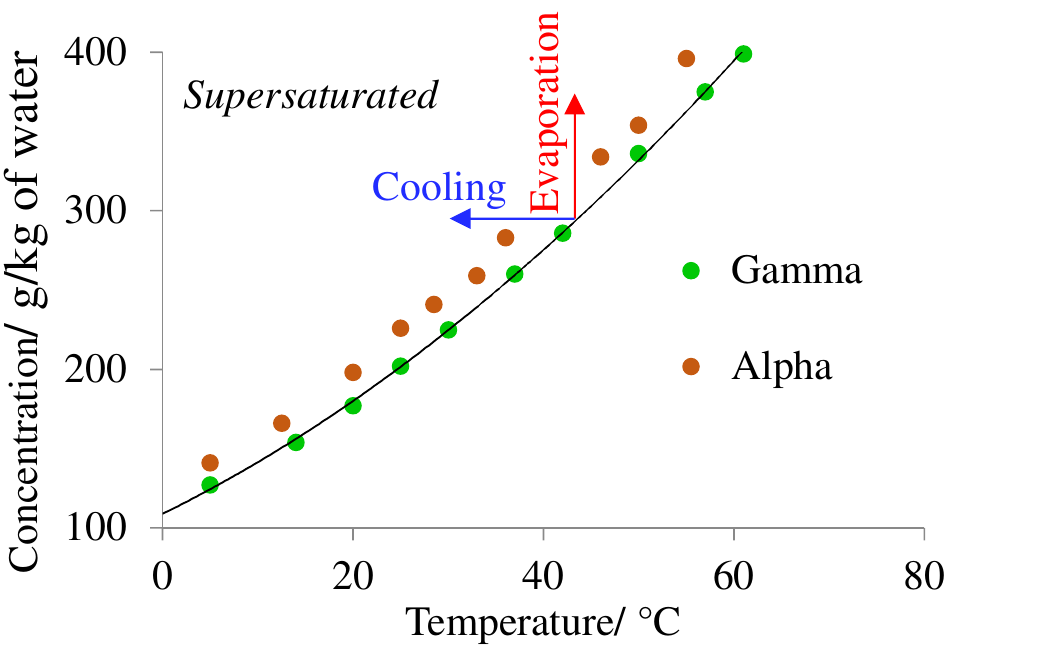}
(b)\includegraphics[ scale=0.75]{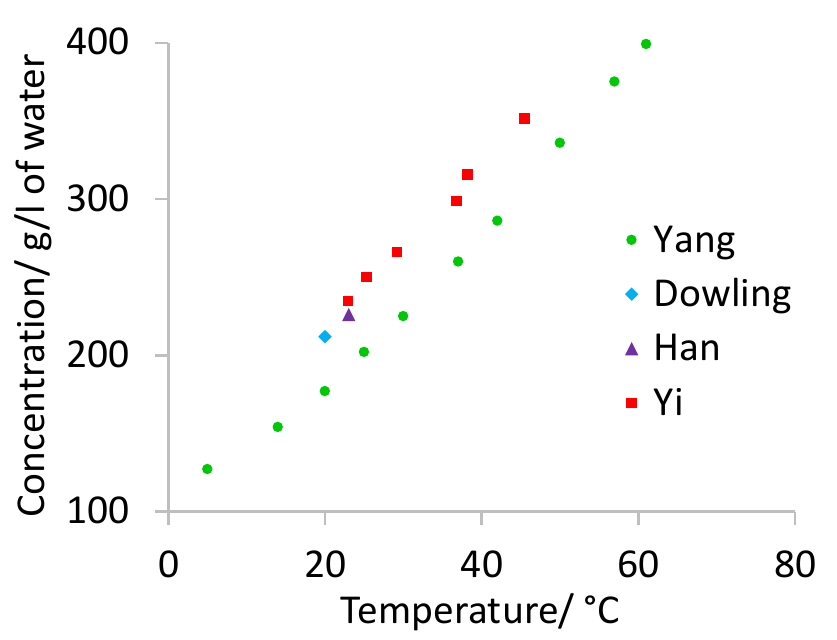}
\caption{Solubility in water
of the $\alpha$ and $\gamma$ polymorphs of glycine as a function of temperature. (a) Data from Yang et al \cite{yang08b}. We have fit a second order polynomial (black curve),
so that the solubility can be approximated between data points.
This fit is solubility (g/kg)~$=0.0301 T^2 + 2.96 T + 109$, with $T$ in $^{\degree}$C. (b) Comparison of $\gamma$ glycine solubility from several
sources: Yang et al. \cite{yang08b}, Dowling et al. \cite{dowling10},
Han et al. \cite{han15} and Yi et al. \cite{yi05}.
} \label{solubility}
\end{figure}

For glycine in water the solubility varies as shown in
Figure \ref{solubility}. We can see that glycine's solubility is very sensitive to temperature. This is beneficial in that it allows us to easily create highly supersaturated solutions (by cooling), but problematic in that to perform an experiment at constant supersaturation, temperature must be controlled very precisely. The curve in Figure \ref{solubility}(a),
is used for our supersaturation calculations. We include solubility data from several sources in
Figure \ref{solubility}(b) to show that values for
glycine solubility in the literature can vary significantly between studies.

\begin{figure}[htb!]
\centering
(a)\includegraphics[scale =0.75]{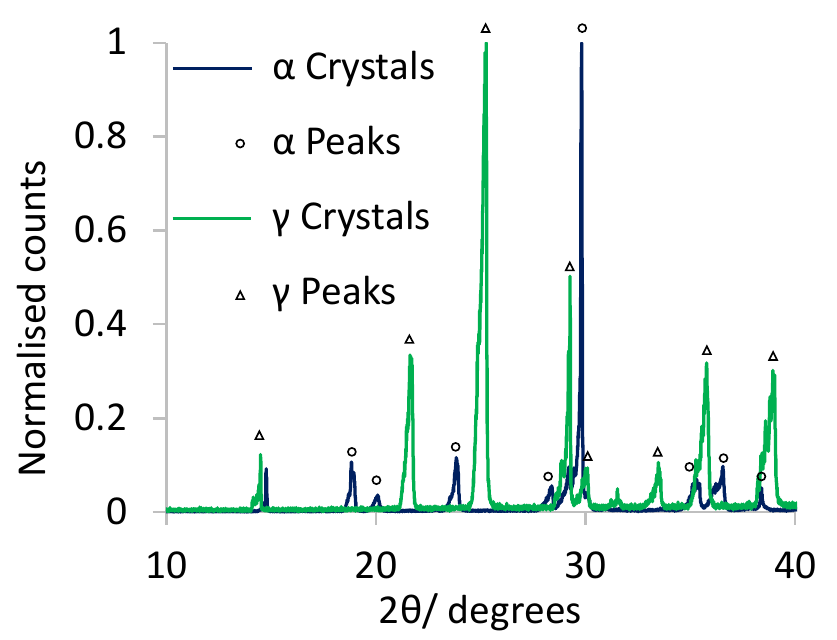}
(b)(i)\includegraphics[scale =0.75]{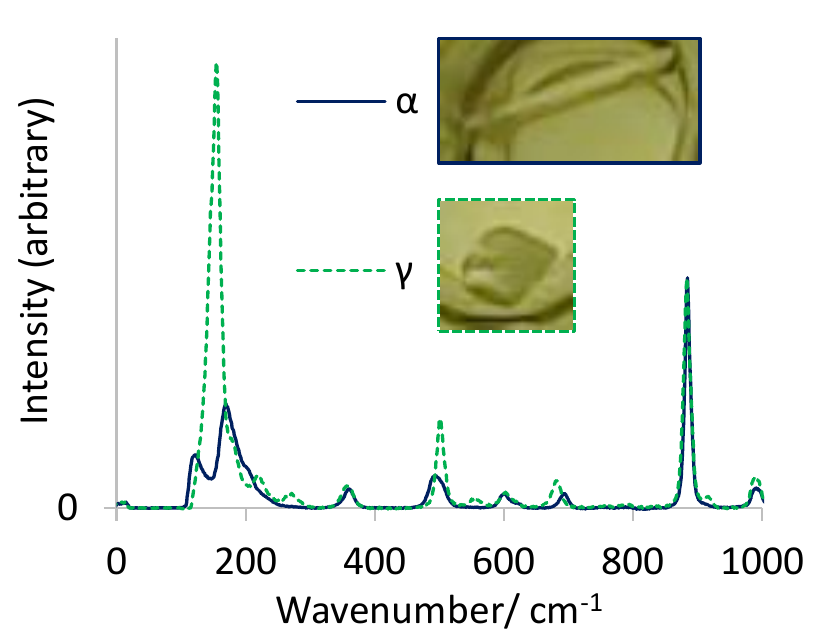}
(b)(ii)\includegraphics[scale =0.75]{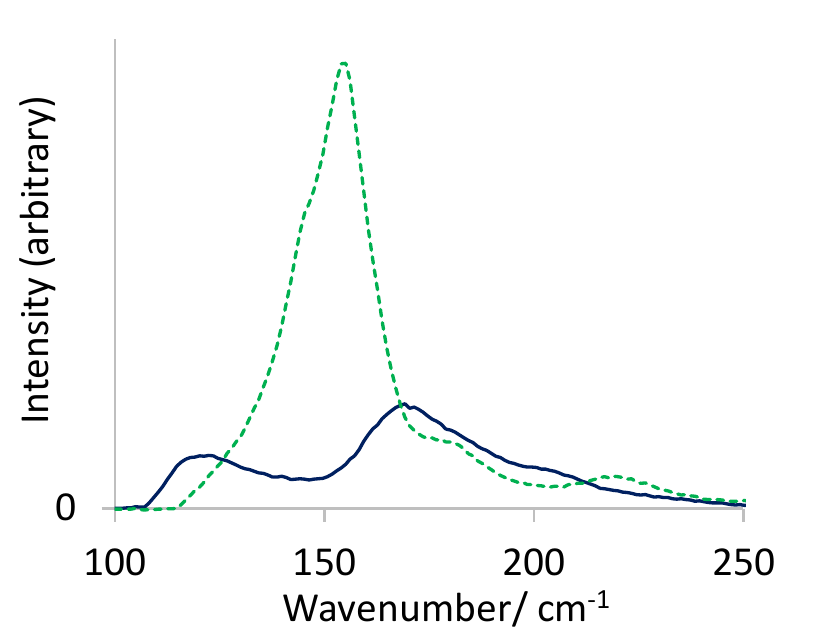}
(b)(iii)\includegraphics[scale =0.75]{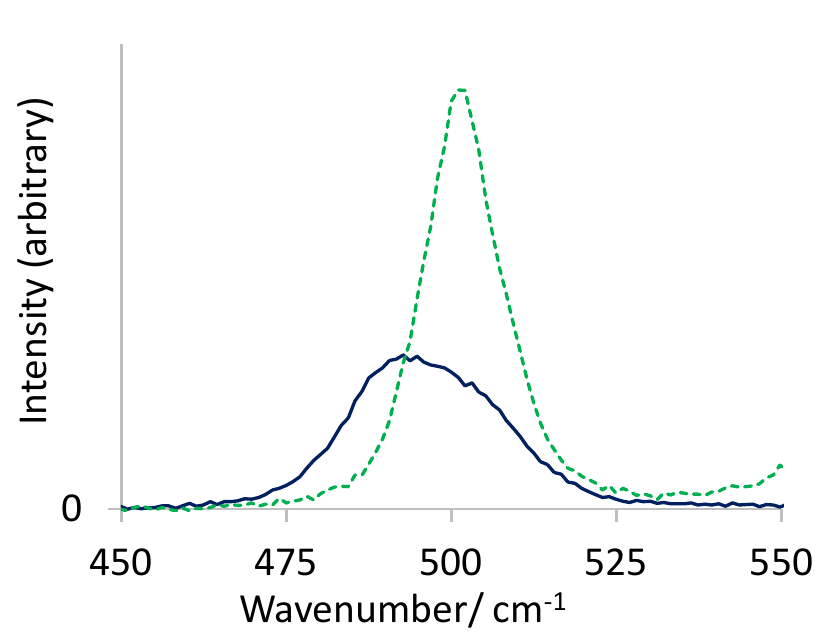}
(c)\includegraphics[scale =0.75]{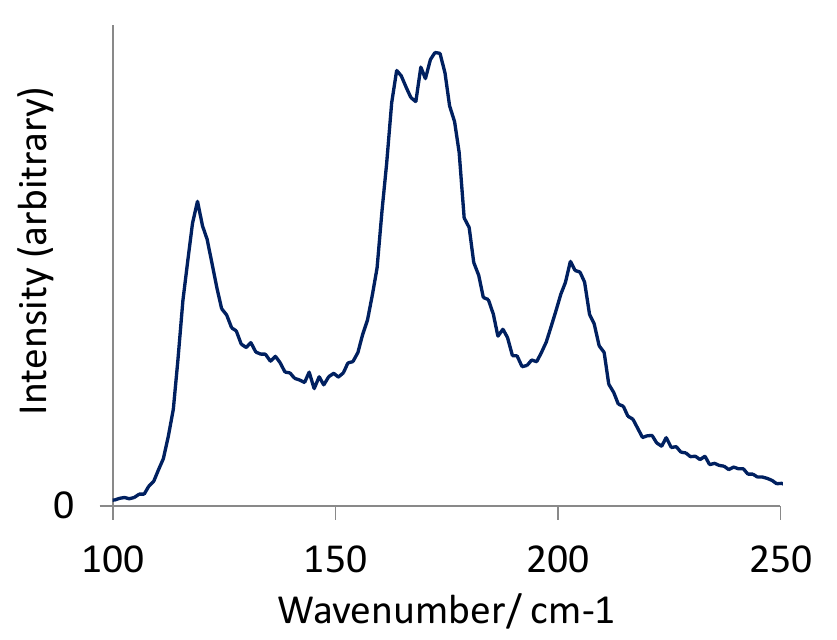}
(d)\includegraphics[scale =0.75]{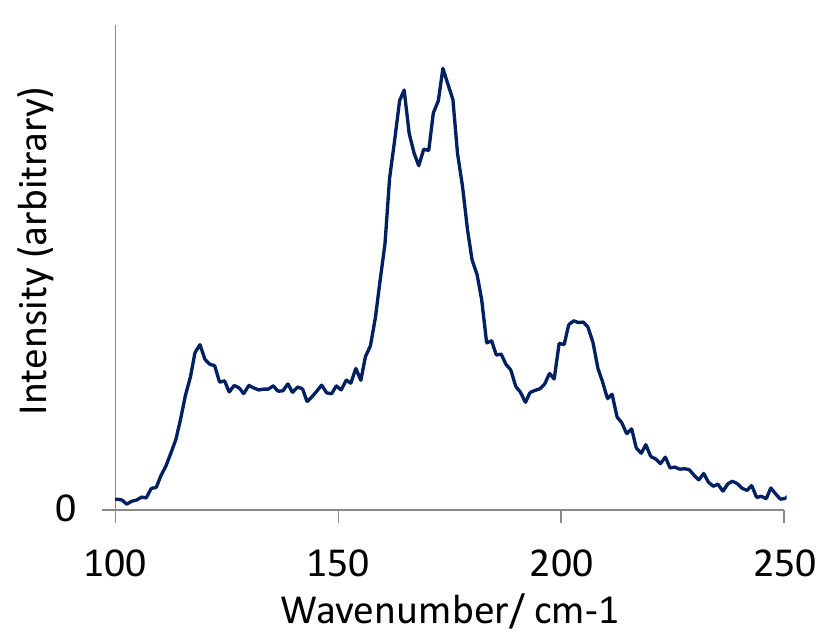}
\caption{\textbf{(a)} XRD patterns of two powdered samples, one of $\alpha$ glycine and one of $\gamma$ glycine as identified with Raman spectroscopy. The glycine identified as $\alpha$ via Raman spectroscopy is shown in blue and glycine identified as $\gamma$ is shown in green. The circles represent known $\alpha$ XRD peaks while the triangles represent known $\gamma$ XRD peaks.
\textbf{(b)}(i) Raman spectra for a typical $\alpha$ glycine crystal and a typical $\gamma$ glycine crystal. The spectra are normalised to the intensity of their highest peak at 886 cm$^{-1}$ \textbf{(b)}(ii) The spectra of (b)(i) in the region 100 - 250 cm $^{-1}$.\textbf{(b)}(ii) The spectra of (b)(i) in the region 450 - 550 cm $^{-1}$. \textbf{(c)} and \textbf{(d)} Two additional $\alpha$ glycine spectra where the characteristic peaks at  164 cm$^{-1}$, 171 cm$^{-1}$ and 203 cm$^{-1}$ (which cannot be easily seen in (b)(ii)) can be more clearly seen.}\label{XRD_Raman}
\end{figure}

\section{Validation of Raman spectra for identifying
the $\alpha$ and $\gamma$ polymorphs of glycine}

The $\alpha$ and $\gamma$ crystals have distinct Raman spectra,
which can be identified by a set of distinctive peaks\cite{shi05}. Examples of spectra from each polymorph  are given
in Figure \ref{XRD_Raman}(b).
Both polymorphs have a very intense peak at 886 cm$^{-1}$.
There are clear differences between the two spectra in the
region 100-200 cm$^{-1}$ where the peaks can be attributed to intermolecular vibrations\cite{shi05}.
The $\alpha$ polymorph has several small peaks
(relative to the 886 cm$^{-1}$ peak)
at 118 cm$^{-1}$, 164 cm$^{-1}$, 171 cm$^{-1}$ and 203 cm$^{-1}$,
which the $\gamma$ polymorph does not have. The 203 cm$^{-1}$ peak is often
of low intensity and the peaks at 164 cm$^{-1}$ and 171 cm$^{-1}$ are often merged. Spectra where these peaks are more clearly visible can be seen in
Figure \ref{XRD_Raman}(c) and (d).
The $\gamma$ polymorph has a very intense peak (comparable in intensity to the 886 cm$^{-1}$ peak) at 157 cm$^{-1}$, which the $\alpha$ polymorph does not have. There are also differences between the polymorphs at $\approx$ 500 cm$^{-1}$. The $\alpha$ polymorph has two low intensity peaks at 492 cm$^{-1}$ (NH$_3$ torsional mode\cite{shi05}) and 502 cm$^{-1}$ (CO$_2$ rocking mode\cite{shi05}) while the $\gamma$ polymorph has just one high intensity peak at 504 cm$^{-1}$.

\begin{figure}[htb!]
\centering
\includegraphics[scale =0.6]{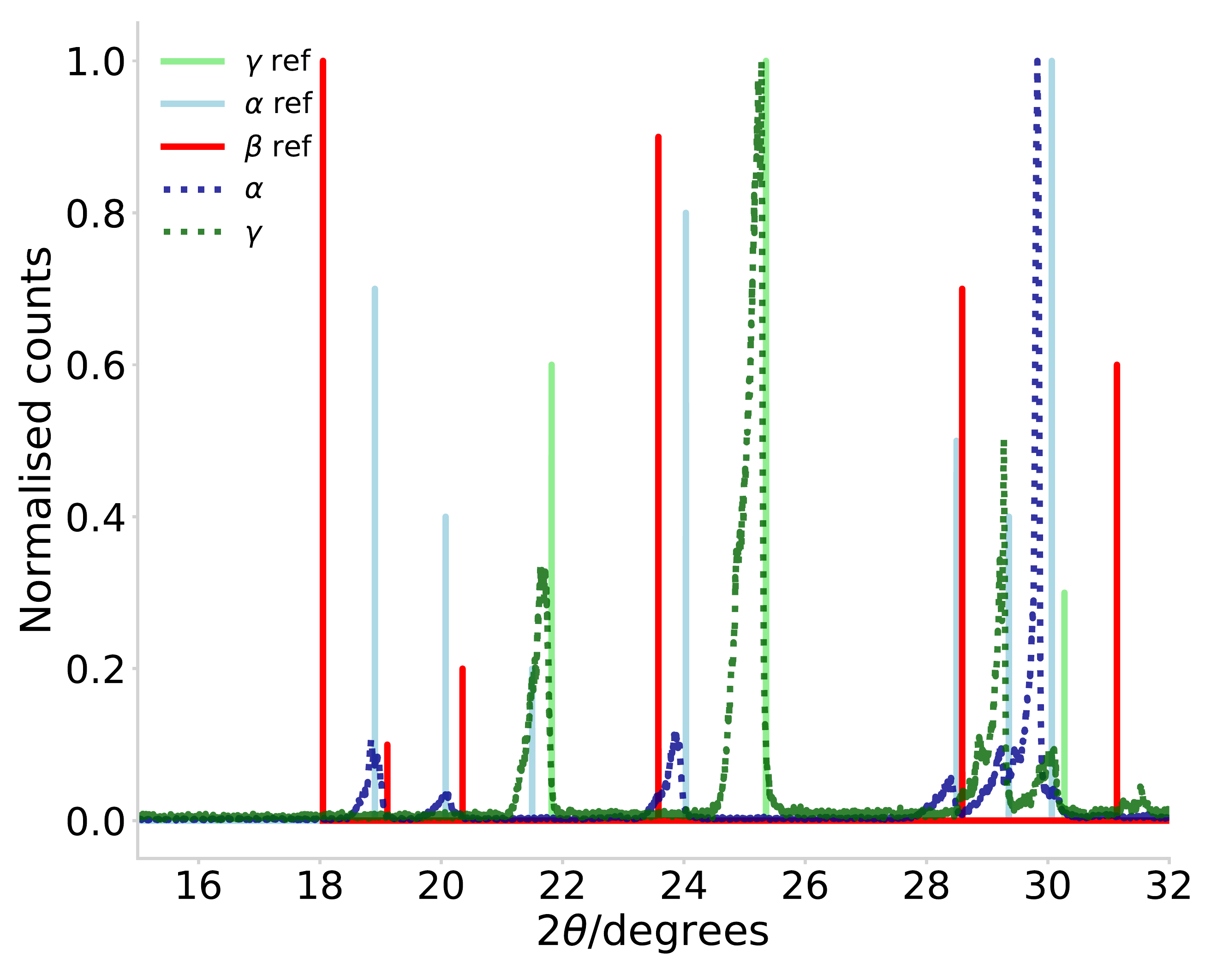}
\caption{ Experimental XRD patterns of two powdered samples,
one of $\alpha$ glycine (blue dotted curve) and one of $\gamma$ glycine (green dotted curve),
as identified with Raman spectroscopy.
The reference peak positions for all three polymorphs are shown as solid vertical lines.
Our PANalytical X’Pert Pro
diffractometer uses JCPDS reference files, numbers 00-032-1702 ($\alpha$), 00-002-0171 ($\beta$)
and 00-006-0230 ($\gamma$).
We show a $2\theta$ range where there are three
strong $\beta$ peaks that are all well separated from any $\alpha$ or $\gamma$ peak.
None of the experimentally observed peaks match the $\beta$ reference file.
All five patterns are normalised such that the highest peak has a height of one.
}\label{XRD_check}
\end{figure}

X-ray diffraction (XRD) was used as a test to confirm the results of our Raman spectroscopy analysis.
A sample of crystals was analysed with Raman spectroscopy (the two example spectra shown
in Figure \ref{XRD_Raman}(b)).
Immediately after the Raman spectroscopy had been carried out, the crystals were removed from the microplate and separated into two groups: Those that had been identified as $\alpha$ glycine and those that had been identified as $\gamma$ glycine. The two samples were then prepared for XRD using
the same procedure as described in earlier work\cite{little15}.
The crystals were ground up into a fine powder before
analysis. Powder XRD was carried out using a PANalytical X’Pert Pro
diffractometer across a $2\theta$ range of $10 - 70\degree$ using Cu $K_\alpha$ radiation.
The resulting XRD patterns can be seen in Figure \ref{XRD_Raman}.
In Figure \ref{XRD_Raman} we have identified peaks as 
belonging to either the $\alpha$ or $\gamma$ polymorphs using the standard JCPDS XRD
patterns of these polymorphs. The $\alpha$ polymorph pattern is JCPDS number
00-032-1702, and the $\gamma$ polymorph is JCPDS number 00-006-0230.

As expected, the set of crystals which we identified as $\alpha$ with Raman spectroscopy are found to show only $\alpha$ glycine
peaks with powder XRD. The Raman spectra and XRD were also in agreement for the set of $\gamma$ crystals.
 We never observed
the third polymorph of glycine, $\beta$ glycine, in our XRD patterns.
The $\beta$ polymorph can form needle-like crystals \cite{weissbuch05}, resembling
those we do see, but the XRD patterns rule out significant
amounts of the $\beta$ polymorph, at the point at the end
of the experiment when XRD patterns
are taken. In Figure \ref{XRD_check} we have plotted our XRD patterns together
with reference $\alpha$, $\gamma$ and $\beta$ patterns. Note that this range of $2\theta$
includes the strongest peak of the $\beta$ polymorph (at $2\theta=18.1^{\circ}$) which
is absent in both our sets of diffractograms.

 During the course of our experiments, we do not observe habit changes of our crystals.
This suggests that once formed, our crystals do not transform into another polymorph.
However, both Raman and XRD analyses are made at the end of our runs,
which is days after many of the crystals formed. Thus, although
it seems unlikely, we cannot rule out a
polymorph-to-polymorph transformation during our experiment that does not affect habit.

\begin{figure}[htb!]
\centering
(a)\includegraphics[scale =0.5]{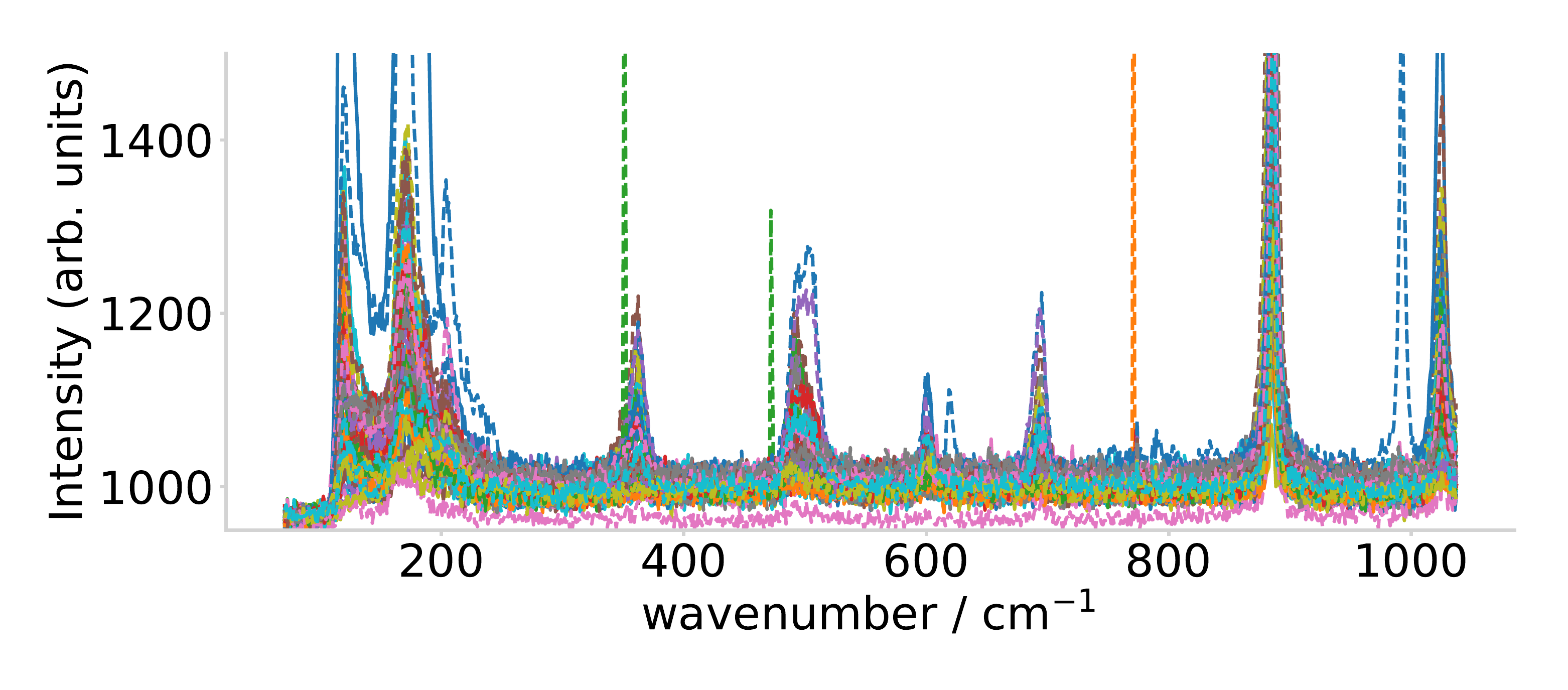}
(b)\includegraphics[scale =0.41]{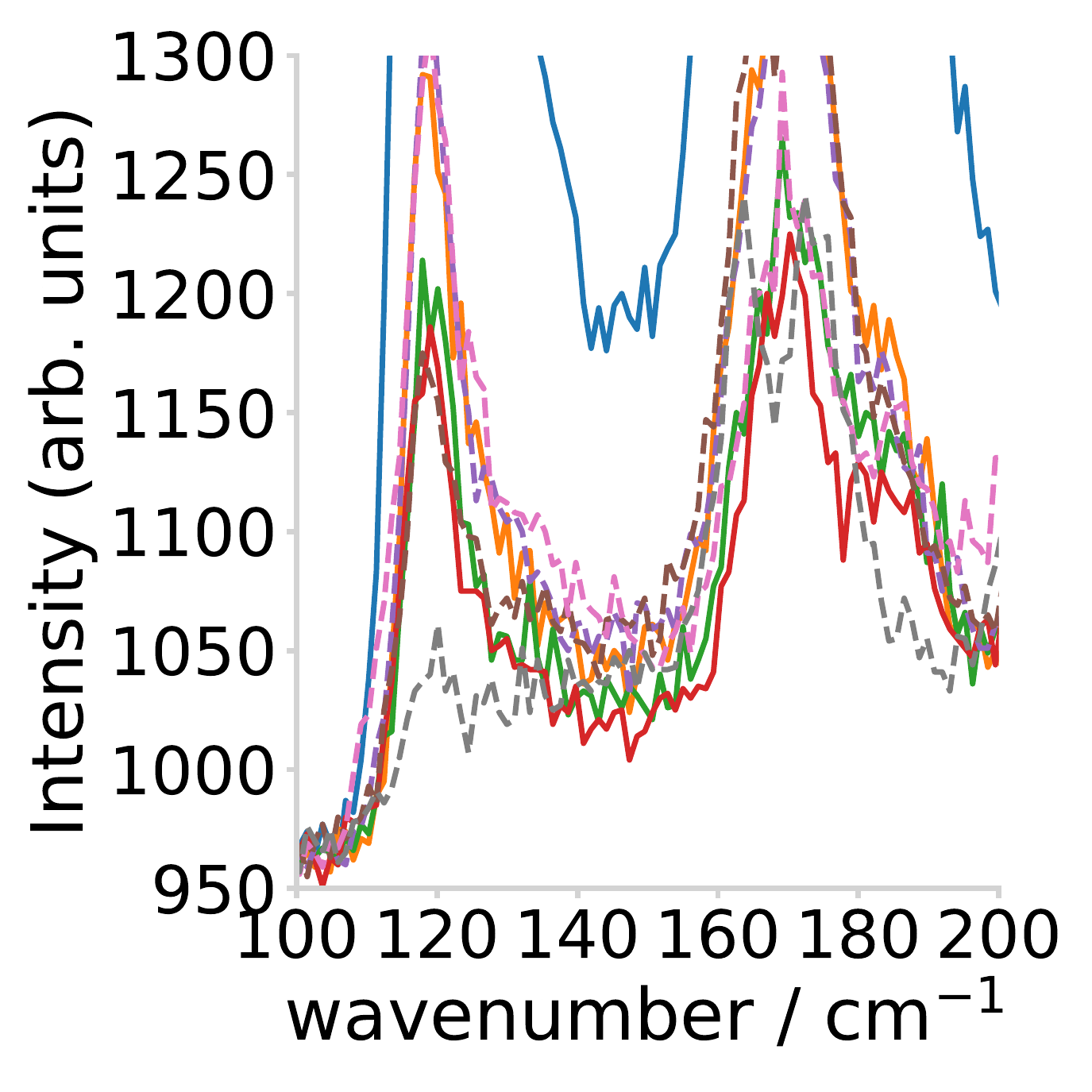}
(c)\includegraphics[scale =0.35]{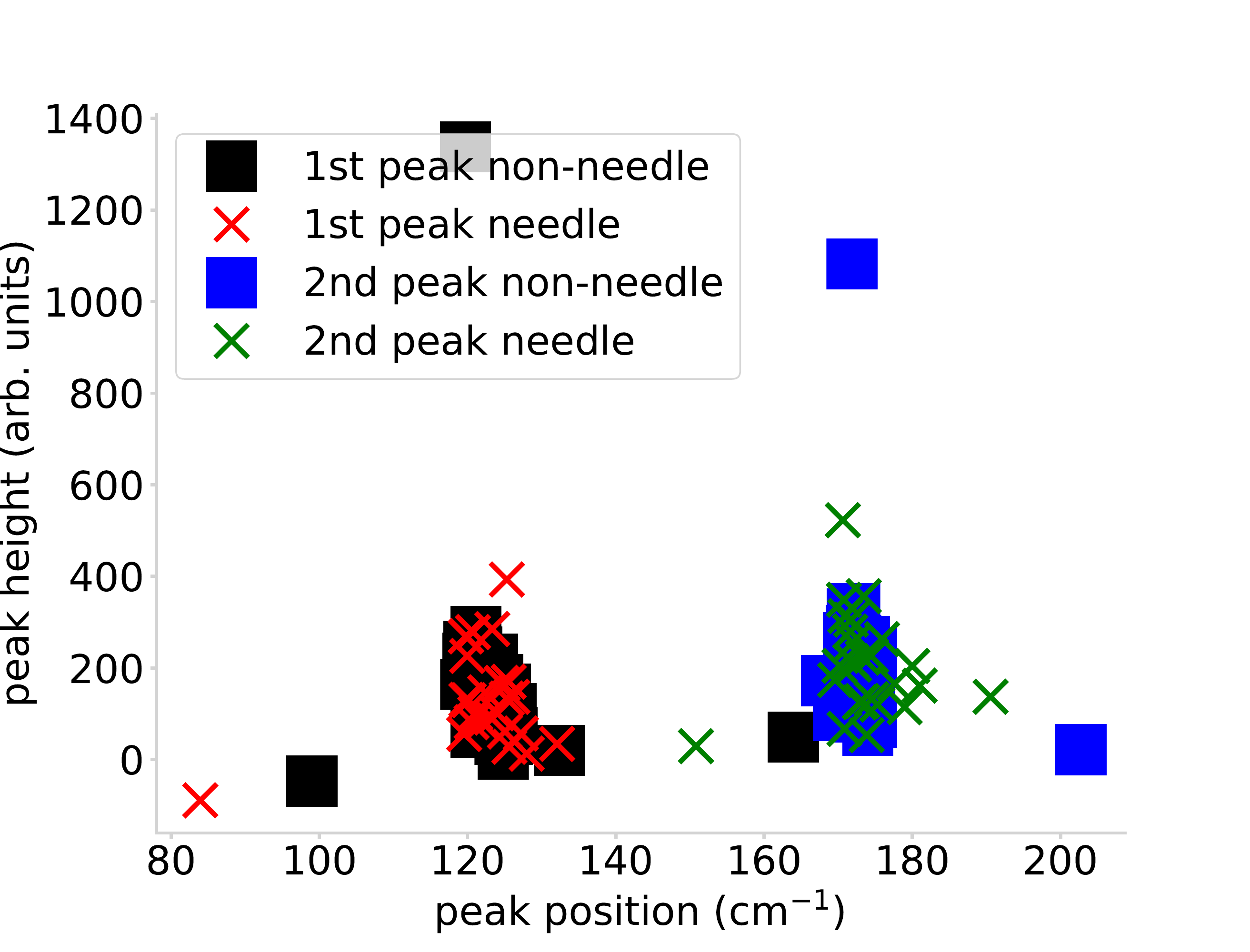}
\caption{\textbf{(a)}
Raman spectra for 24 non-needle like (solid curves) and 26 needle-like
(dashed curves) crystals, all of which we identified
as the $\alpha$ polymorph from its Raman
spectrum. All crystals were
obtained from run number one at a salt concentration of 250 mg/ml.
In two cases (green and orange dashed curves) there are anomalous
spikes at single values in the data, which may be due to energetic
events (cosmic rays)
in the detector.
\textbf{(b)} Four each of the non-needle (solid curves) and needle spectra (dashed curves)
 of (a), plotted
in the range we use for fitting. 
\textbf{(c)} Plot of peak height versus peak position, obtained by a fit
of two Gaussians plus a constant background to the part of the Raman spectra
in (a) between wavenumbers 100 and 200 cm$^{-1}$.
We use two Gaussians as for $\alpha$ we expect \cite{shi05} one
peak at around 118 cm$^{-1}$, plus two peaks at 164 and 171 cm$^{-1}$,
where these two peaks are merged into one in the spectra of most crystals.
Note that in a few instances, a peak is very weak, fitting fails, and
an anomalous point, e.g., slightly negative height, is produced.
}\label{Raman_needle_alpha}
\end{figure}

\begin{figure}[htb!]
\centering
(a)\includegraphics[scale =0.5]{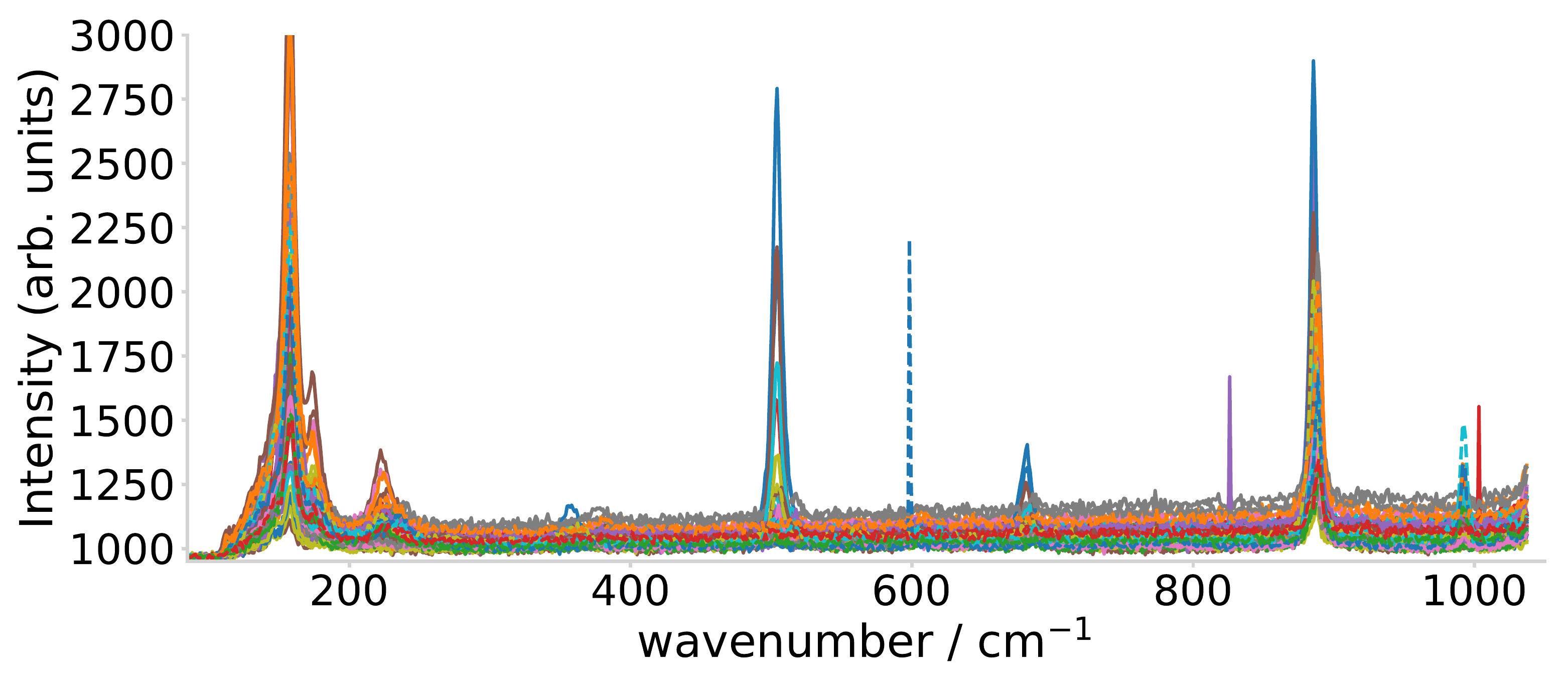}
(b)\includegraphics[scale =0.41]{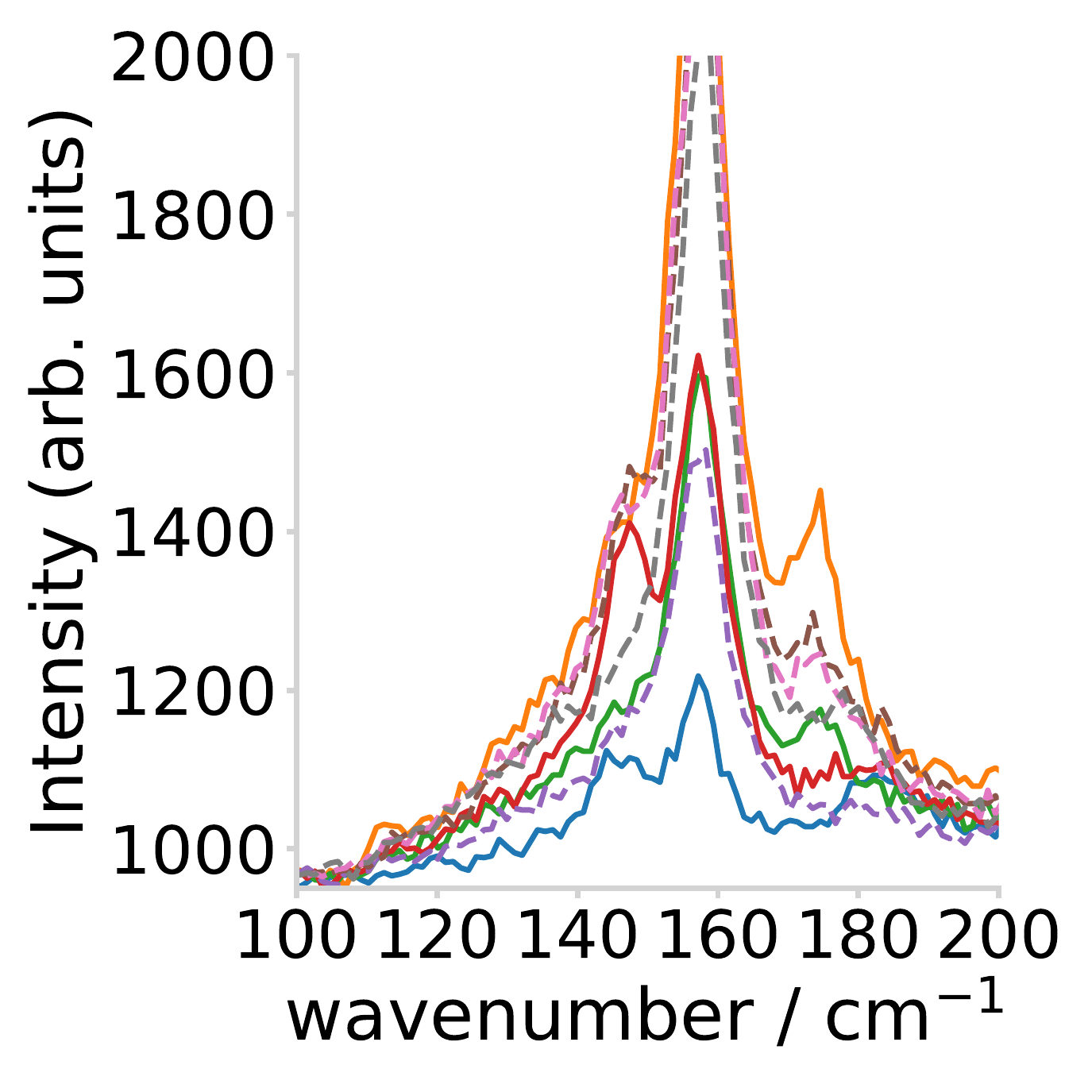}
\hspace*{-0.3cm}(c)\includegraphics[scale =0.35]{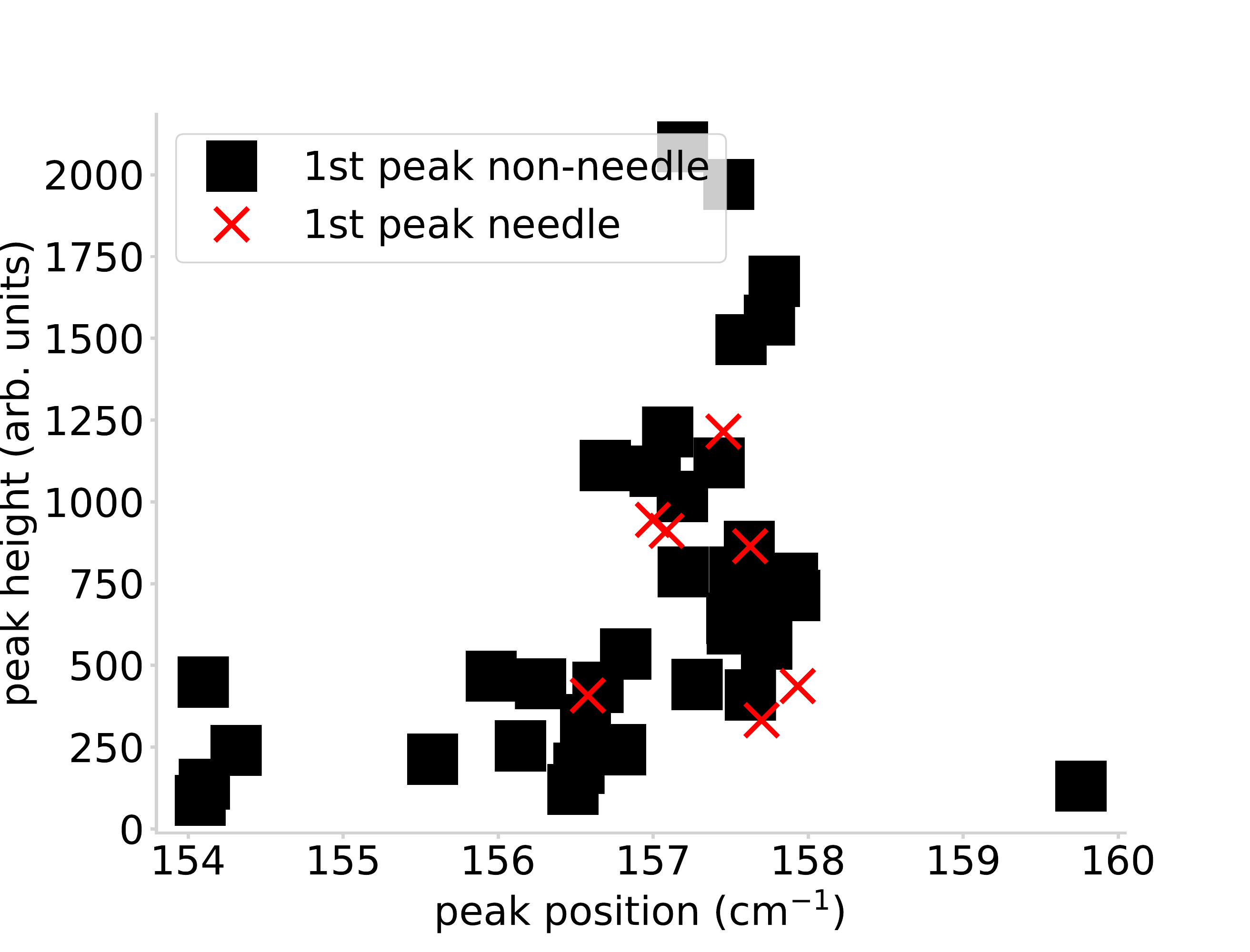}
\caption{\textbf{(a)}
Raman spectra for 37 non-needle like (solid curves) and 7 needle-like
(dashed curves) crystals, all of which we identified
as the $\gamma$ polymorph from their Raman
spectra. All crystals were
obtained from run number one at a salt concentration of 250 mg/ml.
In three cases (blue dashed curve, magenta and red solid curves) there are anomalous
spikes at single values in the data, which may be due to energetic
events (possibly cosmic rays)
in the detector.
\textbf{(b)} Four each of the non-needle (solid curves) and needle (dashed curves)
spectra of (a), plotted
in the range we use for fitting. 
\textbf{(c)} Plot of peak height versus peak position, obtained by a fit
of a Gaussian plus a constant background to the part of the Raman spectra
in (a) between wavenumbers 100 and 200 cm$^{-1}$.
We use one Gaussian as for $\gamma$ we expect \cite{shi05} one
peak at around 157 cm$^{-1}$.
}\label{Raman_needle_gamma}
\end{figure}

\subsection{Raman spectra for needle-like and
non-needle-like crystals}

 We were surprised by our
observation that both polymorphs formed two very
different crystal habits, in each case a needle-like habit 
and a much more compact habit. So we looked carefully at the Raman
spectra we obtained from these crystals.
Raman spectra for needle-like and non-needle crystals are shown
in Figure~\ref{Raman_needle_alpha} for the $\alpha$ polymorph and
in Figure~\ref{Raman_needle_gamma} for the $\gamma$ polymorph.
In Figure~\ref{Raman_needle_alpha}(a) we see Raman spectra from 50
needle and non-needle-like crystals, of the $\alpha$ polymorph.
Figure~\ref{Raman_needle_alpha}(c) is a scatter plot of estimates
of the height of the peak nominally located
at 118 cm$^{-1}$, versus its position,
and the height of the combined 
164 and 171 cm$^{-1}$ peaks, versus their location.
The peak heights and positions are obtained by fitting the sum
of two Gaussians plus a constant background term to
the spectra over the range 100 to 200 cm$^{-1}$.
Note that the Raman spectrum of the $\beta$ polymorph is distinct
from that of the $\alpha$ and $\gamma$ polymorphs \cite{seyedhosseini14,isakov14}.
The Raman spectrum of the $\beta$ is flat betweeen approximately 120 and 200~cm$^{-1}$,
whereas we see peaks in that range, in
both Figures \ref{Raman_needle_alpha} and \ref{Raman_needle_gamma}.

 Note that there is clearly
significant variability in the peak height, but that almost
all estimates of the
peak position are within a narrow range.
The distributions of the peak heights and positions
of the needle and non-needle crystals are indistinguishable.
As far as the Raman spectra are concerned, the needle-like and
non-needle-like both have the characteristic $\alpha$ polymorph
peaks at the same position, and in both cases with variable intensity.
Whatever the cause of the fast growth along one axis of the needles,
it is not apparent in the Raman spectra.
This would be consistent with the difference in growth rates
being due to different
defects, as these defects would be unlikely to show up in the Raman spectra.

The corresponding data for the $\gamma$ polymorph needle-like crystals and
non-needle-like crystals are in Figure \ref{Raman_needle_gamma}.
As with the $\alpha$ polymorph, there is no apparent difference between the
set of needle spectra and the set of non-needle Raman spectra.

\section{Statistics and models for the nucleation of competing polymorphs} \label{sec:med_statsSS}

As we only observe one polymorph or the other in a single well, not both, nucleation of the two polymorphs are mutually exclusive events in our system.
Thus our data on the kinetics of nucleation of single crystals of competing
polymorphs consist of the pair of observations, $(t_{nuc},i=\alpha,\gamma)$, for each well
where crystallisation occurred.
We want to analyse this quantitative data to build the most robust model
with the greatest predictive power.

The statistics of the nucleation of competing polymorphs is subtle, but fortunately, analogous data sets occur in a number of other fields, in particular in medical statistics \cite{tsiatis75,peterson76,slud88,beyersmann09,andersen12}. Typically in mortality studies, there are competing illnesses or causes of death \cite{beyersmann09,dignam12,geskus_book}.
An example might be a study of, say, 100 patients at risk of dying of cancer
or of heart disease, where date and cause of death are recorded.
As the two causes of death, like our competing polymorphs,
are mutually exclusive, the data is also of the form of a pair of observations:
a time, and one of a number of competing outcomes.

\subsection{Models that include only observables}

Our observables are the $I_i$, together with their derivatives $h_i$, and
$P=1-I_{\alpha}-I_{\gamma}$. We can construct models using only these functions.
We can create a model by specifying the two cause specific hazard (CSH) functions.

\subsubsection{Fitting procedure for models that include only observables}

The model we fit is defined by the definitions of $I_i$ and of $P$:
\begin{eqnarray}
\frac{{\rm d}I_i(t)}{{\rm d} t} &=& P(t)h_i(t) ~~~~~~i=\alpha,\gamma \label{fit_model}\\
P(t)& = & 1-I_{\alpha}(t)-I_{\gamma}(t)
\end{eqnarray}
with two boundary conditions $I_i(t=0)=0$.
We can obtain an equation for ${\rm d}P/{\rm d}t$ and integrate it, to obtain
\begin{equation}
P(t)=\exp\left[-\int_{0}^{t}{\rm d}t'\left(h_{\alpha}(t')+h_{\gamma}(t')\right)\right]
\label{eq_pdef}
\end{equation}
which when the $h_i$ are Weibull CSHs, becomes
\begin{equation}
P(t)=\exp\left[-\left(t/\tau_{\alpha}\right)^{\beta_\alpha}
-\left(t/\tau_{\gamma}\right)^{\beta_\gamma}\right]
\label{eq_pdef_weibull}
\end{equation}
Equation (\ref{fit_model}) then becomes
\begin{equation}
\frac{{\rm d}I_i(t)}{{\rm d} t} =\exp\left[-\left(t/\tau_{\alpha}\right)^{\beta_\alpha}
-\left(t/\tau_{\gamma}\right)^{\beta_\gamma}\right]
\beta_i\left(t^{\beta_i-1}/\tau_{i}^{\beta_i}\right) ~~~~~~i=\alpha,\gamma
\label{fit_model_weibull}
\end{equation}

For given values of the four (two $\tau_i$ and two $\beta_i$) parameters,
we integrate the coupled
ordinary differential equations for the $I_i$, Eq.~(\ref{fit_model_weibull}),
to get the two $I_i(t)$ functions.
To fit to data we simply vary the four parameters, and minimize the sum
of the squares of the difference between the modelled $I_i(t)$ and observed $I_i(t)$.
This is done using a Python program \cite{laurie_thesis}.

\subsection{Models with latent nucleation times}

Models can also be constructed that rely on latent nucleation times, i.e.,
on hypothetical nucleation times $t_{nuc,\alpha}$ and $t_{nuc,\gamma}$ for each well.
As we only observe one polymorph for a well, for those wells where we observe nucleation
we only measure the shorter one of these two times, the other one is not observable.
Thus in all cases one of these times is hidden, hence the name latent time.
See Beyersmann {\it et al.}\cite{beyersmann09} and Geskus \cite{geskus_book}
for discussion of the advantages
and disadvantages of models that rely on latent times.
Tsiatis \cite{tsiatis75}, Peterson \cite{peterson76}, and Slud and Byar \cite{slud88} both
discuss the limits of what can be inferred about
$t_{nuc,\alpha}$ and $t_{nuc,\gamma}$, from data of our type.
In general, the two nucleation times for a single well will be correlated, for example
there may be a tendency that if one nucleation time is small in a well, that
the other polymorph's nucleation time in the same well, is also small.

We can write the probability that no nucleation has occurred in a droplet at time
$t$ as
\begin{equation}
P(t)={\rm Pr}(t_{nuc,\alpha}> t,t_{nuc,\gamma}>t)
\label{eq_2p}
\end{equation}
i.e., the probability that both  $t_{nuc,\alpha}$ and $t_{nuc,\gamma}$ are greater than $t$.

\subsubsection{Model with independent latent nucleation times}

If the two nucleation times are independent then Eq.~(\ref{eq_2p}) simplifies to
\begin{equation}
P(t)={\rm Pr}(t_{nuc,\alpha}> t){\rm Pr}(t_{nuc,\gamma}>t)
\label{eq_2p_ind}
\end{equation}
If the two independent latent nucleation times are both modelled by Weibulls, then
for the probability density function for $t_{nuc,i}$, we have
\begin{equation}
p(t_{nuc,i})=\beta_i\left(t^{\beta_i-1}/\tau_{i}^{\beta_i}\right)
\exp\left[-\left(t/\tau_{i}\right)^{\beta_i}\right]~~~~~i=\alpha,\gamma
\end{equation}
and the cumulative probabilities are
\begin{equation}
{\rm Pr}(t_{nuc,i})=\exp\left[-\left(t/\tau_{i}\right)^{\beta_i}\right]~~~~~i=\alpha,\gamma
\end{equation}
When the latent times are independent, the
rate at which $\alpha$ is observed to nucleate is just $p(t_{nuc,\alpha})$ times
the probability that nucleation
of the $\gamma$ polymorph has not yet occurred, so we have
for $\alpha$ nucleation
\begin{eqnarray}
\frac{{\rm d}I_{\alpha}(t)}{{\rm d} t} &=&
\beta_i\left(t^{\beta_{\alpha}-1}/\tau_{\alpha}^{\beta_{\alpha}}\right)
\exp\left[-\left(t/\tau_{\alpha}\right)^{\beta_{\alpha}}\right]
\times
\exp\left[-\left(t/\tau_{\gamma}\right)^{\beta_{\gamma}}\right]\nonumber\\
&=&\exp\left[-\left(t/\tau_{\alpha}\right)^{\beta_{\alpha}}
-\left(t/\tau_{\gamma}\right)^{\beta_{\gamma}}\right]
\beta_{\alpha}\left(t^{\beta_{\alpha}-1}/\tau_{\alpha}^{\beta_{\alpha}}\right)
\label{eqi_ind}
\end{eqnarray}
plus an analogous equation for ${\rm d}I_{\gamma}/{\rm d}t$.

These two equations are identical to Eq.~(\ref{fit_model_weibull}).
So our model in the previous section that modelled the observable CSHs, via
Weibulls (equation (3) of the main text), can be obtained starting
with latent nucleation times, and then assuming that they are both Weibull
distributed, and are independent.
Thus the fits of our model using only observables,
are also what we would obtain from a model of independent latent times.
However, it should be noted that \cite{tsiatis75},
although for every set of observables $I_i$ there is a corresponding
model with independent latent times, for the same $I_i$ there are an infinite
number of models with differently correlated latent times that yields this same $I_i$.

\subsubsection{Computational generation of models with independent latent nucleation times}

As we just discussed, a model with independent nucleation times is the same as our model based on observables. Nonetheless, we now briefly describe how to generate a model based on independent latent nucleation times computationally here. Then in the next sub-section we describe how we introduce correlations. To generate the model with independent times, we simulate the behaviour of a large number samples, and determine the
fraction of those samples at time $t$, where nucleation has not occurred, $N(t)$.
When the number of samples is sufficiently large $N(t) \approx P(t)$. For each well we need both a $t_{nuc,\alpha}$ and $t_{nuc,\gamma}$, we then select the shorter time to be the nucleation time of the well.

For this model we need to generate two sets of variables, $t_{nuc,\alpha}$ and $t_{nuc,\gamma}$, such that both sets of variable are Weibull distributed. We can do this by generating a set of uniform random variables for each polymorph, and putting
each random number through the inverse cumulative distribution function (CDF) of a Weibull distribution.
\begin{equation}\label{eq:inverse_weibull}
y = \tau_i(-\ln(1-u))^{\frac{1}{\beta_i}}~~~~~i=\alpha,\gamma
\end{equation}
Where $u$ is a uniform random variable such that $0 \le u < 1$ and $y$ is the output variable which is Weibull distributed. In this way we can easily create
two independent
sets of Weibull distributed nucleation times. We can then vary the $\beta$ and $\tau$ values to fit our data.

\subsubsection{Model with correlated latent nucleation times}

We do not know if nucleation of the two polymorphs is correlated.
So to understand the role of correlations, it
is useful to have a model with variable
correlations between the nucleation times of the two polymorphs.
To generate the model, we start,
as in the previous section, with two sets of Weibull distributed numbers for $t_{nuc,\alpha}$ and $t_{nuc,\gamma}$. In this case however we want those two sets of numbers to be correlated such that for a well with a short $\alpha$ nucleation time there is a high probability of a short $\gamma$ nucleation time

We set about generating the correlated nucleation times
as follows. We generate two correlated sets of numbers, both Gaussian distributed.
We then put each number into a Gaussian CDF which transforms them to uniformly distributed numbers between zero and one.
After that, we put the uniformly distributed numbers into an inverse Weibull CDF,
Eq.~(\ref{eq:inverse_weibull}).
At this point, we have two Weibull distributed sets of numbers that are correlated.
By generating a large number of nucleation time pairs, we obtain $I_\alpha$ and $I_\gamma$  pairs where the $\alpha$ and $\gamma$ nucleation processes have the desired correlation. We used these generated distributions to fit our data,
in Figure 2 of the main text.

\subsubsection{Measuring correlation}

We use Spearman's rank coefficient \cite{hoffman15} to measure correlation in this model. We briefly explain how Spearman's rank coefficient
works. For each sample we have two times: $t_{nuc,\alpha}$ and $t_{nuc,\gamma}$. We rank all the $t_{nuc,\alpha}$, for all samples in order of increasing length. We do the same with the set of $t_{nuc,\gamma}$. We then measure for each pair of times the difference between the  $t_{nuc,\alpha}$ rank, $r_\alpha$ and the
$t_{nuc,\gamma}$ rank, $r_\gamma$. For example, if for
one sample the nucleation times are $t_{nuc,\alpha}$ and $t_{nuc,\gamma}$, and if
$t_{nuc,\alpha}$ is the 6th longest nucleation time ($r_\alpha$ = 6) and $t_{nuc,\gamma}$ is the 11th longest nucleation time ($r_\gamma$ = 11), then the difference in ranks, $r_\gamma - r_\alpha$ is 5. Note that if $t_{nuc,\alpha}$ and $t_{nuc,\gamma}$ are perfectly correlated all their ranks will be the same.

We measure the difference in rank for each pair of variables. The sum of the square rank differences is the covariance of $r_\alpha$ and $r_\gamma$.  The formula we use to calculate Spearman's rank correlation coefficient, $R_s$, is
\begin{equation}
R_s = \frac{\mbox{cov}(r_\alpha,r_\gamma)}{\sigma_{r_\alpha} \sigma_{r_\gamma}}
\end{equation}
where $\sigma_{r_\alpha}$ and  $\sigma_{r_\gamma}$ are the standard deviations of all $r_\alpha$ and $r_\gamma$ respectively and $\mbox{cov}(r_\alpha,r_\gamma)$ is the covariance of $r_\alpha$ and $r_\gamma$. When data are strongly positively correlated $R_s$ approaches 1, when data are uncorrelated $R_s$ is close to zero and when data are strongly negatively correlated $R_s$ approaches -1.
The Spearman's rank correlation coefficient between the two functions in the model 
used in the fit in Figure 2 of the main text, is 0.95,
i.e, $\alpha$ and $\gamma$ nucleation are strongly correlated.


\section{Additional results on nucleation}

\subsection{CIF plots for the five salt concentrations between 60 and 250 mg/ml}

\begin{figure*}[htb!]
\centering
(a)\includegraphics[scale = 0.85]{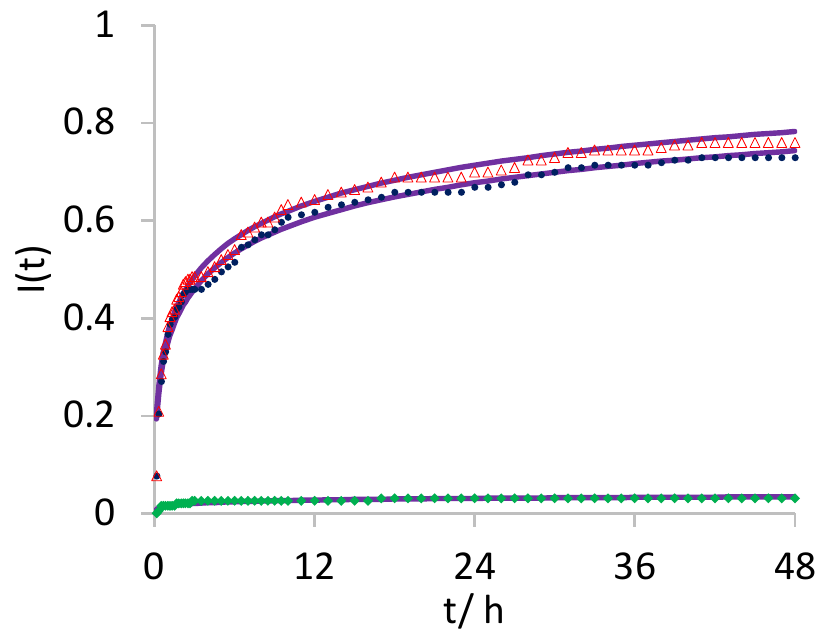}
(b)\includegraphics[scale = 0.85]{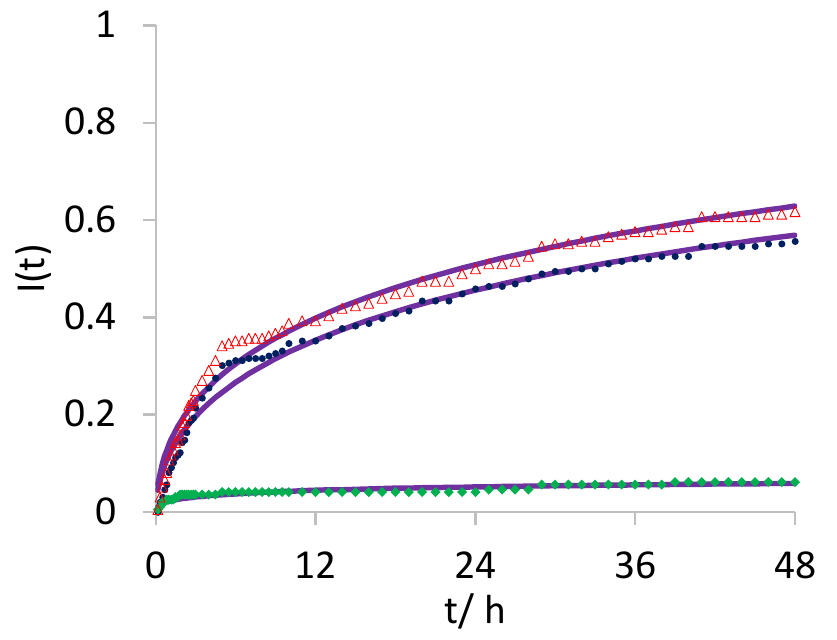}
(c)\includegraphics[scale = 0.85]{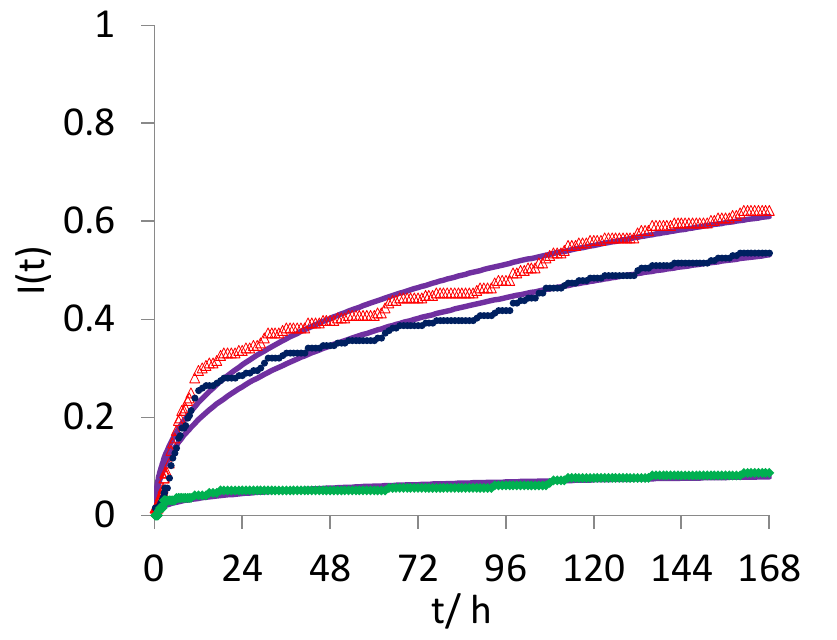}
(d)\includegraphics[scale = 0.85]{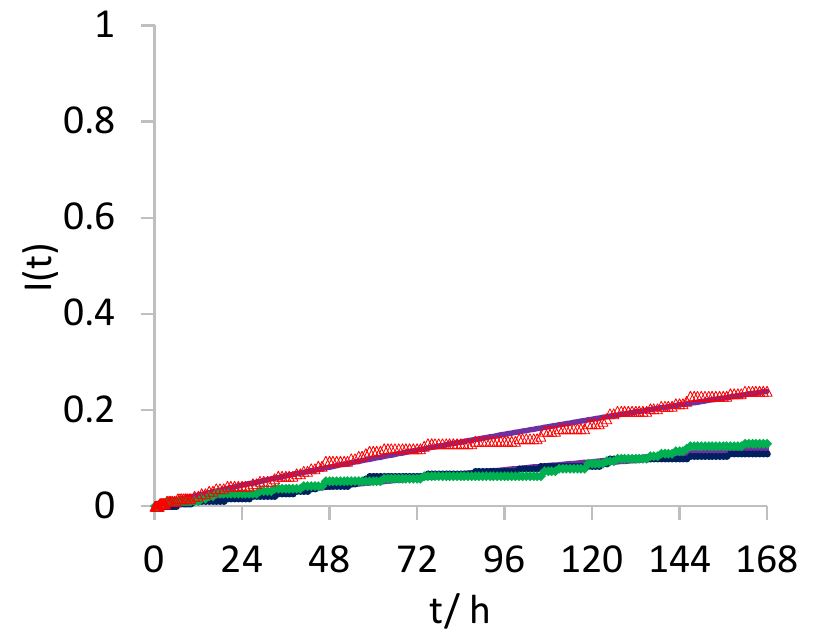}
(e)\includegraphics[scale = 0.85]{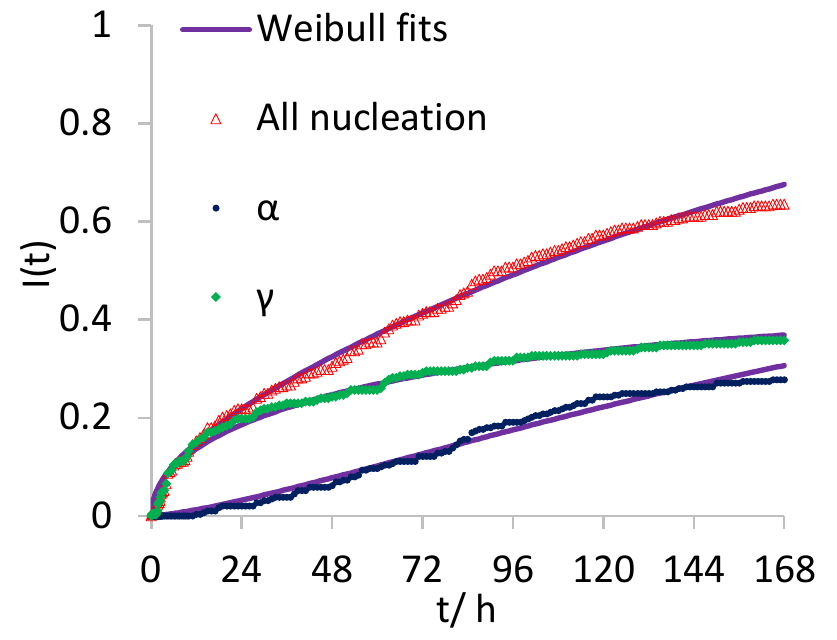}
\caption{Plots of CIFs for five salt concentrations:
(a) 60 mg/ml, (b) 90 mg/ml, (c) 150 mg/ml, (d) 200 mg/ml, and (e) 250 mg/ml.
The legend in (e) applies to all plots. The points are our data:
$I_{\alpha}$ (blue), $I_{\gamma}$ (green) and
$I_{\alpha}+I_{\gamma}$ (red), respectively.
Purple curves are fits of models with Weibull CSHs to the data.
} \label{Induction_curvesSS}
\end{figure*}

In Figure \ref{Induction_curvesSS},
we have plotted the CIFs: $I_{\alpha}$, $I_{\gamma}$ and the sum $I_{\alpha}+I_{\gamma}$,
for five different salt concentrations.
We do not plot results for the experiments at 300 mg/ml NaCl because there is little nucleation  at this salt concentration.
Note the change of time scale between Figure \ref{Induction_curvesSS}(b) and (c). The experimental timescale is longer at higher NaCl concentrations because as salt is added, nucleation slows.
The plots for 250 mg/ml are also shown in Figure 2 of the main text.
The fit parameters for the Weibull fits are in Table \ref{table_weibullfits}.

\begin{table}[ht!]
\centering
\caption{Best-fit values for fits of
models with Weibull CSHs to the CIFS for $\alpha$ and $\gamma$ nucleation.}
    \begin{tabular}{ c| c c c c | c c  }
    NaCl &  $\tau_\alpha$ (h) & $\beta_\alpha$ & $\tau_\gamma$ (h) & $\beta_\gamma$ & $R_\alpha^2$ & $R_\gamma^2$ \\
(mg/ml) & & & & & & \\
\hline
60 & 7.49  & 0.26 & 4.74$\times$ 10$^5$ & 0.28 & 1.00  & 1.00\\
90 & 57.2  & 0.49 & 5.84$\times$ 10$^5$ & 0.26 & 0.99  & 1.00\\
150 & 245   & 0.49 & 3.35$\times$ 10$^4$ & 0.40 & 0.99 & 1.00 \\
200 & 1650  & 0.89 & 1250 & 0.97 & 1.00 & 1.00 \\
250 & 244 & 1.41 & 610 & 0.48 & 0.99 & 0.99    \\ \hline
\end{tabular}\\
\vspace*{0.3cm}
{$R_i^2$, $i=\alpha$, $\gamma$, is 
the $R^2$ value for comparison of the fit $I_i$ to the data.}
\label{table_weibullfits}
\end{table}

\begin{figure}[tb!]
\centering
\includegraphics[width=6cm]{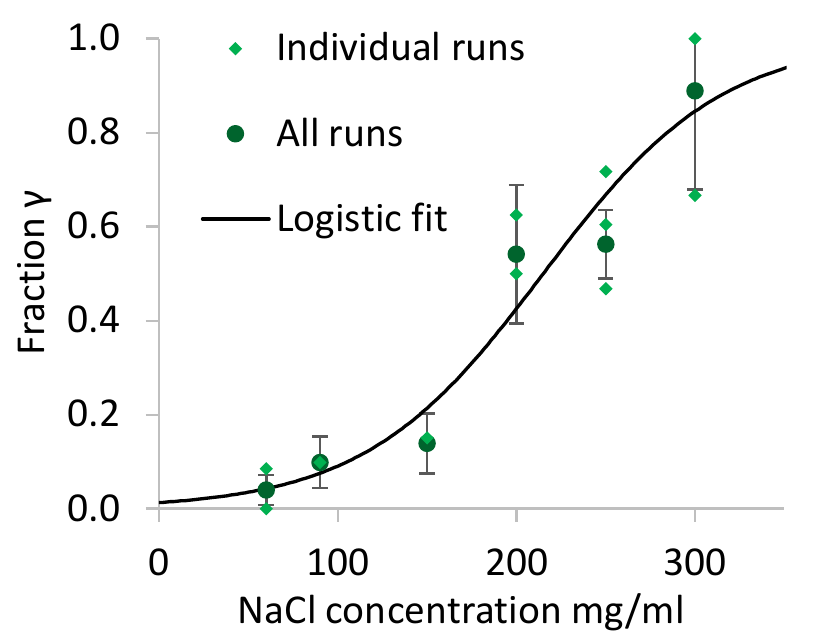}
\caption{The final fraction of wells that
contain the $\gamma$ polymorph, $f_{\gamma}$,
plotted as a function of NaCl concentration.
The dark green circles are the overall fraction averaged over all runs at a single NaCl
concentration. The smaller green diamonds are the fractions in individual runs.
The solid curve is a fit of a logistic function to
$f_{\gamma}=n_\gamma/[n_\alpha + n_\gamma]$,
where $n_\alpha$ and $n_\gamma$ are the number of wells  containing $\alpha$ crystals and $\gamma$ crystals, respectively. 
The error bars shown have a total height of
$2\left[f_{\gamma}(1-f_{\gamma})/(n_\alpha + n_\gamma) \right] ^{1/2}$,
where $n_\alpha$ is the number of wells
containing $\alpha$ crystals and $n_\gamma$ is the number of wells
containing $\gamma$ crystals.
}\label{frac_gamma}
\end{figure}

\subsection{Increasing salt concentration favours the $\gamma$ polymorph}

We plot the final fraction of crystals that are in the $\gamma$ polymorph,
in Figure \ref{frac_gamma}.
This is for
experiments were carried out at NaCl concentrations $c_{NaCl}$ from 60 mg/ml to 300 mg/ml.
At least two runs were carried out at each salt concentration.
We define a run as the set of nucleation times recorded from one 96-well microplate.
On completing each run, the fraction of wells containing each polymorph was determined.

In Figure \ref{frac_gamma}, the variation of the final fraction of crystals in the $\gamma$ polymorph,
$f_{\gamma}$, is modelled using a logistic function.
The fit is shown as a black curve in Figure \ref{frac_gamma}.
The logistic function is
\begin{equation}
f_{\gamma}\left(c_{NaCl}\right)=\frac{1}{1+\exp\left[-(c_{NaCl}-c_{1/2})/c_{SW}\right]}
\label{logistic}
\end{equation}
and it has two parameters: $c_{1/2}$ and $c_{SW}$.
As we can see in Figure \ref{frac_gamma}, this functional form fits our data well.
The best-fit parameters are $c_{1/2}=215$~mg/ml, and $c_{SW}=100$~mg/ml.
The parameter, $c_{1/2}$, is an estimate for the salt concentration at which half the samples are $\alpha$ and
half $\gamma$. There is broad region, a few hundred mg/ml, over which we go from a region with very small, but non-zero, amounts of the $\gamma$, to a large majority of the crystals in the $\gamma$ polymorph. The $c_{SW}$ parameter can be used as an estimator of the width of this region.
The data is in Table \ref{table:frac_gamma}.

\begin{table}[htb!]
\singlespace
\begin{center}
\caption{Values of $f_{\gamma}$ at six NaCl concentrations.}
\label{table:frac_gamma}
    \begin{tabular}{| c | c c c| } \hline
    NaCl (mg/ml) & $f_{\gamma}$ & $n_\alpha$ & $n_\gamma$ \\ \hline
60  & 0.04 & 143 & 6 \\
90 & 0.10 & 109 & 12 \\
150 & 0.14 & 102 & 17 \\
200 & 0.54 & 21 & 25 \\
250 & 0.56 & 80 & 103 \\
300 & 0.89 & 1 & 8\\ \hline
	\end{tabular}\\
\end{center}
\vspace*{0.3cm}
$f_{\gamma}$ is the fraction of wells at the end of the experiment,
where the the crystal(s) are of the $\gamma$ polymorph.
Also shown are the total numbers of wells with $\alpha$ and with
$\gamma$ crystals, $n_{\alpha}$ and $n_{\gamma}$, respectively.
\end{table}

\begin{figure}[htb!]
\singlespace
\centering
(a)\includegraphics[scale = 0.85]{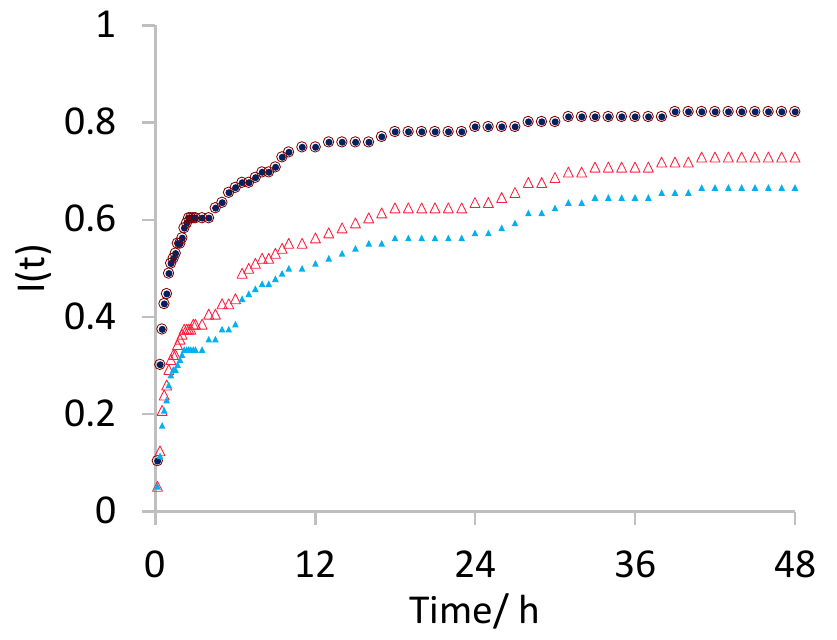}
(b)\includegraphics[scale = 0.85]{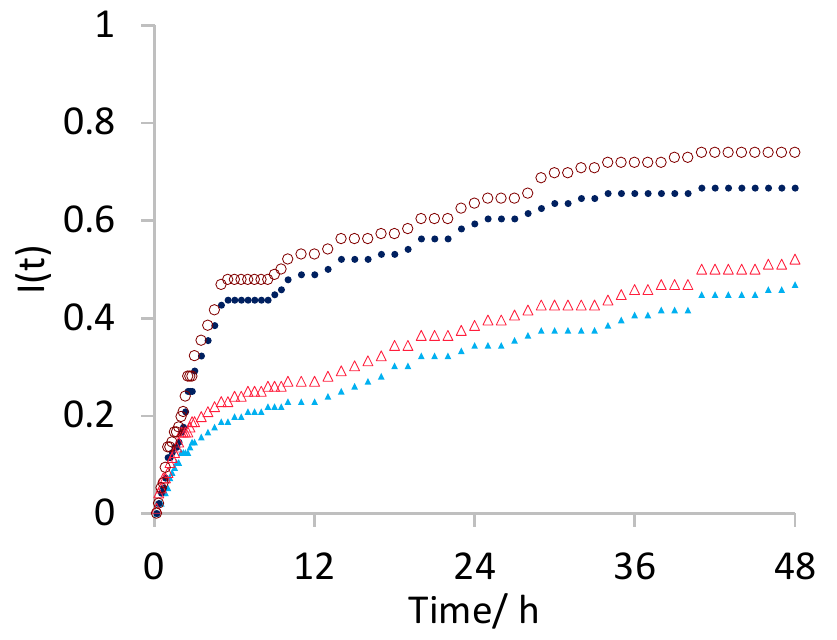}
(c)\includegraphics[scale = 0.85]{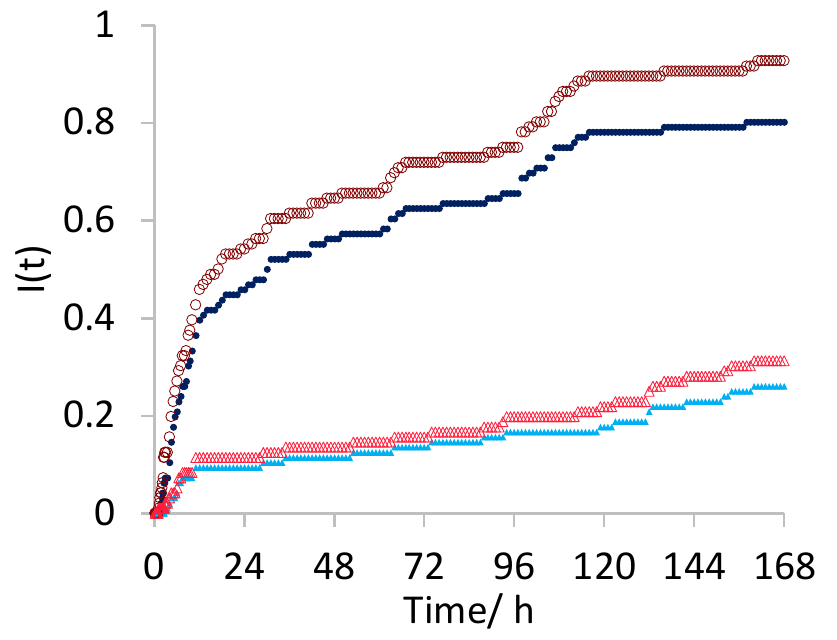}
(d)\includegraphics[scale = 0.85]{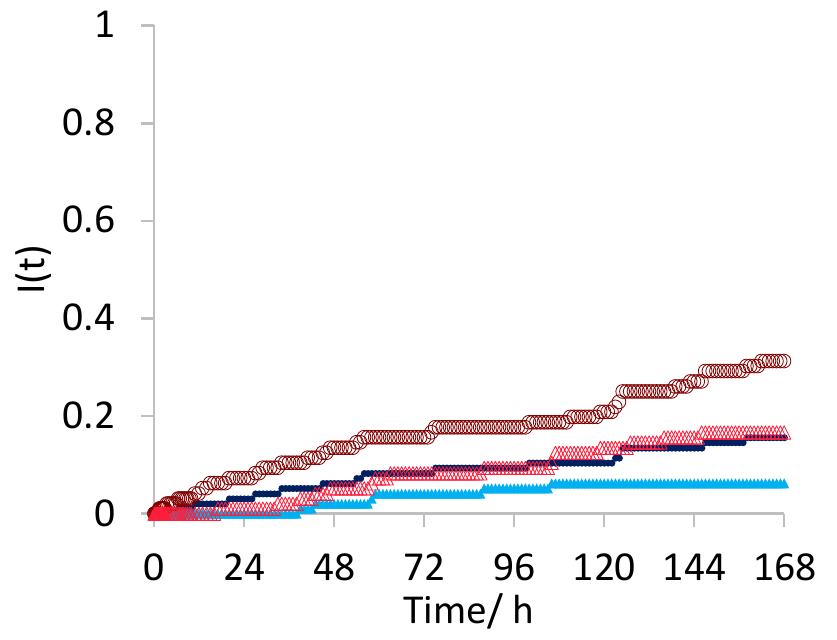}
(e)\includegraphics[scale = 0.85]{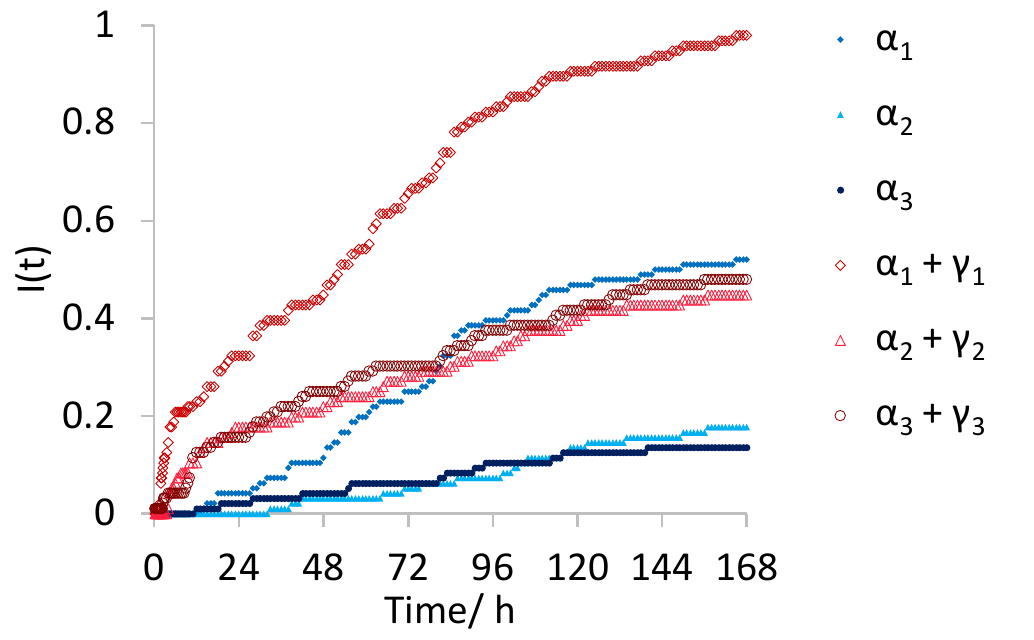}
\caption{CIFs for individual runs at NaCl concentrations:
(a) 60 mg/ml, (b) 90 mg/ml, (c) 150 mg/ml, (d) 200 mg/ml and (e) 250 mg/ml.
The key in (e) applies to all 5 plots. There are 2 runs in (a) to (d), and 3 runs in (e).
For each run we plot $I_{\alpha}$ as closed blue symbols, and $I_{\alpha}+I_{\gamma}$ as open red/brown symbols.
For example, in (a) for the first run no $\gamma$ crystals form, so the solid and open
symbols are on top of each other as $I_{\gamma}=0$, while for the second run a small
number of $\gamma$ crystals start to form after a few hours, and so the pale blue closed
and open red symbols move apart.
} \label{Induction_runs}
\end{figure}

\subsection{Rates of nucleation, and total amount of nucleation are not well reproducible, relative rates and fractions are reproducible}

The nucleation time cumulative incidence functions (CIFs)
are plotted for each individual
run in Figure \ref{Induction_runs}. The CIFs in
Figure \ref{Induction_curvesSS}
were obtained by combining the data in these individual runs.
The number of wells in which nucleation occurs, varies
from run to run, i.e., reproducibility of the amount of nucleation is poor.
What is reproducible, is the fraction of nucleation events of
each polymorph. In
Figure \ref{Induction_runs}, $\gamma$ nucleation corresponds to the difference between the all-nucleation CIFs and the $\alpha$ nucleation CIFs for each run. We see at low NaCl concentrations nucleation is dominated by the $\alpha$ polymorph and this phenomenon is reproducible between runs. Even when two runs have a significantly different amount of nucleation occurring, the relative amount of the nucleation of each polymorph is similar.
For example at 150 mg/ml NaCl, see
Figure \ref{Induction_runs}(c), one run (dark red) has about three times more nucleation events than the other (light red). However, for both runs, about 15\% of the total nucleation events give the $\gamma$ polymorph.

It should also be noted that the distribution of nucleation times is reproducibly different for the two polymorphs at the high salt concentration of 250 mg/ml NaCl,
see Figure \ref{Induction_runs}. The $\gamma$ nucleation time distribution initially has a very fast relative nucleation rate which slows down over time in all of the runs, while the $\alpha$ nucleation time distribution initially has a very slow relative nucleation rate which speeds up over time giving an `s-shaped' curve. This is well illustrated by the median nucleation times of the subpopulations of $\alpha$ nucleating wells and $\gamma$ nucleating wells for each run. The median $\alpha$ nucleation times for the three runs are 77 h, 102 h and 81 h, while median $\gamma$ nucleation times for those runs are 13.5 h, 14 h and 34 h respectively. We can therefore say, in addition to the fraction of nucleation that corresponds to each polymorph being reproducible, the relative change in the effective nucleation rates of the two polymorphs is also reproducible.

\begin{figure}[ht]
\centering
(a)\includegraphics[scale = 0.85]{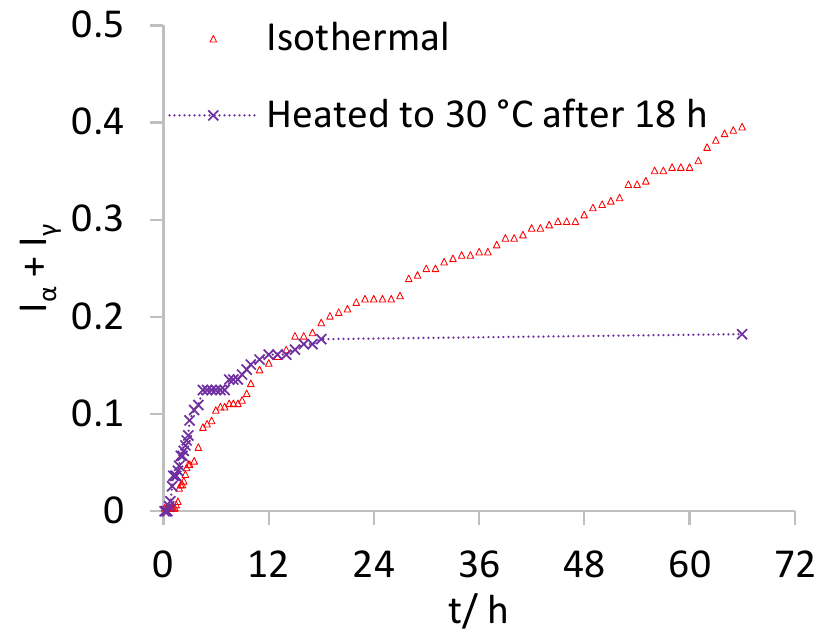}
(b)\includegraphics[scale=0.85]{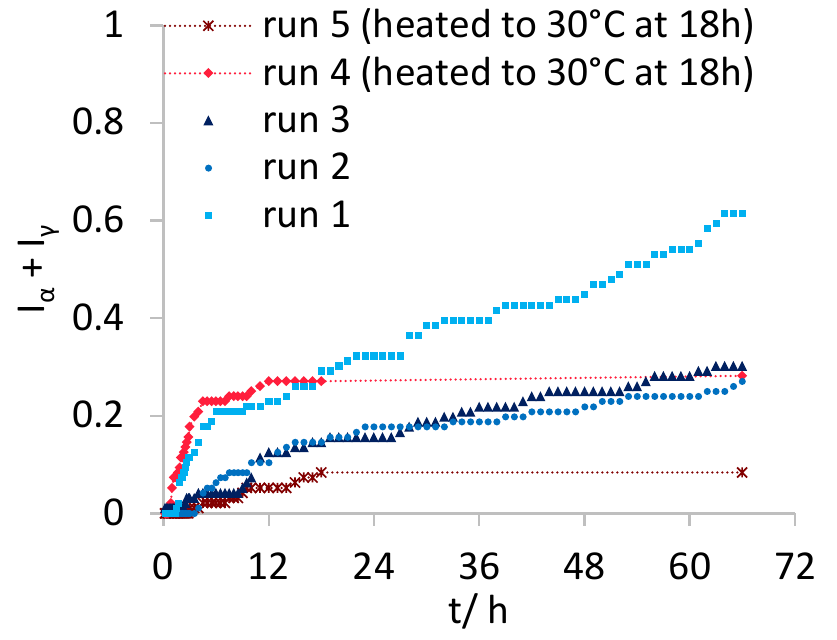}
\caption{Plots of the sum of the two CIF functions, $I_{\alpha}+I_{\gamma}$,
i.e., the fraction of wells where crystallisation has occurred, is plotted
as a function of time. This is at 250 mg/ml salt.
(a) The red triangles are the average $I_{\alpha}+I_{\gamma}$,
for all three isothermal runs, and the purple crosses are
is the average of two runs that are at
21 $\degree$C for the first 18 h,
after which the microplate is maintained at
30.4 $\degree$C for the following 48 h.
(b) Here we have plotted the individual runs of the systems that
were averaged to obtain the data in (a). This is two runs
heated to $30^{\degree}$C after 18 h (runs 4 and 5),
and the isothermal individual runs (runs 1, 2 and 3).
}\label{18h_heating}
\end{figure}

\subsection{Time-dependent supersaturation increases polymorph purity}

In Figure \ref{18h_heating}(a), we have plotted the sum of the two CIFs, $I_{\alpha}+I_{\gamma}$
for both isothermal experiments, and experiments where the temperature
is increased from room temperature (close to 21 $\degree$C \cite{little15}) to 
30.4 $\degree$C, after 18 hours. All runs are at 250 mg/ml NaCl.
Note that for the experiments warmed to
30.4 $\degree$C, nucleation is almost completely stopped.
In Figure \ref{18h_heating}(b), we
show the individual runs that make up
Figure \ref{18h_heating}(a).
We can see the trends observed in the individual runs are the same as we
observe in the datasets where all runs under the same conditions are combined.

\begin{figure}[htb]
\centering
\singlespace
\includegraphics[scale = 1.2]{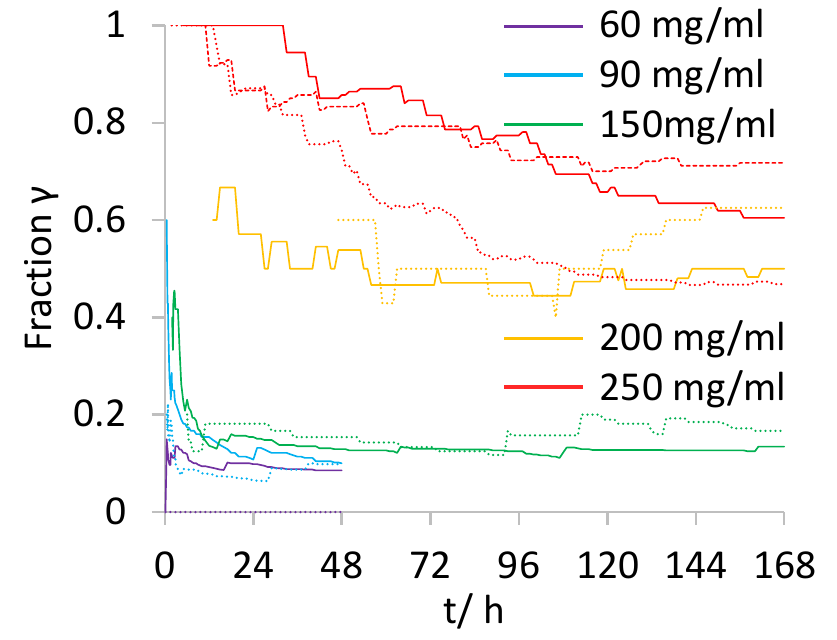}
\caption{Polymorph composition as a function of time. The composition is the fraction of the wells where crystallisation has occured, that contain the $\gamma$ polymorph. The lines start at the point when five nucleation events have occurred. Each curve is one run, and the colour indicates the salt concentration. Different runs at the same concentration are distinguished by being solid, dashed and dotted.} \label{purity_vs_time_individual_runs}
\end{figure}

\subsection{Individual purity vs times}\label{app:indiv_purity_vs_time}

We have looked at how polymorph composition varies over time. Here we show that the runs that make up the datasets plotted in
Figure 4 follow the same trends. The runs are plotted in
Figure \ref{purity_vs_time_individual_runs}, and data is shown in
Table \ref{frac_gamma_table_indiv}.

\begin{table}[htb!]
\begin{center}
\singlespace
\caption{Results of individual runs,
for the fraction of $\gamma$ at early times, and at the end.}
\label{frac_gamma_table_indiv}
    \begin{tabular}{ |c | c c |c c |}
\hline
    NaCl conc / mg/ml & Fraction $\gamma$ at $n_{\ge10}$ & Final $\gamma$ fraction &
 $n_{\ge 10}$ & $t_{n\ge 10}$ / h
    \\ \hline
60  & 0.0 & 0.0 &  10 & 0.16 \\
60  & 0.083 & 0.086 &  12 & 0.33 \\ \hline
90  & 0.30 & 0.10   & 10 & 1.16 \\
90  & 0.15 & 0.099 &  13 & 1.0 \\ \hline
150 & 0.45 & 0.13 &  11 & 2.5 \\
150 & 0.18 & 0.17 &  11 & 11 \\ \hline
200 & 0.50 & 0.50 &  10 & 34 \\
200 & 0.40 & 0.63 &  10 & 106 \\ \hline
250 & 1.0 & 0.47 &  10 & 2.66\\
250 & 1.0 & 0.60 &  10 & 10\\
250 & 1.0 & 0.72 &  11 & 11\\
\hline
\end{tabular}
\end{center}
\vspace*{0.3cm}
The second column is the fraction of crystals in the $\gamma$ polymorph, at a time, $t_{n\ge 10}$, early in the experiment.
$t_{n\ge 10}$ is the earliest observation time at which we have 10 or more nucleation events; the precise number of nucleation events,
$n_{\ge 10}$, is in column four.
The third column is the fraction of the $\gamma$ polymorph
at the end of the experiment.

\end{table}

\section{Additional results on growth rates and  crystal habits}

\begin{figure}[ht]
\centering
\includegraphics[scale=1.1]{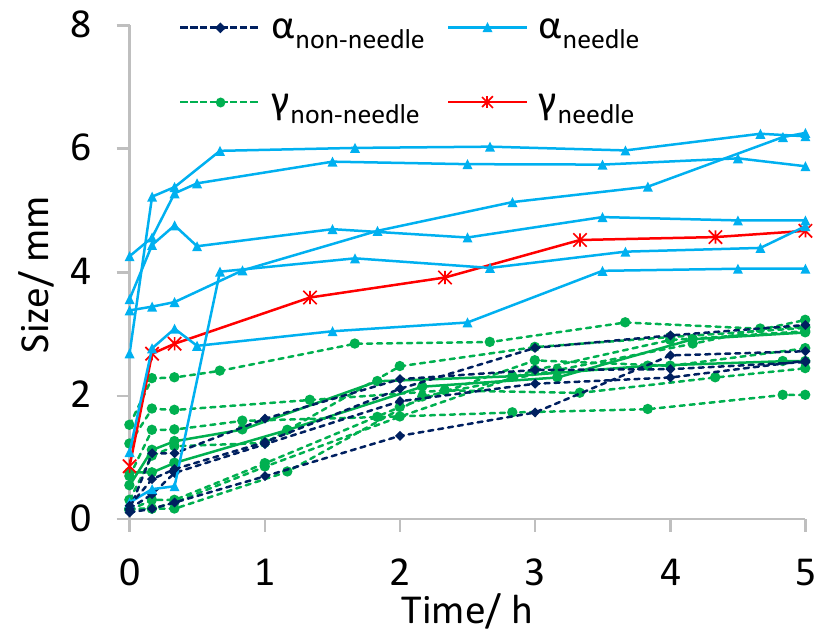}
\caption{Plot of the size of a crystal, as a function of time.
The NaCl concentrations is 90 mg/ml.
Data is shown for 10 $\alpha$ and 10 $\gamma$ crystals.
The sizes of $\alpha$ crystals are shown as
light blue (needles) and black (non-needles) lines-and-points.
The sizes of $\gamma$ crystals are shown as
red (needles) and green (non-needles) lines-and-points.
For each crystal $t = 0$ is defined as the time of the first image in which there is a visible crystal.
} \label{growthsSS}
\end{figure}

\subsection{The effect of growth rates on the error in nucleation time measurements}

Our measurements for nucleation times are only accurate if the time for nucleation,
i.e., for the crystal to cross the nucleation barrier and start growing
irreversibly, is much larger than the time taken for the crystal
to grow from just past the barrier, to a size large enough to be visible.

We have plotted the sizes of 10 $\alpha$ and 10 $\gamma$ crystals, at 90 mg/ml NaCl in Figure \ref{growthsSS}. Data for 250 mg/ml NaCl are in Figure 5 of the main text.
Note that growth rates vary with salt concentration.
The crystals grow faster at 90 mg/ml of salt than 250 mg/ml.
We focus on the $\gamma$ crystals for estimating nucleaction time errors as their growth is slightly slower than that of the $\alpha$ crystals. Here the growth rate for the $\gamma$ crystals is around 2 mm/h. This means the error in our nucleation time measurements should be less than 10 minutes in most cases.
This is a small error for all but the very shortest nucleation times.

For a salt concentration of 250 mg/ml, we plotted the sizes of 10 $\alpha$ and 10 $\gamma$ crystals,
as a function of time, in Figure 5 of the main text.
Most but not all of the $\gamma$ crystals are growing at rate of
around 0.2 mm/h. This implies that our measured nucleation times are on average just over 30 mins too long, i.e., we first see a $\gamma$ crystal about 30 mins after it
nucleated as a microscopic nucleus.
A few $\alpha$ crystals are also growing at around 0.2 mm/h, but most are
initially growing of order 10 times faster.
Our runs at 250 mg/ml salt are of 168 h duration, so except for nucleation
at early times, a 30 mins error is a relatively small error.

\begin{table}[ht]
\begin{center}
\caption{Mean glycine sizes at initial detection
and at two subsequent times, for both $\alpha$ and $\gamma$ polymorphs.}
\label{growth_IQR_table}
    \begin{tabular}{| c | c | c c | c c|}
\hline
  NaCl conc / mg/ml &  	Time / mins  & \multicolumn{4}{c|}{Size/ mm} \\
            & & \multicolumn{2}{c|}{mean} &\multicolumn{2}{c|}{Interquartile range} \\
&     & $\alpha$ & $\gamma$ & $\alpha$ & $\gamma$ \\ \hline
90    & 0   & 1.6 $\pm$ 0.5 & 0.6  $\pm$ 0.1  &  3.2 & 0.69 \\
& 20     & 2.5 $\pm$ 0.7 & 1.2  $\pm$ 0.3 & 4.0 & 1.5  \\
   & 300  & 4.3 $\pm$ 0.5 & 3.0  $\pm$ 0.2 & 3.0  & 0.56  \\ \hline
250  & 0    & 0.5 $\pm$ 0.1 & 0.17 $\pm$ 0.02  & 0.74 & 0.11 \\
& 20    & 0.9 $\pm$ 0.2 & 0.23 $\pm$ 0.01 & 1.2  & 0.025 \\
  & 300 & 1.8 $\pm$ 0.3 & 0.9  $\pm$ 0.1  & 2.3 & 0.23 \\ \hline

	\end{tabular}
\end{center}
We estimate the uncertainties of the mean values with error
estimates that are the standard deviation of the measured sizes,
divided by the square root of the number of crystals measured (10 crystals of each polymorph at each concentration).
We measure the width of the distribution of sizes of the crystals
via the interquartile range.
\end{table}

\subsection{Variation in growth rates between crystals}

In Table \ref{growth_IQR_table}, we present crystal sizes at three times,
from the same data set as used in Figure 5 and Figure \ref{growthsSS}.
In this table, the interquartile range (IQR) is $Q_3-Q_1$, where $Q_1$ is the first quartile
of the distribution of crystal sizes, i.e., the size at which 25\% of the
crystals are smaller and 75\% are larger, and $Q_3$ is the size
where 75\% are smaller and 25\% are larger. The IQR is a convenient
measure of the spread in crystal sizes as it is relatively insensitive
to outliers (unlike the standard deviation), and we have
outliers in the crystal size, see Figure \ref{growthsSS}.
It is clear that the growth rates of both polymorphs vary widely between one crystal and another.
This is especially true for the $\alpha$ polymorph where at early times the IQR is larger than the mean.

\begin{figure}[ht]
\centering
(a)\includegraphics[scale=0.9]{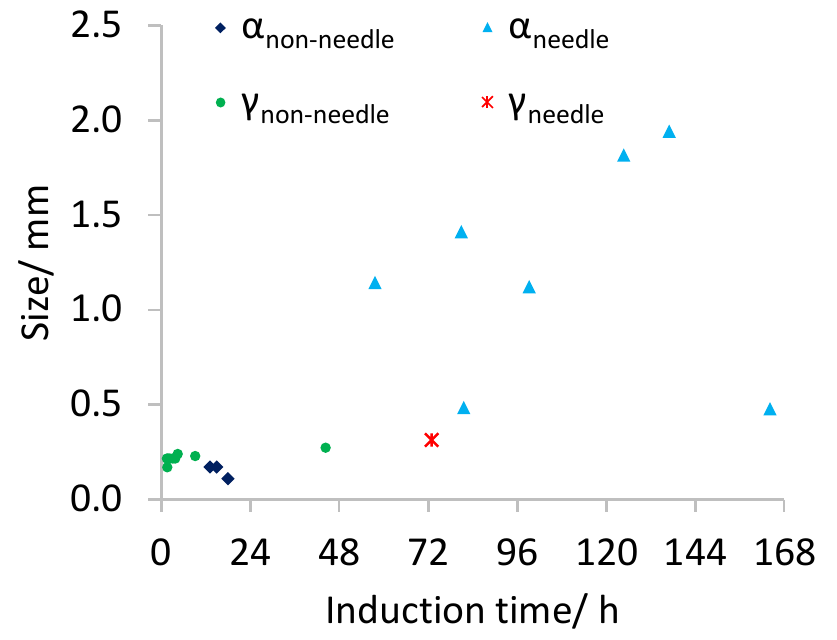}
(b)\includegraphics[scale=0.9]{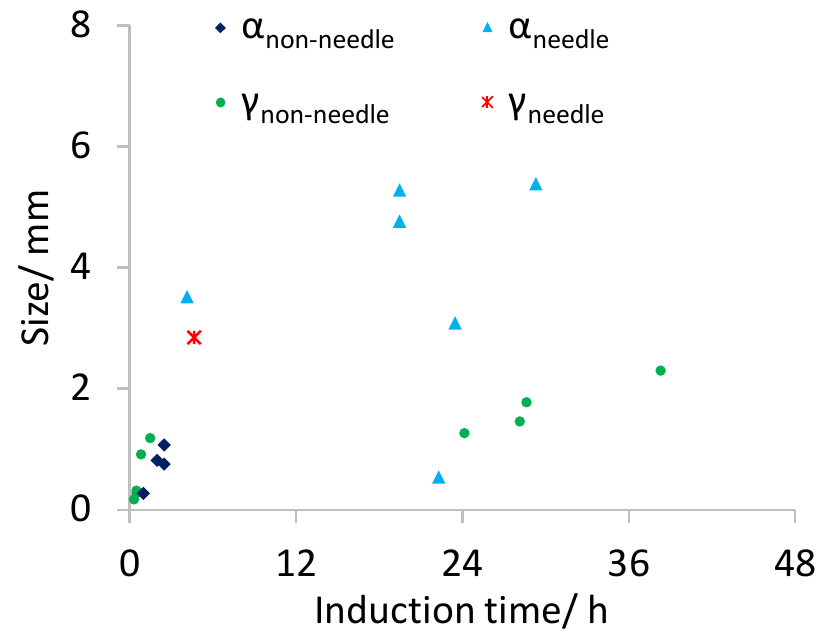}
\caption{Here we plot the size of crystals 20 mins after they were initially detected,
as a function of the time at which they nucleated. In (a) the salt
concentration is 250 mg/ml, and in (b) it is 90 mg/ml.}
\label{habit_vs_nucleation_times}
\end{figure}

\subsection{Growth rate and nucleation time}

In the main text we showed that needle-like crystals tend to have faster growth rates. We have also showed that crystals that nucleate at later times are more likely to be needle-like. It follows that the crystals with longer nucleation times generally have faster growth rates. As we have discussed, the growth rate of crystals is difficult to quantify for our data because the growth rate changes with time. Here we plot the size of a crystal 20 mins after it is initially detected against nucleation time. This is shown
in Figure \ref{habit_vs_nucleation_times}.
We see that at both salt concentrations, the earliest nucleating crystals
are small at 20 mins.

\begin{figure}[ht]
\centering
\includegraphics[scale=0.9]{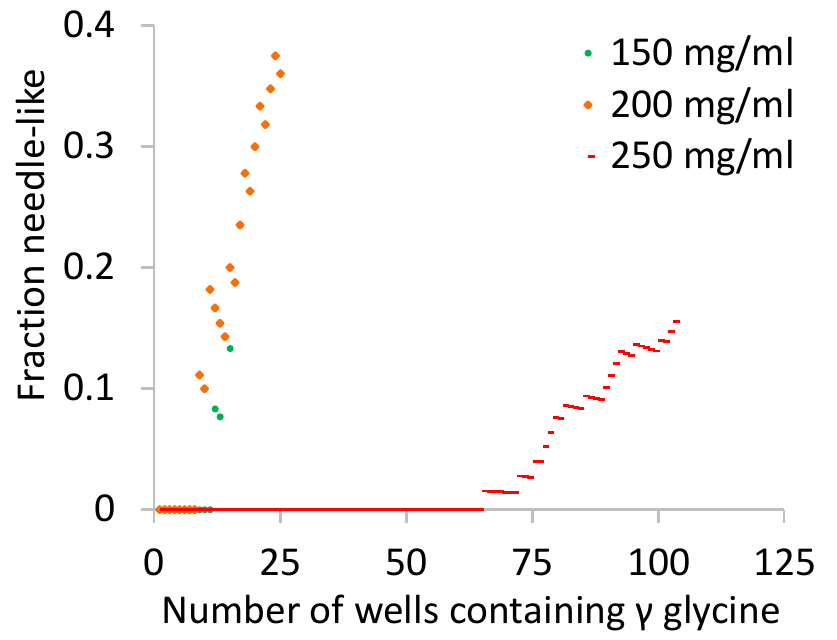}
\caption{For the $\gamma$ polymorph, we plot the fraction of crystals
 with a needle-like habit, as a function of the total
number of wells that contain crystals of that polymorph.
} \label{needle-like_induction_timesSS}
\end{figure}

\subsection{Nucleation time and crystal habit}

For $\gamma$ crystals at high salt concentrations, we observe
the same correlation between crystal habit and nucleation time, as we did
for the $\alpha$ polymorph (Figure 7). The data for the $\gamma$ polymorph are shown in
Figure \ref{needle-like_induction_timesSS}.
We see the same trend as for the $\alpha$ polymorph, with non-needle-like
crystals nucleating at early times, and needle-like crystals at late times, although
the fraction of needles is lower for the $\gamma$ polymorph.
At low salt concentrations, there article too few $\gamma$ crystals to make clear statements.
For example, at 60 mg/ml and 90 mg/ml there are only six and twelve $\gamma$ nucleation events in total, respectively.


\begin{thebibliography}{0}%
\makeatletter
\providecommand \@ifxundefined [1]{%
 \@ifx{#1\undefined}
}%
\providecommand \@ifnum [1]{%
 \ifnum #1\expandafter \@firstoftwo
 \else \expandafter \@secondoftwo
 \fi
}%
\providecommand \@ifx [1]{%
 \ifx #1\expandafter \@firstoftwo
 \else \expandafter \@secondoftwo
 \fi
}%
\providecommand \natexlab [1]{#1}%
\providecommand \enquote  [1]{``#1''}%
\providecommand \bibnamefont  [1]{#1}%
\providecommand \bibfnamefont [1]{#1}%
\providecommand \citenamefont [1]{#1}%
\providecommand \href@noop [0]{\@secondoftwo}%
\providecommand \href [0]{\begingroup \@sanitize@url \@href}%
\providecommand \@href[1]{\@@startlink{#1}\@@href}%
\providecommand \@@href[1]{\endgroup#1\@@endlink}%
\providecommand \@sanitize@url [0]{\catcode `\\12\catcode `\$12\catcode
  `\&12\catcode `\#12\catcode `\^12\catcode `\_12\catcode `\%12\relax}%
\providecommand \@@startlink[1]{}%
\providecommand \@@endlink[0]{}%
\providecommand \url  [0]{\begingroup\@sanitize@url \@url }%
\providecommand \@url [1]{\endgroup\@href {#1}{\urlprefix }}%
\providecommand \urlprefix  [0]{URL }%
\providecommand \Eprint [0]{\href }%
\providecommand \doibase [0]{http://dx.doi.org/}%
\providecommand \selectlanguage [0]{\@gobble}%
\providecommand \bibinfo  [0]{\@secondoftwo}%
\providecommand \bibfield  [0]{\@secondoftwo}%
\providecommand \translation [1]{[#1]}%
\providecommand \BibitemOpen [0]{}%
\providecommand \bibitemStop [0]{}%
\providecommand \bibitemNoStop [0]{.\EOS\space}%
\providecommand \EOS [0]{\spacefactor3000\relax}%
\providecommand \BibitemShut  [1]{\csname bibitem#1\endcsname}%
\let\auto@bib@innerbib\@empty
\end{thebibliography}%


\begin{thebibliography}{49}%
\makeatletter
\providecommand \@ifxundefined [1]{%
 \@ifx{#1\undefined}
}%
\providecommand \@ifnum [1]{%
 \ifnum #1\expandafter \@firstoftwo
 \else \expandafter \@secondoftwo
 \fi
}%
\providecommand \@ifx [1]{%
 \ifx #1\expandafter \@firstoftwo
 \else \expandafter \@secondoftwo
 \fi
}%
\providecommand \natexlab [1]{#1}%
\providecommand \enquote  [1]{``#1''}%
\providecommand \bibnamefont  [1]{#1}%
\providecommand \bibfnamefont [1]{#1}%
\providecommand \citenamefont [1]{#1}%
\providecommand \href@noop [0]{\@secondoftwo}%
\providecommand \href [0]{\begingroup \@sanitize@url \@href}%
\providecommand \@href[1]{\@@startlink{#1}\@@href}%
\providecommand \@@href[1]{\endgroup#1\@@endlink}%
\providecommand \@sanitize@url [0]{\catcode `\\12\catcode `\$12\catcode
  `\&12\catcode `\#12\catcode `\^12\catcode `\_12\catcode `\%12\relax}%
\providecommand \@@startlink[1]{}%
\providecommand \@@endlink[0]{}%
\providecommand \url  [0]{\begingroup\@sanitize@url \@url }%
\providecommand \@url [1]{\endgroup\@href {#1}{\urlprefix }}%
\providecommand \urlprefix  [0]{URL }%
\providecommand \Eprint [0]{\href }%
\providecommand \doibase [0]{http://dx.doi.org/}%
\providecommand \selectlanguage [0]{\@gobble}%
\providecommand \bibinfo  [0]{\@secondoftwo}%
\providecommand \bibfield  [0]{\@secondoftwo}%
\providecommand \translation [1]{[#1]}%
\providecommand \BibitemOpen [0]{}%
\providecommand \bibitemStop [0]{}%
\providecommand \bibitemNoStop [0]{.\EOS\space}%
\providecommand \EOS [0]{\spacefactor3000\relax}%
\providecommand \BibitemShut  [1]{\csname bibitem#1\endcsname}%
\let\auto@bib@innerbib\@empty
\bibitem [{\citenamefont {Bauer}\ \emph {et~al.}(2001)\citenamefont {Bauer},
  \citenamefont {Spanton}, \citenamefont {Henry}, \citenamefont {Quick},
  \citenamefont {Dziki}, \citenamefont {Porter},\ and\ \citenamefont
  {Morris}}]{bauer01}%
  \BibitemOpen
  \bibfield  {author} {\bibinfo {author} {\bibfnamefont {J.}~\bibnamefont
  {Bauer}}, \bibinfo {author} {\bibfnamefont {S.}~\bibnamefont {Spanton}},
  \bibinfo {author} {\bibfnamefont {R.}~\bibnamefont {Henry}}, \bibinfo
  {author} {\bibfnamefont {J.}~\bibnamefont {Quick}}, \bibinfo {author}
  {\bibfnamefont {W.}~\bibnamefont {Dziki}}, \bibinfo {author} {\bibfnamefont
  {W.}~\bibnamefont {Porter}}, \ and\ \bibinfo {author} {\bibfnamefont
  {J.}~\bibnamefont {Morris}},\ }\href {\doibase 10.1023/A:1011052932607}
  {\bibfield  {journal} {\bibinfo  {journal} {Pharma. Res.}\ }\textbf {\bibinfo
  {volume} {18}},\ \bibinfo {pages} {859} (\bibinfo {year} {2001})}\BibitemShut
  {NoStop}%
\bibitem [{\citenamefont {Morissette}\ \emph {et~al.}(2003)\citenamefont
  {Morissette}, \citenamefont {Soukasene}, \citenamefont {Levinson},
  \citenamefont {Cima},\ and\ \citenamefont {Almarsson}}]{morissette03}%
  \BibitemOpen
  \bibfield  {author} {\bibinfo {author} {\bibfnamefont {S.~L.}\ \bibnamefont
  {Morissette}}, \bibinfo {author} {\bibfnamefont {S.}~\bibnamefont
  {Soukasene}}, \bibinfo {author} {\bibfnamefont {D.}~\bibnamefont {Levinson}},
  \bibinfo {author} {\bibfnamefont {M.~J.}\ \bibnamefont {Cima}}, \ and\
  \bibinfo {author} {\bibfnamefont {O.}~\bibnamefont {Almarsson}},\ }\href@noop
  {} {\bibfield  {journal} {\bibinfo  {journal} {Proc. Nat. Acad. Sci.}\
  }\textbf {\bibinfo {volume} {100}},\ \bibinfo {pages} {2180} (\bibinfo {year}
  {2003})}\BibitemShut {NoStop}%
\bibitem [{\citenamefont {Little}, \citenamefont {Sear},\ and\ \citenamefont
  {Keddie}(2015)}]{little15}%
  \BibitemOpen
  \bibfield  {author} {\bibinfo {author} {\bibfnamefont {L.~J.}\ \bibnamefont
  {Little}}, \bibinfo {author} {\bibfnamefont {R.~P.}\ \bibnamefont {Sear}}, \
  and\ \bibinfo {author} {\bibfnamefont {J.~L.}\ \bibnamefont {Keddie}},\
  }\href@noop {} {\bibfield  {journal} {\bibinfo  {journal} {Crys. Growth
  Design}\ }\textbf {\bibinfo {volume} {15}},\ \bibinfo {pages} {5345}
  (\bibinfo {year} {2015})}\BibitemShut {NoStop}%
\bibitem [{\citenamefont {{R. P. Sear}}(2014)}]{sear_rev14}%
  \BibitemOpen
  \bibfield  {author} {\bibinfo {author} {\bibnamefont {{R. P. Sear}}},\ }\href
  {\doibase 10.1039/C4CE00344F} {\bibfield  {journal} {\bibinfo  {journal}
  {CrystEngComm}\ }\textbf {\bibinfo {volume} {16}},\ \bibinfo {pages} {6506}
  (\bibinfo {year} {2014})}\BibitemShut {NoStop}%
\bibitem [{\citenamefont {Diao}\ \emph {et~al.}(2012)\citenamefont {Diao},
  \citenamefont {Whaley}, \citenamefont {Helgeson}, \citenamefont {Woldeyes},
  \citenamefont {Doyle}, \citenamefont {Myerson}, \citenamefont {Hatton},\ and\
  \citenamefont {Trout}}]{diao12}%
  \BibitemOpen
  \bibfield  {author} {\bibinfo {author} {\bibfnamefont {Y.}~\bibnamefont
  {Diao}}, \bibinfo {author} {\bibfnamefont {K.~E.}\ \bibnamefont {Whaley}},
  \bibinfo {author} {\bibfnamefont {M.~E.}\ \bibnamefont {Helgeson}}, \bibinfo
  {author} {\bibfnamefont {M.~A.}\ \bibnamefont {Woldeyes}}, \bibinfo {author}
  {\bibfnamefont {P.~S.}\ \bibnamefont {Doyle}}, \bibinfo {author}
  {\bibfnamefont {A.~S.}\ \bibnamefont {Myerson}}, \bibinfo {author}
  {\bibfnamefont {T.~A.}\ \bibnamefont {Hatton}}, \ and\ \bibinfo {author}
  {\bibfnamefont {B.~L.}\ \bibnamefont {Trout}},\ }\href {\doibase
  10.1021/ja210006t} {\bibfield  {journal} {\bibinfo  {journal} {J. Am. Chem.
  Soc.}\ }\textbf {\bibinfo {volume} {134}},\ \bibinfo {pages} {673} (\bibinfo
  {year} {2012})}\BibitemShut {NoStop}%
\bibitem [{\citenamefont {Toldy}\ \emph {et~al.}(2012)\citenamefont {Toldy},
  \citenamefont {Badruddoza}, \citenamefont {Zheng}, \citenamefont {Hatton},
  \citenamefont {Gunawan}, \citenamefont {Rajagopalan},\ and\ \citenamefont
  {Khan}}]{toldy12}%
  \BibitemOpen
  \bibfield  {author} {\bibinfo {author} {\bibfnamefont {A.~I.}\ \bibnamefont
  {Toldy}}, \bibinfo {author} {\bibfnamefont {A.~Z.~M.}\ \bibnamefont
  {Badruddoza}}, \bibinfo {author} {\bibfnamefont {L.}~\bibnamefont {Zheng}},
  \bibinfo {author} {\bibfnamefont {T.~A.}\ \bibnamefont {Hatton}}, \bibinfo
  {author} {\bibfnamefont {R.}~\bibnamefont {Gunawan}}, \bibinfo {author}
  {\bibfnamefont {R.}~\bibnamefont {Rajagopalan}}, \ and\ \bibinfo {author}
  {\bibfnamefont {S.~A.}\ \bibnamefont {Khan}},\ }\href@noop {} {\bibfield
  {journal} {\bibinfo  {journal} {Crys. Growth Des.}\ }\textbf {\bibinfo
  {volume} {12}},\ \bibinfo {pages} {3977} (\bibinfo {year}
  {2012})}\BibitemShut {NoStop}%
\bibitem [{\citenamefont {Kim}\ \emph {et~al.}(2013)\citenamefont {Kim},
  \citenamefont {Park}, \citenamefont {Shim},\ and\ \citenamefont
  {Koo}}]{kim13}%
  \BibitemOpen
  \bibfield  {author} {\bibinfo {author} {\bibfnamefont {J.-W.}\ \bibnamefont
  {Kim}}, \bibinfo {author} {\bibfnamefont {J.-H.}\ \bibnamefont {Park}},
  \bibinfo {author} {\bibfnamefont {H.-M.}\ \bibnamefont {Shim}}, \ and\
  \bibinfo {author} {\bibfnamefont {K.-K.}\ \bibnamefont {Koo}},\ }\href
  {\doibase 10.1021/cg4006423} {\bibfield  {journal} {\bibinfo  {journal}
  {Crys. Growth Des.}\ }\textbf {\bibinfo {volume} {13}},\ \bibinfo {pages}
  {4688} (\bibinfo {year} {2013})}\BibitemShut {NoStop}%
\bibitem [{\citenamefont {Brandel}\ and\ \citenamefont {ter
  Horst}(2015)}]{brandel15}%
  \BibitemOpen
  \bibfield  {author} {\bibinfo {author} {\bibfnamefont {C.}~\bibnamefont
  {Brandel}}\ and\ \bibinfo {author} {\bibfnamefont {J.~H.}\ \bibnamefont {ter
  Horst}},\ }\href {\doibase 10.1039/C4FD00230J} {\bibfield  {journal}
  {\bibinfo  {journal} {Faraday Discuss.}\ }\textbf {\bibinfo {volume} {179}},\
  \bibinfo {pages} {199} (\bibinfo {year} {2015})}\BibitemShut {NoStop}%
\bibitem [{\citenamefont {Akella}\ \emph {et~al.}(2014)\citenamefont {Akella},
  \citenamefont {Mowitz}, \citenamefont {Heymann},\ and\ \citenamefont
  {Fraden}}]{akella14}%
  \BibitemOpen
  \bibfield  {author} {\bibinfo {author} {\bibfnamefont {S.}~\bibnamefont
  {Akella}}, \bibinfo {author} {\bibfnamefont {A.}~\bibnamefont {Mowitz}},
  \bibinfo {author} {\bibfnamefont {M.}~\bibnamefont {Heymann}}, \ and\
  \bibinfo {author} {\bibfnamefont {S.}~\bibnamefont {Fraden}},\ }\href@noop {}
  {\bibfield  {journal} {\bibinfo  {journal} {Crys. Growth Design}\ }\textbf
  {\bibinfo {volume} {14}},\ \bibinfo {pages} {4487} (\bibinfo {year}
  {2014})}\BibitemShut {NoStop}%
\bibitem [{\citenamefont {Duft}\ and\ \citenamefont {Leisner}(2004)}]{duft04}%
  \BibitemOpen
  \bibfield  {author} {\bibinfo {author} {\bibfnamefont {D.}~\bibnamefont
  {Duft}}\ and\ \bibinfo {author} {\bibfnamefont {T.}~\bibnamefont {Leisner}},\
  }\href@noop {} {\bibfield  {journal} {\bibinfo  {journal} {Atmos. Chem.
  Phys.}\ }\textbf {\bibinfo {volume} {4}},\ \bibinfo {pages} {1997} (\bibinfo
  {year} {2004})}\BibitemShut {NoStop}%
\bibitem [{\citenamefont {Laval}, \citenamefont {Crombez},\ and\ \citenamefont
  {Salmon}(2009)}]{laval09}%
  \BibitemOpen
  \bibfield  {author} {\bibinfo {author} {\bibfnamefont {P.}~\bibnamefont
  {Laval}}, \bibinfo {author} {\bibfnamefont {A.}~\bibnamefont {Crombez}}, \
  and\ \bibinfo {author} {\bibfnamefont {J.-B.}\ \bibnamefont {Salmon}},\
  }\href@noop {} {\bibfield  {journal} {\bibinfo  {journal} {Langmuir}\
  }\textbf {\bibinfo {volume} {25}},\ \bibinfo {pages} {1836} (\bibinfo {year}
  {2009})}\BibitemShut {NoStop}%
\bibitem [{\citenamefont {Javid}\ \emph {et~al.}(2016)\citenamefont {Javid},
  \citenamefont {Kendall}, \citenamefont {Burns},\ and\ \citenamefont
  {Sefcik}}]{javid16}%
  \BibitemOpen
  \bibfield  {author} {\bibinfo {author} {\bibfnamefont {N.}~\bibnamefont
  {Javid}}, \bibinfo {author} {\bibfnamefont {T.}~\bibnamefont {Kendall}},
  \bibinfo {author} {\bibfnamefont {I.~S.}\ \bibnamefont {Burns}}, \ and\
  \bibinfo {author} {\bibfnamefont {J.}~\bibnamefont {Sefcik}},\ }\href@noop {}
  {\bibfield  {journal} {\bibinfo  {journal} {Crystal Growth \& Design}\
  }\textbf {\bibinfo {volume} {16}},\ \bibinfo {pages} {4196} (\bibinfo {year}
  {2016})}\BibitemShut {NoStop}%
\bibitem [{\citenamefont {Yang}\ \emph {et~al.}(2008)\citenamefont {Yang},
  \citenamefont {Lu}, \citenamefont {Wang},\ and\ \citenamefont
  {Ching}}]{yang08a}%
  \BibitemOpen
  \bibfield  {author} {\bibinfo {author} {\bibfnamefont {X.}~\bibnamefont
  {Yang}}, \bibinfo {author} {\bibfnamefont {J.}~\bibnamefont {Lu}}, \bibinfo
  {author} {\bibfnamefont {X.-J.}\ \bibnamefont {Wang}}, \ and\ \bibinfo
  {author} {\bibfnamefont {C.-B.}\ \bibnamefont {Ching}},\ }\href {\doibase
  http://dx.doi.org/10.1016/j.jcrysgro.2007.11.072} {\bibfield  {journal}
  {\bibinfo  {journal} {J. Cryst. Growth}\ }\textbf {\bibinfo {volume} {310}},\
  \bibinfo {pages} {604 } (\bibinfo {year} {2008})}\BibitemShut {NoStop}%
\bibitem [{\citenamefont {Han}, \citenamefont {Chow},\ and\ \citenamefont
  {Tan}(2012)}]{han12}%
  \BibitemOpen
  \bibfield  {author} {\bibinfo {author} {\bibfnamefont {G.}~\bibnamefont
  {Han}}, \bibinfo {author} {\bibfnamefont {P.~S.}\ \bibnamefont {Chow}}, \
  and\ \bibinfo {author} {\bibfnamefont {R.~B.~H.}\ \bibnamefont {Tan}},\
  }\href@noop {} {\bibfield  {journal} {\bibinfo  {journal} {Crys. Growth
  Design}\ }\textbf {\bibinfo {volume} {12}},\ \bibinfo {pages} {2213}
  (\bibinfo {year} {2012})}\BibitemShut {NoStop}%
\bibitem [{\citenamefont {Han}, \citenamefont {Chow},\ and\ \citenamefont
  {Tan}(2015)}]{han15}%
  \BibitemOpen
  \bibfield  {author} {\bibinfo {author} {\bibfnamefont {G.}~\bibnamefont
  {Han}}, \bibinfo {author} {\bibfnamefont {P.~S.}\ \bibnamefont {Chow}}, \
  and\ \bibinfo {author} {\bibfnamefont {R.~B.~H.}\ \bibnamefont {Tan}},\
  }\href@noop {} {\bibfield  {journal} {\bibinfo  {journal} {Cryst. Growth
  Design}\ }\textbf {\bibinfo {volume} {15}},\ \bibinfo {pages} {1082}
  (\bibinfo {year} {2015})}\BibitemShut {NoStop}%
\bibitem [{\citenamefont {Srinivasan}\ and\ \citenamefont
  {Arumugam}(2007)}]{srinivasan07}%
  \BibitemOpen
  \bibfield  {author} {\bibinfo {author} {\bibfnamefont {K.}~\bibnamefont
  {Srinivasan}}\ and\ \bibinfo {author} {\bibfnamefont {J.}~\bibnamefont
  {Arumugam}},\ }\href {\doibase
  http://dx.doi.org/10.1016/j.optmat.2006.11.049} {\bibfield  {journal}
  {\bibinfo  {journal} {Optical Materials}\ }\textbf {\bibinfo {volume} {30}},\
  \bibinfo {pages} {40 } (\bibinfo {year} {2007})}\BibitemShut {NoStop}%
\bibitem [{\citenamefont {Srinivasan}(2008)}]{srinivasan08}%
  \BibitemOpen
  \bibfield  {author} {\bibinfo {author} {\bibfnamefont {K.}~\bibnamefont
  {Srinivasan}},\ }\href {\doibase
  http://dx.doi.org/10.1016/j.jcrysgro.2008.10.084} {\bibfield  {journal}
  {\bibinfo  {journal} {J. Cryst. Growth}\ }\textbf {\bibinfo {volume} {311}},\
  \bibinfo {pages} {156 } (\bibinfo {year} {2008})}\BibitemShut {NoStop}%
\bibitem [{\citenamefont {Kim}\ \emph {et~al.}(2011)\citenamefont {Kim},
  \citenamefont {Centrone}, \citenamefont {Hatton},\ and\ \citenamefont
  {Myerson}}]{kim11myerson}%
  \BibitemOpen
  \bibfield  {author} {\bibinfo {author} {\bibfnamefont {K.}~\bibnamefont
  {Kim}}, \bibinfo {author} {\bibfnamefont {A.}~\bibnamefont {Centrone}},
  \bibinfo {author} {\bibfnamefont {T.~A.}\ \bibnamefont {Hatton}}, \ and\
  \bibinfo {author} {\bibfnamefont {A.~S.}\ \bibnamefont {Myerson}},\ }\href
  {\doibase 10.1039/C0CE00394H} {\bibfield  {journal} {\bibinfo  {journal}
  {CrystEngComm}\ }\textbf {\bibinfo {volume} {13}},\ \bibinfo {pages} {1127}
  (\bibinfo {year} {2011})}\BibitemShut {NoStop}%
\bibitem [{\citenamefont {Forsyth}\ \emph {et~al.}(2016)\citenamefont
  {Forsyth}, \citenamefont {Burns}, \citenamefont {Mulheran},\ and\
  \citenamefont {Sefcik}}]{forsyth16}%
  \BibitemOpen
  \bibfield  {author} {\bibinfo {author} {\bibfnamefont {C.}~\bibnamefont
  {Forsyth}}, \bibinfo {author} {\bibfnamefont {I.~S.}\ \bibnamefont {Burns}},
  \bibinfo {author} {\bibfnamefont {P.~A.}\ \bibnamefont {Mulheran}}, \ and\
  \bibinfo {author} {\bibfnamefont {J.}~\bibnamefont {Sefcik}},\ }\href@noop {}
  {\bibfield  {journal} {\bibinfo  {journal} {Cryst. Growth Design}\ }\textbf
  {\bibinfo {volume} {16}},\ \bibinfo {pages} {136} (\bibinfo {year}
  {2016})}\BibitemShut {NoStop}%
\bibitem [{\citenamefont {Poornachary}, \citenamefont {Chow},\ and\
  \citenamefont {Tan}(2008)}]{poornachary08}%
  \BibitemOpen
  \bibfield  {author} {\bibinfo {author} {\bibfnamefont {S.~K.}\ \bibnamefont
  {Poornachary}}, \bibinfo {author} {\bibfnamefont {P.~S.}\ \bibnamefont
  {Chow}}, \ and\ \bibinfo {author} {\bibfnamefont {R.~B.~H.}\ \bibnamefont
  {Tan}},\ }\href@noop {} {\bibfield  {journal} {\bibinfo  {journal} {Crys.
  Growth Des.}\ }\textbf {\bibinfo {volume} {8}},\ \bibinfo {pages} {179}
  (\bibinfo {year} {2008})}\BibitemShut {NoStop}%
\bibitem [{\citenamefont {Chew}\ \emph {et~al.}(2007)\citenamefont {Chew},
  \citenamefont {Black}, \citenamefont {Chow}, \citenamefont {Tan},\ and\
  \citenamefont {Carpenter}}]{chew07}%
  \BibitemOpen
  \bibfield  {author} {\bibinfo {author} {\bibfnamefont {J.~W.}\ \bibnamefont
  {Chew}}, \bibinfo {author} {\bibfnamefont {S.~N.}\ \bibnamefont {Black}},
  \bibinfo {author} {\bibfnamefont {P.~S.}\ \bibnamefont {Chow}}, \bibinfo
  {author} {\bibfnamefont {R.~B.~H.}\ \bibnamefont {Tan}}, \ and\ \bibinfo
  {author} {\bibfnamefont {K.~J.}\ \bibnamefont {Carpenter}},\ }\href@noop {}
  {\bibfield  {journal} {\bibinfo  {journal} {CrystEngComm}\ }\textbf {\bibinfo
  {volume} {9}},\ \bibinfo {pages} {128} (\bibinfo {year} {2007})}\BibitemShut
  {NoStop}%
\bibitem [{\citenamefont {Chen}\ \emph {et~al.}(2011)\citenamefont {Chen},
  \citenamefont {Cook}, \citenamefont {Nicholson},\ and\ \citenamefont
  {Cooper}}]{chen11}%
  \BibitemOpen
  \bibfield  {author} {\bibinfo {author} {\bibfnamefont {C.}~\bibnamefont
  {Chen}}, \bibinfo {author} {\bibfnamefont {O.}~\bibnamefont {Cook}}, \bibinfo
  {author} {\bibfnamefont {C.~E.}\ \bibnamefont {Nicholson}}, \ and\ \bibinfo
  {author} {\bibfnamefont {S.~J.}\ \bibnamefont {Cooper}},\ }\href@noop {}
  {\bibfield  {journal} {\bibinfo  {journal} {Cryst. Growth Design}\ }\textbf
  {\bibinfo {volume} {11}},\ \bibinfo {pages} {2228} (\bibinfo {year}
  {2011})}\BibitemShut {NoStop}%
\bibitem [{\citenamefont {Jawor-Baczynska}\ \emph {et~al.}(2013)\citenamefont
  {Jawor-Baczynska}, \citenamefont {Moore}, \citenamefont {Lee}, \citenamefont
  {McCormick},\ and\ \citenamefont {Sefcik}}]{jawor13}%
  \BibitemOpen
  \bibfield  {author} {\bibinfo {author} {\bibfnamefont {A.}~\bibnamefont
  {Jawor-Baczynska}}, \bibinfo {author} {\bibfnamefont {B.~D.}\ \bibnamefont
  {Moore}}, \bibinfo {author} {\bibfnamefont {H.~S.}\ \bibnamefont {Lee}},
  \bibinfo {author} {\bibfnamefont {A.~V.}\ \bibnamefont {McCormick}}, \ and\
  \bibinfo {author} {\bibfnamefont {J.}~\bibnamefont {Sefcik}},\ }\href
  {\doibase 10.1039/C3FD00066D} {\bibfield  {journal} {\bibinfo  {journal}
  {Faraday Discuss.}\ }\textbf {\bibinfo {volume} {167}},\ \bibinfo {pages}
  {425} (\bibinfo {year} {2013})}\BibitemShut {NoStop}%
\bibitem [{\citenamefont {Han}\ \emph {et~al.}(2013)\citenamefont {Han},
  \citenamefont {Thirunahari}, \citenamefont {Shan~Chow},\ and\ \citenamefont
  {Tan}}]{han13}%
  \BibitemOpen
  \bibfield  {author} {\bibinfo {author} {\bibfnamefont {G.}~\bibnamefont
  {Han}}, \bibinfo {author} {\bibfnamefont {S.}~\bibnamefont {Thirunahari}},
  \bibinfo {author} {\bibfnamefont {P.}~\bibnamefont {Shan~Chow}}, \ and\
  \bibinfo {author} {\bibfnamefont {R.~B.~H.}\ \bibnamefont {Tan}},\
  }\href@noop {} {\bibfield  {journal} {\bibinfo  {journal} {CrystEngComm}\
  }\textbf {\bibinfo {volume} {15}},\ \bibinfo {pages} {1218} (\bibinfo {year}
  {2013})}\BibitemShut {NoStop}%
\bibitem [{\citenamefont {He}\ \emph {et~al.}(2006)\citenamefont {He},
  \citenamefont {Bhamidi}, \citenamefont {Wilson}, \citenamefont {Tan},
  \citenamefont {Kenis},\ and\ \citenamefont {Zukoski}}]{he06}%
  \BibitemOpen
  \bibfield  {author} {\bibinfo {author} {\bibfnamefont {G.}~\bibnamefont
  {He}}, \bibinfo {author} {\bibfnamefont {V.}~\bibnamefont {Bhamidi}},
  \bibinfo {author} {\bibfnamefont {S.~R.}\ \bibnamefont {Wilson}}, \bibinfo
  {author} {\bibfnamefont {R.~B.~H.}\ \bibnamefont {Tan}}, \bibinfo {author}
  {\bibfnamefont {P.~J.~A.}\ \bibnamefont {Kenis}}, \ and\ \bibinfo {author}
  {\bibfnamefont {C.~F.}\ \bibnamefont {Zukoski}},\ }\href@noop {} {\bibfield
  {journal} {\bibinfo  {journal} {Crys. Growth Design}\ }\textbf {\bibinfo
  {volume} {6}},\ \bibinfo {pages} {1746} (\bibinfo {year} {2006})}\BibitemShut
  {NoStop}%
\bibitem [{\citenamefont {Kim}\ \emph {et~al.}(2009)\citenamefont {Kim},
  \citenamefont {Lee}, \citenamefont {Centrone}, \citenamefont {Hatton},\ and\
  \citenamefont {Myerson}}]{kim09glycine}%
  \BibitemOpen
  \bibfield  {author} {\bibinfo {author} {\bibfnamefont {K.}~\bibnamefont
  {Kim}}, \bibinfo {author} {\bibfnamefont {I.~s.}\ \bibnamefont {Lee}},
  \bibinfo {author} {\bibfnamefont {A.}~\bibnamefont {Centrone}}, \bibinfo
  {author} {\bibfnamefont {T.~A.}\ \bibnamefont {Hatton}}, \ and\ \bibinfo
  {author} {\bibfnamefont {A.~S.}\ \bibnamefont {Myerson}},\ }\href {\doibase
  10.1021/ja908055y} {\bibfield  {journal} {\bibinfo  {journal} {J. Am. Chem.
  Soc.}\ }\textbf {\bibinfo {volume} {131}},\ \bibinfo {pages} {18212}
  (\bibinfo {year} {2009})}\BibitemShut {NoStop}%
\bibitem [{\citenamefont {Nicholson}\ \emph {et~al.}(2005)\citenamefont
  {Nicholson}, \citenamefont {Cooper}, \citenamefont {Marcellin},\ and\
  \citenamefont {Jamieson}}]{nicholson05}%
  \BibitemOpen
  \bibfield  {author} {\bibinfo {author} {\bibfnamefont {C.~E.}\ \bibnamefont
  {Nicholson}}, \bibinfo {author} {\bibfnamefont {S.~J.}\ \bibnamefont
  {Cooper}}, \bibinfo {author} {\bibfnamefont {C.}~\bibnamefont {Marcellin}}, \
  and\ \bibinfo {author} {\bibfnamefont {M.~J.}\ \bibnamefont {Jamieson}},\
  }\href@noop {} {\bibfield  {journal} {\bibinfo  {journal} {J. Am. Chem.
  Soc.}\ }\textbf {\bibinfo {volume} {127}},\ \bibinfo {pages} {11894}
  (\bibinfo {year} {2005})}\BibitemShut {NoStop}%
\bibitem [{\citenamefont {Nicholson}\ \emph {et~al.}(2011)\citenamefont
  {Nicholson}, \citenamefont {Chen}, \citenamefont {Mendis},\ and\
  \citenamefont {Cooper}}]{nicholson11}%
  \BibitemOpen
  \bibfield  {author} {\bibinfo {author} {\bibfnamefont {C.~E.}\ \bibnamefont
  {Nicholson}}, \bibinfo {author} {\bibfnamefont {C.}~\bibnamefont {Chen}},
  \bibinfo {author} {\bibfnamefont {B.}~\bibnamefont {Mendis}}, \ and\ \bibinfo
  {author} {\bibfnamefont {S.~J.}\ \bibnamefont {Cooper}},\ }\href@noop {}
  {\bibfield  {journal} {\bibinfo  {journal} {Crys. Growth Design}\ }\textbf
  {\bibinfo {volume} {11}},\ \bibinfo {pages} {363} (\bibinfo {year}
  {2011})}\BibitemShut {NoStop}%
\bibitem [{\citenamefont {Rivera}, \citenamefont {Allis},\ and\ \citenamefont
  {Hudson}(2008)}]{rivera08}%
  \BibitemOpen
  \bibfield  {author} {\bibinfo {author} {\bibfnamefont {S.~A.}\ \bibnamefont
  {Rivera}}, \bibinfo {author} {\bibfnamefont {D.~G.}\ \bibnamefont {Allis}}, \
  and\ \bibinfo {author} {\bibfnamefont {B.~S.}\ \bibnamefont {Hudson}},\
  }\href@noop {} {\bibfield  {journal} {\bibinfo  {journal} {Cryst. Growth
  Design}\ }\textbf {\bibinfo {volume} {8}},\ \bibinfo {pages} {3905} (\bibinfo
  {year} {2008})}\BibitemShut {NoStop}%
\bibitem [{\citenamefont {Duff}\ \emph {et~al.}(2014)\citenamefont {Duff},
  \citenamefont {Dahal}, \citenamefont {Schmit},\ and\ \citenamefont
  {Peters}}]{duff14}%
  \BibitemOpen
  \bibfield  {author} {\bibinfo {author} {\bibfnamefont {N.}~\bibnamefont
  {Duff}}, \bibinfo {author} {\bibfnamefont {Y.~R.}\ \bibnamefont {Dahal}},
  \bibinfo {author} {\bibfnamefont {J.~D.}\ \bibnamefont {Schmit}}, \ and\
  \bibinfo {author} {\bibfnamefont {B.}~\bibnamefont {Peters}},\ }\href@noop {}
  {\bibfield  {journal} {\bibinfo  {journal} {J. Chem. Phys.}\ }\textbf
  {\bibinfo {volume} {140}},\ \bibinfo {eid} {014501} (\bibinfo {year}
  {2014})}\BibitemShut {NoStop}%
\bibitem [{\citenamefont {Shi}\ and\ \citenamefont {Wang}(2005)}]{shi05}%
  \BibitemOpen
  \bibfield  {author} {\bibinfo {author} {\bibfnamefont {Y.}~\bibnamefont
  {Shi}}\ and\ \bibinfo {author} {\bibfnamefont {L.}~\bibnamefont {Wang}},\
  }\href {http://stacks.iop.org/0022-3727/38/i=19/a=024} {\bibfield  {journal}
  {\bibinfo  {journal} {J. Phys. D App. Phys.}\ }\textbf {\bibinfo {volume}
  {38}},\ \bibinfo {pages} {3741} (\bibinfo {year} {2005})}\BibitemShut
  {NoStop}%
\bibitem [{\citenamefont {Sultana}\ and\ \citenamefont
  {Jensen}(2012)}]{sultana12}%
  \BibitemOpen
  \bibfield  {author} {\bibinfo {author} {\bibfnamefont {M.}~\bibnamefont
  {Sultana}}\ and\ \bibinfo {author} {\bibfnamefont {K.~F.}\ \bibnamefont
  {Jensen}},\ }\href@noop {} {\bibfield  {journal} {\bibinfo  {journal} {Cryst.
  Growth Design}\ }\textbf {\bibinfo {volume} {12}},\ \bibinfo {pages} {6260}
  (\bibinfo {year} {2012})}\BibitemShut {NoStop}%
\bibitem [{\citenamefont {Baran}\ and\ \citenamefont
  {Ratajczak}(2005)}]{baran05}%
  \BibitemOpen
  \bibfield  {author} {\bibinfo {author} {\bibfnamefont {J.}~\bibnamefont
  {Baran}}\ and\ \bibinfo {author} {\bibfnamefont {H.}~\bibnamefont
  {Ratajczak}},\ }\href {\doibase http://dx.doi.org/10.1016/j.saa.2004.11.064}
  {\bibfield  {journal} {\bibinfo  {journal} {Spectrochimica Acta A}\ }\textbf
  {\bibinfo {volume} {61}},\ \bibinfo {pages} {1611 } (\bibinfo {year}
  {2005})}\BibitemShut {NoStop}%
\bibitem [{\citenamefont {Cui}\ \emph {et~al.}(2016)\citenamefont {Cui},
  \citenamefont {Stojakovic}, \citenamefont {Kijima},\ and\ \citenamefont
  {Myerson}}]{cui16}%
  \BibitemOpen
  \bibfield  {author} {\bibinfo {author} {\bibfnamefont {Y.}~\bibnamefont
  {Cui}}, \bibinfo {author} {\bibfnamefont {J.}~\bibnamefont {Stojakovic}},
  \bibinfo {author} {\bibfnamefont {H.}~\bibnamefont {Kijima}}, \ and\ \bibinfo
  {author} {\bibfnamefont {A.~S.}\ \bibnamefont {Myerson}},\ }\href@noop {}
  {\bibfield  {journal} {\bibinfo  {journal} {Crys. Growth Design}\ }\textbf
  {\bibinfo {volume} {16}},\ \bibinfo {pages} {6131} (\bibinfo {year}
  {2016})}\BibitemShut {NoStop}%
\bibitem [{\citenamefont {Han}\ \emph {et~al.}(2010)\citenamefont {Han},
  \citenamefont {Poornachary}, \citenamefont {Chow},\ and\ \citenamefont
  {Tan}}]{han10}%
  \BibitemOpen
  \bibfield  {author} {\bibinfo {author} {\bibfnamefont {G.}~\bibnamefont
  {Han}}, \bibinfo {author} {\bibfnamefont {S.~K.}\ \bibnamefont
  {Poornachary}}, \bibinfo {author} {\bibfnamefont {P.~S.}\ \bibnamefont
  {Chow}}, \ and\ \bibinfo {author} {\bibfnamefont {R.~B.~H.}\ \bibnamefont
  {Tan}},\ }\href@noop {} {\bibfield  {journal} {\bibinfo  {journal} {Cryst.
  Growth Design}\ }\textbf {\bibinfo {volume} {10}},\ \bibinfo {pages} {4883}
  (\bibinfo {year} {2010})}\BibitemShut {NoStop}%
\bibitem [{\citenamefont {Li}\ and\ \citenamefont
  {Rodríguez-Hornedo}(1992)}]{li92}%
  \BibitemOpen
  \bibfield  {author} {\bibinfo {author} {\bibfnamefont {L.}~\bibnamefont
  {Li}}\ and\ \bibinfo {author} {\bibfnamefont {N.}~\bibnamefont
  {Rodríguez-Hornedo}},\ }\href {\doibase
  http://dx.doi.org/10.1016/0022-0248(92)90172-F} {\bibfield  {journal}
  {\bibinfo  {journal} {J. Cryst. Growth}\ }\textbf {\bibinfo {volume} {121}},\
  \bibinfo {pages} {33 } (\bibinfo {year} {1992})}\BibitemShut {NoStop}%
\bibitem [{\citenamefont {Dowling}\ \emph {et~al.}(2010)\citenamefont
  {Dowling}, \citenamefont {Davey}, \citenamefont {Curtis}, \citenamefont
  {Han}, \citenamefont {Poornachary}, \citenamefont {Chow},\ and\ \citenamefont
  {Tan}}]{dowling10}%
  \BibitemOpen
  \bibfield  {author} {\bibinfo {author} {\bibfnamefont {R.}~\bibnamefont
  {Dowling}}, \bibinfo {author} {\bibfnamefont {R.~J.}\ \bibnamefont {Davey}},
  \bibinfo {author} {\bibfnamefont {R.~A.}\ \bibnamefont {Curtis}}, \bibinfo
  {author} {\bibfnamefont {G.}~\bibnamefont {Han}}, \bibinfo {author}
  {\bibfnamefont {S.~K.}\ \bibnamefont {Poornachary}}, \bibinfo {author}
  {\bibfnamefont {P.~S.}\ \bibnamefont {Chow}}, \ and\ \bibinfo {author}
  {\bibfnamefont {R.~B.~H.}\ \bibnamefont {Tan}},\ }\href {\doibase
  10.1039/C0CC00336K} {\bibfield  {journal} {\bibinfo  {journal} {Chem.
  Commun.}\ }\textbf {\bibinfo {volume} {46}},\ \bibinfo {pages} {5924}
  (\bibinfo {year} {2010})}\BibitemShut {NoStop}%
\bibitem [{\citenamefont {Tsiatis}(1975)}]{tsiatis75}%
  \BibitemOpen
  \bibfield  {author} {\bibinfo {author} {\bibfnamefont {A.}~\bibnamefont
  {Tsiatis}},\ }\href@noop {} {\bibfield  {journal} {\bibinfo  {journal} {Proc.
  Nat. Acad. Sci.}\ }\textbf {\bibinfo {volume} {72}},\ \bibinfo {pages} {20}
  (\bibinfo {year} {1975})}\BibitemShut {NoStop}%
\bibitem [{\citenamefont {Peterson}(1976)}]{peterson76}%
  \BibitemOpen
  \bibfield  {author} {\bibinfo {author} {\bibfnamefont {A.~V.}\ \bibnamefont
  {Peterson}},\ }\href@noop {} {\bibfield  {journal} {\bibinfo  {journal}
  {Proc. Nat. Acad. Sci.}\ }\textbf {\bibinfo {volume} {73}},\ \bibinfo {pages}
  {11} (\bibinfo {year} {1976})}\BibitemShut {NoStop}%
\bibitem [{\citenamefont {Slud}\ and\ \citenamefont {Byar}(1988)}]{slud88}%
  \BibitemOpen
  \bibfield  {author} {\bibinfo {author} {\bibfnamefont {E.}~\bibnamefont
  {Slud}}\ and\ \bibinfo {author} {\bibfnamefont {D.}~\bibnamefont {Byar}},\
  }\href@noop {} {\bibfield  {journal} {\bibinfo  {journal} {Biometrics}\
  }\textbf {\bibinfo {volume} {44}},\ \bibinfo {pages} {265} (\bibinfo {year}
  {1988})}\BibitemShut {NoStop}%
\bibitem [{\citenamefont {Beyersmann}\ \emph {et~al.}(2009)\citenamefont
  {Beyersmann}, \citenamefont {Latouche}, \citenamefont {Buchhol},\ and\
  \citenamefont {Schumacher}}]{beyersmann09}%
  \BibitemOpen
  \bibfield  {author} {\bibinfo {author} {\bibfnamefont {J.}~\bibnamefont
  {Beyersmann}}, \bibinfo {author} {\bibfnamefont {A.}~\bibnamefont
  {Latouche}}, \bibinfo {author} {\bibfnamefont {A.}~\bibnamefont {Buchhol}}, \
  and\ \bibinfo {author} {\bibfnamefont {M.}~\bibnamefont {Schumacher}},\
  }\href@noop {} {\bibfield  {journal} {\bibinfo  {journal} {Statist. Med.}\
  }\textbf {\bibinfo {volume} {28}},\ \bibinfo {pages} {956} (\bibinfo {year}
  {2009})}\BibitemShut {NoStop}%
\bibitem [{\citenamefont {Cox}\ and\ \citenamefont {Oakes}(1984)}]{cox_book}%
  \BibitemOpen
  \bibfield  {author} {\bibinfo {author} {\bibfnamefont {D.~R.}\ \bibnamefont
  {Cox}}\ and\ \bibinfo {author} {\bibfnamefont {D.}~\bibnamefont {Oakes}},\
  }\href@noop {} {\emph {\bibinfo {title} {Analysis of Survival Data}}}\
  (\bibinfo  {publisher} {Chapman and Hall},\ \bibinfo {year}
  {1984})\BibitemShut {NoStop}%
\bibitem [{\citenamefont {Lee}(1992)}]{lee_book}%
  \BibitemOpen
  \bibfield  {author} {\bibinfo {author} {\bibfnamefont {E.~T.}\ \bibnamefont
  {Lee}},\ }\href@noop {} {\emph {\bibinfo {title} {Statistical Methods for
  Survival Data Analysis}}},\ \bibinfo {edition} {2nd}\ ed.\ (\bibinfo
  {publisher} {Wiley},\ \bibinfo {year} {1992})\BibitemShut {NoStop}%
\bibitem [{\citenamefont {Geskus}(2015)}]{geskus_book}%
  \BibitemOpen
  \bibfield  {author} {\bibinfo {author} {\bibfnamefont {R.~B.}\ \bibnamefont
  {Geskus}},\ }\href@noop {} {\emph {\bibinfo {title} {Data Analysis with
  Competing Risks and Intermediate States}}}\ (\bibinfo  {publisher} {Chapman
  and Hall / CRC},\ \bibinfo {year} {2015})\BibitemShut {NoStop}%
\bibitem [{\citenamefont {Dignam}, \citenamefont {Zhang},\ and\ \citenamefont
  {Kocherginsky}(2012)}]{dignam12}%
  \BibitemOpen
  \bibfield  {author} {\bibinfo {author} {\bibfnamefont {J.~J.}\ \bibnamefont
  {Dignam}}, \bibinfo {author} {\bibfnamefont {Q.}~\bibnamefont {Zhang}}, \
  and\ \bibinfo {author} {\bibfnamefont {M.~N.}\ \bibnamefont {Kocherginsky}},\
  }\href@noop {} {\bibfield  {journal} {\bibinfo  {journal} {Clin Cancer Res}\
  ,\ \bibinfo {pages} {2301}} (\bibinfo {year} {2012})}\BibitemShut {NoStop}%
\bibitem [{\citenamefont {Davey}\ and\ \citenamefont
  {Garside}(2000)}]{davey_book}%
  \BibitemOpen
  \bibfield  {author} {\bibinfo {author} {\bibfnamefont {R.}~\bibnamefont
  {Davey}}\ and\ \bibinfo {author} {\bibfnamefont {J.}~\bibnamefont
  {Garside}},\ }\href@noop {} {\emph {\bibinfo {title} {From Molecules to
  Crystallizers}}}\ (\bibinfo  {publisher} {Oxford University Press},\ \bibinfo
  {address} {Oxford},\ \bibinfo {year} {2000})\BibitemShut {NoStop}%
\bibitem [{\citenamefont {Ferrari}\ \emph {et~al.}(2003)\citenamefont
  {Ferrari}, \citenamefont {Davey}, \citenamefont {Cross}, \citenamefont
  {Gillon},\ and\ \citenamefont {Towler}}]{ferrari03}%
  \BibitemOpen
  \bibfield  {author} {\bibinfo {author} {\bibfnamefont {E.~S.}\ \bibnamefont
  {Ferrari}}, \bibinfo {author} {\bibfnamefont {R.~J.}\ \bibnamefont {Davey}},
  \bibinfo {author} {\bibfnamefont {W.~I.}\ \bibnamefont {Cross}}, \bibinfo
  {author} {\bibfnamefont {A.~L.}\ \bibnamefont {Gillon}}, \ and\ \bibinfo
  {author} {\bibfnamefont {C.~S.}\ \bibnamefont {Towler}},\ }\href@noop {}
  {\bibfield  {journal} {\bibinfo  {journal} {Cryst. Growth Design}\ }\textbf
  {\bibinfo {volume} {3}},\ \bibinfo {pages} {53} (\bibinfo {year}
  {2003})}\BibitemShut {NoStop}%
\bibitem [{\citenamefont {Botsaris}(1976)}]{botsaris76}%
  \BibitemOpen
  \bibfield  {author} {\bibinfo {author} {\bibfnamefont {G.~D.}\ \bibnamefont
  {Botsaris}},\ }in\ \href {\doibase 10.1007/978-1-4615-7258-9_1} {\emph
  {\bibinfo {booktitle} {Industrial Crystallization}}},\ \bibinfo {editor}
  {edited by\ \bibinfo {editor} {\bibfnamefont {J.}~\bibnamefont {Mullin}}}\
  (\bibinfo  {publisher} {Springer US},\ \bibinfo {year} {1976})\ pp.\ \bibinfo
  {pages} {3--22}\BibitemShut {NoStop}%
\bibitem [{\citenamefont {Agrawal}\ and\ \citenamefont
  {Paterson}(2015)}]{agrawal15}%
  \BibitemOpen
  \bibfield  {author} {\bibinfo {author} {\bibfnamefont {S.~G.}\ \bibnamefont
  {Agrawal}}\ and\ \bibinfo {author} {\bibfnamefont {A.~H.~J.}\ \bibnamefont
  {Paterson}},\ }\href@noop {} {\bibfield  {journal} {\bibinfo  {journal}
  {Chem. Eng. Comm.}\ }\textbf {\bibinfo {volume} {202}},\ \bibinfo {pages}
  {698} (\bibinfo {year} {2015})}\BibitemShut {NoStop}%
\end{thebibliography}

\begin{thebibliography}{22}%
\makeatletter
\providecommand \@ifxundefined [1]{%
 \@ifx{#1\undefined}
}%
\providecommand \@ifnum [1]{%
 \ifnum #1\expandafter \@firstoftwo
 \else \expandafter \@secondoftwo
 \fi
}%
\providecommand \@ifx [1]{%
 \ifx #1\expandafter \@firstoftwo
 \else \expandafter \@secondoftwo
 \fi
}%
\providecommand \natexlab [1]{#1}%
\providecommand \enquote  [1]{``#1''}%
\providecommand \bibnamefont  [1]{#1}%
\providecommand \bibfnamefont [1]{#1}%
\providecommand \citenamefont [1]{#1}%
\providecommand \href@noop [0]{\@secondoftwo}%
\providecommand \href [0]{\begingroup \@sanitize@url \@href}%
\providecommand \@href[1]{\@@startlink{#1}\@@href}%
\providecommand \@@href[1]{\endgroup#1\@@endlink}%
\providecommand \@sanitize@url [0]{\catcode `\\12\catcode `\$12\catcode
  `\&12\catcode `\#12\catcode `\^12\catcode `\_12\catcode `\%12\relax}%
\providecommand \@@startlink[1]{}%
\providecommand \@@endlink[0]{}%
\providecommand \url  [0]{\begingroup\@sanitize@url \@url }%
\providecommand \@url [1]{\endgroup\@href {#1}{\urlprefix }}%
\providecommand \urlprefix  [0]{URL }%
\providecommand \Eprint [0]{\href }%
\providecommand \doibase [0]{http://dx.doi.org/}%
\providecommand \selectlanguage [0]{\@gobble}%
\providecommand \bibinfo  [0]{\@secondoftwo}%
\providecommand \bibfield  [0]{\@secondoftwo}%
\providecommand \translation [1]{[#1]}%
\providecommand \BibitemOpen [0]{}%
\providecommand \bibitemStop [0]{}%
\providecommand \bibitemNoStop [0]{.\EOS\space}%
\providecommand \EOS [0]{\spacefactor3000\relax}%
\providecommand \BibitemShut  [1]{\csname bibitem#1\endcsname}%
\let\auto@bib@innerbib\@empty
\bibitem [{\citenamefont {Li}\ and\ \citenamefont
  {Rodríguez-Hornedo}(1992)}]{li92}%
  \BibitemOpen
  \bibfield  {author} {\bibinfo {author} {\bibfnamefont {L.}~\bibnamefont
  {Li}}\ and\ \bibinfo {author} {\bibfnamefont {N.}~\bibnamefont
  {Rodríguez-Hornedo}},\ }\href {\doibase
  http://dx.doi.org/10.1016/0022-0248(92)90172-F} {\bibfield  {journal}
  {\bibinfo  {journal} {J. Cryst. Growth}\ }\textbf {\bibinfo {volume} {121}},\
  \bibinfo {pages} {33 } (\bibinfo {year} {1992})}\BibitemShut {NoStop}%
\bibitem [{\citenamefont {Han}, \citenamefont {Chow},\ and\ \citenamefont
  {Tan}(2012)}]{han12}%
  \BibitemOpen
  \bibfield  {author} {\bibinfo {author} {\bibfnamefont {G.}~\bibnamefont
  {Han}}, \bibinfo {author} {\bibfnamefont {P.~S.}\ \bibnamefont {Chow}}, \
  and\ \bibinfo {author} {\bibfnamefont {R.~B.~H.}\ \bibnamefont {Tan}},\
  }\href@noop {} {\bibfield  {journal} {\bibinfo  {journal} {Crys. Growth
  Design}\ }\textbf {\bibinfo {volume} {12}},\ \bibinfo {pages} {2213}
  (\bibinfo {year} {2012})}\BibitemShut {NoStop}%
\bibitem [{\citenamefont {Han}, \citenamefont {Chow},\ and\ \citenamefont
  {Tan}(2015)}]{han15}%
  \BibitemOpen
  \bibfield  {author} {\bibinfo {author} {\bibfnamefont {G.}~\bibnamefont
  {Han}}, \bibinfo {author} {\bibfnamefont {P.~S.}\ \bibnamefont {Chow}}, \
  and\ \bibinfo {author} {\bibfnamefont {R.~B.~H.}\ \bibnamefont {Tan}},\
  }\href@noop {} {\bibfield  {journal} {\bibinfo  {journal} {Cryst. Growth
  Design}\ }\textbf {\bibinfo {volume} {15}},\ \bibinfo {pages} {1082}
  (\bibinfo {year} {2015})}\BibitemShut {NoStop}%
\bibitem [{\citenamefont {Dowling}\ \emph {et~al.}(2010)\citenamefont
  {Dowling}, \citenamefont {Davey}, \citenamefont {Curtis}, \citenamefont
  {Han}, \citenamefont {Poornachary}, \citenamefont {Chow},\ and\ \citenamefont
  {Tan}}]{dowling10}%
  \BibitemOpen
  \bibfield  {author} {\bibinfo {author} {\bibfnamefont {R.}~\bibnamefont
  {Dowling}}, \bibinfo {author} {\bibfnamefont {R.~J.}\ \bibnamefont {Davey}},
  \bibinfo {author} {\bibfnamefont {R.~A.}\ \bibnamefont {Curtis}}, \bibinfo
  {author} {\bibfnamefont {G.}~\bibnamefont {Han}}, \bibinfo {author}
  {\bibfnamefont {S.~K.}\ \bibnamefont {Poornachary}}, \bibinfo {author}
  {\bibfnamefont {P.~S.}\ \bibnamefont {Chow}}, \ and\ \bibinfo {author}
  {\bibfnamefont {R.~B.~H.}\ \bibnamefont {Tan}},\ }\href {\doibase
  10.1039/C0CC00336K} {\bibfield  {journal} {\bibinfo  {journal} {Chem.
  Commun.}\ }\textbf {\bibinfo {volume} {46}},\ \bibinfo {pages} {5924}
  (\bibinfo {year} {2010})}\BibitemShut {NoStop}%
\bibitem [{\citenamefont {Sultana}\ and\ \citenamefont
  {Jensen}(2012)}]{sultana12}%
  \BibitemOpen
  \bibfield  {author} {\bibinfo {author} {\bibfnamefont {M.}~\bibnamefont
  {Sultana}}\ and\ \bibinfo {author} {\bibfnamefont {K.~F.}\ \bibnamefont
  {Jensen}},\ }\href@noop {} {\bibfield  {journal} {\bibinfo  {journal} {Cryst.
  Growth Design}\ }\textbf {\bibinfo {volume} {12}},\ \bibinfo {pages} {6260}
  (\bibinfo {year} {2012})}\BibitemShut {NoStop}%
\bibitem [{\citenamefont {Toldy}\ \emph {et~al.}(2012)\citenamefont {Toldy},
  \citenamefont {Badruddoza}, \citenamefont {Zheng}, \citenamefont {Hatton},
  \citenamefont {Gunawan}, \citenamefont {Rajagopalan},\ and\ \citenamefont
  {Khan}}]{toldy12}%
  \BibitemOpen
  \bibfield  {author} {\bibinfo {author} {\bibfnamefont {A.~I.}\ \bibnamefont
  {Toldy}}, \bibinfo {author} {\bibfnamefont {A.~Z.~M.}\ \bibnamefont
  {Badruddoza}}, \bibinfo {author} {\bibfnamefont {L.}~\bibnamefont {Zheng}},
  \bibinfo {author} {\bibfnamefont {T.~A.}\ \bibnamefont {Hatton}}, \bibinfo
  {author} {\bibfnamefont {R.}~\bibnamefont {Gunawan}}, \bibinfo {author}
  {\bibfnamefont {R.}~\bibnamefont {Rajagopalan}}, \ and\ \bibinfo {author}
  {\bibfnamefont {S.~A.}\ \bibnamefont {Khan}},\ }\href@noop {} {\bibfield
  {journal} {\bibinfo  {journal} {Crys. Growth Des.}\ }\textbf {\bibinfo
  {volume} {12}},\ \bibinfo {pages} {3977} (\bibinfo {year}
  {2012})}\BibitemShut {NoStop}%
\bibitem [{\citenamefont {Yang}, \citenamefont {Wang},\ and\ \citenamefont
  {Ching}(2008)}]{yang08b}%
  \BibitemOpen
  \bibfield  {author} {\bibinfo {author} {\bibfnamefont {X.}~\bibnamefont
  {Yang}}, \bibinfo {author} {\bibfnamefont {X.}~\bibnamefont {Wang}}, \ and\
  \bibinfo {author} {\bibfnamefont {C.~B.}\ \bibnamefont {Ching}},\ }\href@noop
  {} {\bibfield  {journal} {\bibinfo  {journal} {J. Chem. Eng. Data}\ }\textbf
  {\bibinfo {volume} {53}},\ \bibinfo {pages} {1133} (\bibinfo {year}
  {2008})}\BibitemShut {NoStop}%
\bibitem [{\citenamefont {Yi}\ \emph {et~al.}(2005)\citenamefont {Yi},
  \citenamefont {Hatziavramidis}, \citenamefont {Myerson}, \citenamefont
  {Waldo}, \citenamefont {Beylin},\ and\ \citenamefont {Mustakis}}]{yi05}%
  \BibitemOpen
  \bibfield  {author} {\bibinfo {author} {\bibfnamefont {Y.}~\bibnamefont
  {Yi}}, \bibinfo {author} {\bibfnamefont {D.}~\bibnamefont {Hatziavramidis}},
  \bibinfo {author} {\bibfnamefont {A.~S.}\ \bibnamefont {Myerson}}, \bibinfo
  {author} {\bibfnamefont {M.}~\bibnamefont {Waldo}}, \bibinfo {author}
  {\bibfnamefont {V.~G.}\ \bibnamefont {Beylin}}, \ and\ \bibinfo {author}
  {\bibfnamefont {J.}~\bibnamefont {Mustakis}},\ }\href@noop {} {\bibfield
  {journal} {\bibinfo  {journal} {Ind. Eng. Chem. Res.}\ }\textbf {\bibinfo
  {volume} {44}},\ \bibinfo {pages} {5427} (\bibinfo {year}
  {2005})}\BibitemShut {NoStop}%
\bibitem [{\citenamefont {Shi}\ and\ \citenamefont {Wang}(2005)}]{shi05}%
  \BibitemOpen
  \bibfield  {author} {\bibinfo {author} {\bibfnamefont {Y.}~\bibnamefont
  {Shi}}\ and\ \bibinfo {author} {\bibfnamefont {L.}~\bibnamefont {Wang}},\
  }\href {http://stacks.iop.org/0022-3727/38/i=19/a=024} {\bibfield  {journal}
  {\bibinfo  {journal} {J. Phys. D App. Phys.}\ }\textbf {\bibinfo {volume}
  {38}},\ \bibinfo {pages} {3741} (\bibinfo {year} {2005})}\BibitemShut
  {NoStop}%
\bibitem [{\citenamefont {Little}, \citenamefont {Sear},\ and\ \citenamefont
  {Keddie}(2015)}]{little15}%
  \BibitemOpen
  \bibfield  {author} {\bibinfo {author} {\bibfnamefont {L.~J.}\ \bibnamefont
  {Little}}, \bibinfo {author} {\bibfnamefont {R.~P.}\ \bibnamefont {Sear}}, \
  and\ \bibinfo {author} {\bibfnamefont {J.~L.}\ \bibnamefont {Keddie}},\
  }\href@noop {} {\bibfield  {journal} {\bibinfo  {journal} {Crys. Growth
  Design}\ }\textbf {\bibinfo {volume} {15}},\ \bibinfo {pages} {5345}
  (\bibinfo {year} {2015})}\BibitemShut {NoStop}%
\bibitem [{\citenamefont {Weissbuch}\ \emph {et~al.}(2005)\citenamefont
  {Weissbuch}, \citenamefont {Torbeev}, \citenamefont {Leiserowitz},\ and\
  \citenamefont {Lahav}}]{weissbuch05}%
  \BibitemOpen
  \bibfield  {author} {\bibinfo {author} {\bibfnamefont {I.}~\bibnamefont
  {Weissbuch}}, \bibinfo {author} {\bibfnamefont {V.~Y.}\ \bibnamefont
  {Torbeev}}, \bibinfo {author} {\bibfnamefont {L.}~\bibnamefont
  {Leiserowitz}}, \ and\ \bibinfo {author} {\bibfnamefont {M.}~\bibnamefont
  {Lahav}},\ }\href@noop {} {\bibfield  {journal} {\bibinfo  {journal} {Ang.
  Chem. Int. Ed.}\ }\textbf {\bibinfo {volume} {44}},\ \bibinfo {pages} {3226}
  (\bibinfo {year} {2005})}\BibitemShut {NoStop}%
\bibitem [{\citenamefont {Seyedhosseini}\ \emph {et~al.}(2014)\citenamefont
  {Seyedhosseini}, \citenamefont {Ivanov}, \citenamefont {Bystrov},
  \citenamefont {Bdikin}, \citenamefont {Zelenovskiy}, \citenamefont {Shur},
  \citenamefont {Kudryavtsev}, \citenamefont {Mishina}, \citenamefont {Sigov},\
  and\ \citenamefont {Kholkin}}]{seyedhosseini14}%
  \BibitemOpen
  \bibfield  {author} {\bibinfo {author} {\bibfnamefont {E.}~\bibnamefont
  {Seyedhosseini}}, \bibinfo {author} {\bibfnamefont {M.}~\bibnamefont
  {Ivanov}}, \bibinfo {author} {\bibfnamefont {V.}~\bibnamefont {Bystrov}},
  \bibinfo {author} {\bibfnamefont {I.}~\bibnamefont {Bdikin}}, \bibinfo
  {author} {\bibfnamefont {P.}~\bibnamefont {Zelenovskiy}}, \bibinfo {author}
  {\bibfnamefont {V.~Y.}\ \bibnamefont {Shur}}, \bibinfo {author}
  {\bibfnamefont {A.}~\bibnamefont {Kudryavtsev}}, \bibinfo {author}
  {\bibfnamefont {E.~D.}\ \bibnamefont {Mishina}}, \bibinfo {author}
  {\bibfnamefont {A.~S.}\ \bibnamefont {Sigov}}, \ and\ \bibinfo {author}
  {\bibfnamefont {A.~L.}\ \bibnamefont {Kholkin}},\ }\href@noop {} {\bibfield
  {journal} {\bibinfo  {journal} {Cryst. Growth Design}\ }\textbf {\bibinfo
  {volume} {14}},\ \bibinfo {pages} {2831} (\bibinfo {year}
  {2014})}\BibitemShut {NoStop}%
\bibitem [{\citenamefont {Isakov}\ \emph {et~al.}(2014)\citenamefont {Isakov},
  \citenamefont {Petukhova}, \citenamefont {Vasilev}, \citenamefont {Nuraeva},
  \citenamefont {Khazamov}, \citenamefont {Seyedhosseini}, \citenamefont
  {Zelenovskiy}, \citenamefont {Shur},\ and\ \citenamefont
  {Kholkin}}]{isakov14}%
  \BibitemOpen
  \bibfield  {author} {\bibinfo {author} {\bibfnamefont {D.}~\bibnamefont
  {Isakov}}, \bibinfo {author} {\bibfnamefont {D.}~\bibnamefont {Petukhova}},
  \bibinfo {author} {\bibfnamefont {S.}~\bibnamefont {Vasilev}}, \bibinfo
  {author} {\bibfnamefont {A.}~\bibnamefont {Nuraeva}}, \bibinfo {author}
  {\bibfnamefont {T.}~\bibnamefont {Khazamov}}, \bibinfo {author}
  {\bibfnamefont {E.}~\bibnamefont {Seyedhosseini}}, \bibinfo {author}
  {\bibfnamefont {P.}~\bibnamefont {Zelenovskiy}}, \bibinfo {author}
  {\bibfnamefont {V.~Y.}\ \bibnamefont {Shur}}, \ and\ \bibinfo {author}
  {\bibfnamefont {A.~L.}\ \bibnamefont {Kholkin}},\ }\href {\doibase
  10.1021/cg500747x} {\bibfield  {journal} {\bibinfo  {journal} {Cryst. Growth
  Design}\ }\textbf {\bibinfo {volume} {14}},\ \bibinfo {pages} {4138}
  (\bibinfo {year} {2014})}\BibitemShut {NoStop}%
\bibitem [{\citenamefont {Tsiatis}(1975)}]{tsiatis75}%
  \BibitemOpen
  \bibfield  {author} {\bibinfo {author} {\bibfnamefont {A.}~\bibnamefont
  {Tsiatis}},\ }\href@noop {} {\bibfield  {journal} {\bibinfo  {journal} {Proc.
  Nat. Acad. Sci.}\ }\textbf {\bibinfo {volume} {72}},\ \bibinfo {pages} {20}
  (\bibinfo {year} {1975})}\BibitemShut {NoStop}%
\bibitem [{\citenamefont {Peterson}(1976)}]{peterson76}%
  \BibitemOpen
  \bibfield  {author} {\bibinfo {author} {\bibfnamefont {A.~V.}\ \bibnamefont
  {Peterson}},\ }\href@noop {} {\bibfield  {journal} {\bibinfo  {journal}
  {Proc. Nat. Acad. Sci.}\ }\textbf {\bibinfo {volume} {73}},\ \bibinfo {pages}
  {11} (\bibinfo {year} {1976})}\BibitemShut {NoStop}%
\bibitem [{\citenamefont {Slud}\ and\ \citenamefont {Byar}(1988)}]{slud88}%
  \BibitemOpen
  \bibfield  {author} {\bibinfo {author} {\bibfnamefont {E.}~\bibnamefont
  {Slud}}\ and\ \bibinfo {author} {\bibfnamefont {D.}~\bibnamefont {Byar}},\
  }\href@noop {} {\bibfield  {journal} {\bibinfo  {journal} {Biometrics}\
  }\textbf {\bibinfo {volume} {44}},\ \bibinfo {pages} {265} (\bibinfo {year}
  {1988})}\BibitemShut {NoStop}%
\bibitem [{\citenamefont {Beyersmann}\ \emph {et~al.}(2009)\citenamefont
  {Beyersmann}, \citenamefont {Latouche}, \citenamefont {Buchhol},\ and\
  \citenamefont {Schumacher}}]{beyersmann09}%
  \BibitemOpen
  \bibfield  {author} {\bibinfo {author} {\bibfnamefont {J.}~\bibnamefont
  {Beyersmann}}, \bibinfo {author} {\bibfnamefont {A.}~\bibnamefont
  {Latouche}}, \bibinfo {author} {\bibfnamefont {A.}~\bibnamefont {Buchhol}}, \
  and\ \bibinfo {author} {\bibfnamefont {M.}~\bibnamefont {Schumacher}},\
  }\href@noop {} {\bibfield  {journal} {\bibinfo  {journal} {Statist. Med.}\
  }\textbf {\bibinfo {volume} {28}},\ \bibinfo {pages} {956} (\bibinfo {year}
  {2009})}\BibitemShut {NoStop}%
\bibitem [{\citenamefont {Andersen}\ \emph {et~al.}(2012)\citenamefont
  {Andersen}, \citenamefont {Geskus}, \citenamefont {de~Witte},\ and\
  \citenamefont {Putter}}]{andersen12}%
  \BibitemOpen
  \bibfield  {author} {\bibinfo {author} {\bibfnamefont {P.~K.}\ \bibnamefont
  {Andersen}}, \bibinfo {author} {\bibfnamefont {R.~B.}\ \bibnamefont
  {Geskus}}, \bibinfo {author} {\bibfnamefont {T.}~\bibnamefont {de~Witte}}, \
  and\ \bibinfo {author} {\bibfnamefont {H.}~\bibnamefont {Putter}},\ }\href
  {\doibase 10.1093/ije/dyr213} {\bibfield  {journal} {\bibinfo  {journal}
  {Int. J. Epidemiology}\ }\textbf {\bibinfo {volume} {41}},\ \bibinfo {pages}
  {861} (\bibinfo {year} {2012})}\BibitemShut {NoStop}%
\bibitem [{\citenamefont {Dignam}, \citenamefont {Zhang},\ and\ \citenamefont
  {Korcherginsky}(2012)}]{dignam12}%
  \BibitemOpen
  \bibfield  {author} {\bibinfo {author} {\bibfnamefont {J.~J.}\ \bibnamefont
  {Dignam}}, \bibinfo {author} {\bibfnamefont {Q.}~\bibnamefont {Zhang}}, \
  and\ \bibinfo {author} {\bibfnamefont {M.~N.}\ \bibnamefont
  {Korcherginsky}},\ }\href@noop {} {\bibfield  {journal} {\bibinfo  {journal}
  {Clin Cancer Res}\ ,\ \bibinfo {pages} {2301}} (\bibinfo {year}
  {2012})}\BibitemShut {NoStop}%
\bibitem [{\citenamefont {Geskus}(2015)}]{geskus_book}%
  \BibitemOpen
  \bibfield  {author} {\bibinfo {author} {\bibfnamefont {R.~B.}\ \bibnamefont
  {Geskus}},\ }\href@noop {} {\emph {\bibinfo {title} {Data Analysis with
  Competing Risks and Intermediate States}}}\ (\bibinfo  {publisher} {Chapman
  and Hall / CRC},\ \bibinfo {year} {2015})\BibitemShut {NoStop}%
\bibitem [{\citenamefont {Little}(2017)}]{laurie_thesis}%
  \BibitemOpen
  \bibfield  {author} {\bibinfo {author} {\bibfnamefont {L.~J.}\ \bibnamefont
  {Little}},\ }\href@noop {} {\emph {\bibinfo {title} {A tortoise and the hare
  story: The relationship between induction time and polymorphism in glycine
  crystallisation}}}\ (\bibinfo  {publisher} {PhD thesis, University of
  Surrey},\ \bibinfo {year} {2017})\BibitemShut {NoStop}%
\bibitem [{\citenamefont {Hoffman}(2015)}]{hoffman15}%
  \BibitemOpen
  \bibfield  {author} {\bibinfo {author} {\bibfnamefont {J.~I.}\ \bibnamefont
  {Hoffman}},\ }in\ \href {\doibase
  http://dx.doi.org/10.1016/B978-0-12-802387-7.00029-9} {\emph {\bibinfo
  {booktitle} {Biostatistics for Medical and Biomedical Practitioners}}},\
  \bibinfo {editor} {edited by\ \bibinfo {editor} {\bibfnamefont {J.~I.}\
  \bibnamefont {Hoffman}}}\ (\bibinfo  {publisher} {Academic Press},\ \bibinfo
  {year} {2015})\ pp.\ \bibinfo {pages} {513 -- 536}\BibitemShut {NoStop}%
\end{thebibliography}

%

\end{document}